\RequirePackage{fixltx2e}
\documentclass[nofootinbib,reprint,showpacs,noeprint,
  prc,aps,10pt,superscriptaddress,altaffilletter,floatfix]{revtex4-1}
\pdfoutput=1
\usepackage{graphicx}
\usepackage{hyperref}
\usepackage{epstopdf}
\usepackage{amsmath}
\usepackage{amssymb}
\usepackage[amsmath,thmmarks]{ntheorem}
\usepackage{bm}
\usepackage{cleveref}
\usepackage{subfigure}
\usepackage{epstopdf}
\usepackage{array}
\usepackage{ifthen}
\usepackage{booktabs}
\usepackage{textgreek}
\usepackage{multirow}
\usepackage{enumitem}
\usepackage{dcolumn}
\newcolumntype{C}[1]{>{\centering\let\newline\\\arraybackslash\hspace{0pt}}m{#1}}

\usepackage{color}

\graphicspath{{fig/Detector/}{fig/DataQuality/}{fig/Reconstruction/}{fig/MonteCarlo/}{fig/CalibCorrection/}{fig/Physics/}}

\RequirePackage{lineno}

\begin{document}

\newcommand{\isot}[2]{$^{#2}$#1}
\newcommand{\xeiso}{\isot{Xe}{136}}
\newcommand{\thsrc}{\isot{Th}{228}}
\newcommand{\cosrc}{\isot{Co}{60}}
\newcommand{\nonubb}  {$0\nu \beta \beta$}
\newcommand{\twonubb} {$2\nu \beta \beta$}
\newcommand{\vadc} {ADC$_\text{V}$}
\newcommand{\uadc} {ADC$_\text{U}$}
\newcommand{\mus} {\textmu{}s}
\newcommand{\chisq} {$\chi^2$}
\newcommand{\mum} {\textmu{}m}
\newcommand{\checkit}[1]{{\color{red}#1}}
\newcommand{\RunTwoA}{Run 2a}
\newcommand{\SP}[1]{\textsuperscript{#1}}
\newcommand{\SB}[1]{\textsubscript{#1}}
\newcommand{\SPSB}[2]{\rlap{\textsuperscript{#1}}\SB{#2}}
\newcommand{\pmasy}[3]{#1\SPSB{$+$#2}{$-$#3}}
\newcommand{\matel}{$M^{2\nu}$}
\newcommand{\psfac}{$G^{2\nu}$}
\newcommand{\exomeasval}[1][true]{$2.165 \pm 0.016 \ifthenelse{\equal{#1}{true}}{{\rm (stat)}}{} \pm 0.059 \ifthenelse{\equal{#1}{true}}{{\rm (sys)}}{} \cdot 10^{21}$}
\newcommand{\exomeasurement}{T$_{1/2}^{2\nu\beta\beta}$ = \exomeasval{}~years}
\newcommand{\U}{\text{U}}
\newcommand{\V}{\text{V}}
\newcommand{\X}{\text{X}}
\newcommand{\Y}{\text{Y}}
\newcommand{\Z}{\text{Z}}


\title{An improved measurement of the \texorpdfstring{\twonubb{}}{2nubb} half-life of \texorpdfstring{$^{136}$Xe}{Xe-136} with EXO-200}

\newcommand{\IHEP}{\affiliation{Institute of High Energy Physics, Beijing, China}}
\newcommand{\Seoul}{\affiliation{Department of Physics, University of Seoul, Seoul, Korea}}
\newcommand{\Stanford}{\affiliation{Physics Department, Stanford University, Stanford CA, USA}}
\newcommand{\WIPP}{\affiliation{Waste Isolation Pilot Plant, Carlsbad NM, USA}}
\newcommand{\Laurentian}{\affiliation{Department of Physics, Laurentian University, Sudbury ON, Canada}}
\newcommand{\ITEP}{\affiliation{Institute for Theoretical and Experimental Physics, Moscow, Russia}}
\newcommand{\Illinois}{\affiliation{Physics Department, University of Illinois, Urbana-Champaign IL, USA}}
\newcommand{\Caltech}{\affiliation{Kellogg Lab, Caltech, Pasadena, CA, USA}}
\newcommand{\UMass}{\affiliation{Physics Department, University of Massachusetts, Amherst MA, USA}}
\newcommand{\TUM}{\affiliation{Technische Universit\"at M\"unchen, Physikdepartment and Excellence Cluster Universe, Garching, Germany}}
\newcommand{\Maryland}{\affiliation{Physics Department, University of Maryland, College Park MD, USA}}
\newcommand{\LHEP}{\affiliation{LHEP, Albert Einstein Center, University of Bern, Bern, Switzerland}}
\newcommand{\CSU}{\affiliation{Physics Department, Colorado State University, Fort Collins CO, USA}}
\newcommand{\Indiana}{\affiliation{Physics Department and CEEM, Indiana University, Bloomington IN, USA}}
\newcommand{\Alabama}{\affiliation{Department of Physics and Astronomy, University of Alabama, Tuscaloosa AL, USA}}
\newcommand{\Drexel}{\affiliation{Department of Physics, Drexel University, Philadelphia PA, USA}}
\newcommand{\SLAC}{\affiliation{SLAC National Accelerator Laboratory, Stanford CA, USA}}
\newcommand{\Carleton}{\affiliation{Physics Department, Carleton University, Ottawa ON, Canada}}
\author{J.B.~Albert}\Indiana
\author{M.~Auger}\LHEP
\author{D.J.~Auty}\Alabama
\author{P.S.~Barbeau}\Stanford
\author{E.~Beauchamp}\Laurentian
\author{D.~Beck}\Illinois
\author{V.~Belov}\ITEP
\author{C.~Benitez-Medina}\CSU
\author{J.~Bonatt}\UMass\Stanford
\author{M.~Breidenbach}\SLAC
\author{T.~Brunner}\Stanford
\author{A.~Burenkov}\ITEP
\author{G.F.~Cao}\IHEP
\author{C.~Chambers}\CSU
\author{J.~Chaves}\Stanford
\author{B.~Cleveland}\altaffiliation{Also SNOLAB, Sudbury ON, Canada}\Laurentian
\author{S.~Cook}\CSU
\author{A.~Craycraft}\CSU
\author{T.~Daniels}\UMass
\author{M.~Danilov}\ITEP
\author{S.J.~Daugherty}\Indiana
\author{C.G.~Davis}\Maryland
\author{J.~Davis}\Stanford
\author{R.~DeVoe}\Stanford
\author{S.~Delaquis}\LHEP
\author{A.~Dobi}\Maryland
\author{A.~Dolgolenko}\ITEP
\author{M.J.~Dolinski}\Drexel
\author{M.~Dunford}\Carleton
\author{W.~Fairbank Jr.}\CSU
\author{J.~Farine}\Laurentian
\author{W.~Feldmeier}\TUM
\author{P.~Fierlinger}\TUM
\author{D.~Franco}\LHEP
\author{D.~Fudenberg}\Stanford
\author{G.~Giroux}\LHEP
\author{R.~Gornea}\LHEP
\author{K.~Graham}\Carleton
\author{G.~Gratta}\Stanford
\author{C.~Hall}\Maryland
\author{K.~Hall}\CSU
\author{C.~Hargrove}\Carleton
\author{S.~Herrin}\SLAC
\author{M.~Hughes}\Alabama
\author{X.S.~Jiang}\IHEP
\author{A.~Johnson}\SLAC
\author{T.N.~Johnson}\Indiana
\author{S.~Johnston}\UMass
\author{A.~Karelin}\ITEP
\author{L.J.~Kaufman}\Indiana
\author{R.~Killick}\Carleton
\author{S.~Kravitz}\Stanford
\author{A.~Kuchenkov}\ITEP
\author{K.S.~Kumar}\UMass
\author{D.S.~Leonard}\Seoul
\author{F.~Leonard}\Carleton
\author{C.~Licciardi}\Carleton
\author{R.~MacLellan}\SLAC
\author{M.G.~Marino}\altaffiliation{Corresponding author: \href{mailto:michael.marino@mytum.de}{michael.marino.mytum.de}}\TUM
\author{B.~Mong}\Laurentian
\author{M.~Montero D\'{i}ez}\Stanford
\author{D.~Moore}\Stanford
\author{R.~Nelson}\WIPP
\author{K.~O'Sullivan}\altaffiliation{Now at Yale University, New Haven, CT, USA}\Stanford
\author{A.~Odian}\SLAC
\author{I.~Ostrovskiy}\Stanford
\author{C.~Ouellet}\Carleton
\author{A.~Piepke}\Alabama
\author{A.~Pocar}\UMass
\author{C.Y.~Prescott}\SLAC
\author{A.~Rivas}\Stanford
\author{P.C.~Rowson}\SLAC
\author{M.P.~Rozo}\Carleton
\author{J.J.~Russell}\SLAC
\author{A.~Sabourov}\altaffiliation{Now at Air Force Technical Applications Center, Patrick AFB, FL, USA}\Stanford
\author{D.~Sinclair}\altaffiliation{Also TRIUMF, Vancouver BC, Canada}\Carleton
\author{K.~Skarpaas}\SLAC
\author{S.~Slutsky}\Maryland
\author{V.~Stekhanov}\ITEP
\author{V.~Strickland}\altaffiliation{Also TRIUMF, Vancouver BC, Canada}\Carleton
\author{M.~Tarka}\Illinois
\author{T.~Tolba}\LHEP
\author{D.~Tosi}\Stanford
\author{K.~Twelker}\Stanford
\author{P.~Vogel}\Caltech
\author{J.-L.~Vuilleumier}\LHEP
\author{A.~Waite}\SLAC
\author{J.~Walton}\Illinois
\author{T.~Walton}\CSU
\author{M.~Weber}\LHEP
\author{L.J.~Wen}\Stanford
\author{U.~Wichoski}\Laurentian
\author{J.~Wodin}\altaffiliation{Now at SRI International, Menlo Park, CA, USA}\SLAC
\author{J.D.~Wright}\UMass
\author{L.~Yang}\Illinois
\author{Y.-R.~Yen}\Maryland
\author{O.Ya.~Zeldovich}\ITEP
\author{Y.B.~Zhao}\IHEP

\collaboration{EXO Collaboration}
\noaffiliation

\date{\today}

\begin{abstract}

We report on an improved measurement of the \twonubb{} half-life of $^{136}$Xe
performed by EXO-200.  The use of a large and homogeneous time
projection chamber allows for the precise estimate of the fiducial mass used
for the measurement, resulting in a small systematic uncertainty.  We also
discuss in detail the data analysis methods used for double-beta decay
searches with EXO-200, while emphasizing those directly related to the present
measurement.   The $^{136}$Xe \twonubb{} half-life is found to be
\exomeasurement{}. This is the most precisely measured half-life of any
\twonubb{} decay to date.

\end{abstract}


\maketitle


\section{Introduction}\label{sec:Introduction}

Nuclear $\beta\beta$ decay is a well known second-order weak transition which may occur in a number of even-even nuclei. The two-neutrino decay mode (\twonubb{}) has been directly observed in nine nuclei with half-lives ranging between $10^{18}$ and $10^{21}$ years~\cite{PhysRevD.86.010001,Barabash:2010ie,Ackerman:2011gz,KamLANDZen:2012aa}, and half-lives as long as $10^{24}$ years have been established indirectly through radiochemical and geochemical means (see Ref.~\cite{Barabash:2010ie} for a review). 
Following Ref.~\cite{Kotila:2012zza}, the \twonubb{}  half-life $T^{2\nu}_{1/2}$ can be related to the nuclear matrix element $M^{2\nu}$ and the known phase space factor $G^{2\nu}$ according to

\begin{equation}
	\frac{1}{T^{2\nu}_{1/2}} = G^{2\nu} g_A^4 m_e^2 |M^{2\nu}|^2 
  \label{eqn:halflifetwonu}
\end{equation}

\noindent
where $g_A = 1.2701$ and $m_e$ is the mass of the electron. We see from \cref{eqn:halflifetwonu} 
that measurements of \twonubb{} half-lives effectively measure 
the $M^{2\nu}$ matrix elements for $\beta\beta$ source isotopes. 

The interest in $\beta\beta$ decay is, of course, largely driven by the possibility of discovering lepton-number violation via the exotic neutrinoless mode (\nonubb{}).   The observation of the \nonubb{} decay would profoundly alter our understanding of the neutrino sector by demonstrating the Majorana nature of neutrinos and by providing information about the absolute scale of the neutrino mass spectrum.

Perhaps the simplest and most promising mechanism for the \nonubb{} mode is the virtual exchange of light but massive Majorana neutrinos.  The half-life for the neutrinoless mode, when mediated by such massive Majorana neutrino exchange, is given by 

\begin{equation}
	\frac{1}{T^{0\nu}_{1/2}}= G^{0\nu} | M^{0\nu} |^2 | \langle m_{\beta\beta}  \rangle | ^2  
  \label{eqn:halflifenonu}
\end{equation}

\noindent 
where $G^{0\nu}$ is the phase space factor, $M^{0\nu}$ is the nuclear matrix element, and $\langle m_{\beta\beta} \rangle$  is the effective neutrino mass. While the $G^{0\nu}$ are known for all nuclei of interest, the corresponding nuclear matrix elements need be calculated and  have a substantial theoretical uncertainty~\cite{Vogel:2012ja}. 
Since the experimental data are available across a variety of complex nuclei, measurements of  $2\nu\beta\beta$ half-lives provide an important challenge to all types of nuclear structure models and thus also to their ability to correctly evaluate the $M^{0\nu}$ matrix elements.
Moreover, some authors have argued that, although the physics of the two $\beta\beta$ decay modes are quite different, nevertheless measurements of the $2\nu\beta\beta$ half-lives can constrain the particle-particle coupling constant $g_{pp}$  within the  quasiparticle random-phase approximation (QRPA)~\cite{Rodin:2007fz}. This mitigates theoretical instabilities and uncertainties. 

The $M^{2\nu}$ matrix elements  are perhaps even more difficult to evaluate theoretically than $M^{0\nu}$ because the momentum transfer in $2\nu\beta\beta$ decay is comparable to the typical $Q$ value involved, whereas for $0\nu\beta\beta$ decay it is on average considerably larger. As a consequence, the $M^{2\nu}$ of the various nuclei are known to vary by up to an order of magnitude,  whereas the $M^{0\nu}$ appear to be  rather similar to each other. As a result, establishing or substantially constraining the \nonubb{} decay rate in a given nucleus allows one to project the corresponding rate in other nuclei. 

Apart from these physics considerations, precise determinations of the \twonubb{} half-life, such as the one reported here, provide, to a large extent, a validation of the techniques that are also used for the measurement of the \nonubb{} decay and the exceedingly low backgrounds obtained in today's experiments.  

Among the several isotopes being most actively pursued for 
$\beta\beta$ experiments at the multi-kilogram scale 
and beyond, $^{136}$Xe was the last to have its \twonubb{} decay 
observed~\cite{Ackerman:2011gz,KamLANDZen:2012aa}, 
in part because its small matrix element leads to a long
half-life. In this article we report on a substantially improved measurement 
of the \twonubb{} half-life of $^{136}$Xe by the EXO-200 experiment. 
EXO-200 has published results from two datasets, the first collected
between May 21, 2011 and July 9, 2011 (Run 1), and the second collected between
September 22, 2011 and April 15, 2012 (Run 2a). Run 1 data produced 
the first measurement of the \twonubb{} half-life of $^{136}$Xe~\cite{Ackerman:2011gz}
while Run 2a data placed a new limit on the existence of the 
\nonubb{} decay~\cite{Auger:2012ar} in substantial tension with the
observation claim reported in Ref.~\cite{KlapdorKleingrothaus:2006ff}. 
These results have been confirmed and complemented by KamLAND-Zen~\cite{KamLANDZen:2012aa,Gando:2012zm}.

In the interim period between Run 1 and Run 2a 
the EXO-200 front-end electronics were improved, the lead shield was completed, and 
the electronegative impurity content of the xenon was reduced by a factor of ten. 
Because of these improvements, the results presented in 
this article are based upon the more powerful Run 2a data set
and take advantage of significant additional improvements to the event reconstruction 
methods, the Monte Carlo simulation, and a greater understanding of the 
detector response and behavior acquired since the publication of 
Ref.~\cite{Auger:2012ar}.

\section{Detector description} \label{sec:Detector}
\subsection{EXO-200 detector} 

The EXO-200 detector has been described in detail elsewhere~\cite{Auger:2012gs}. 
Here we review those features of the detector which are of particular relevance 
for the measurement of the \twonubb{} decay rate. We give special attention to the
detector geometry, readout scheme, and shielding.

\begin{figure}
    \includegraphics[width=0.46\textwidth]{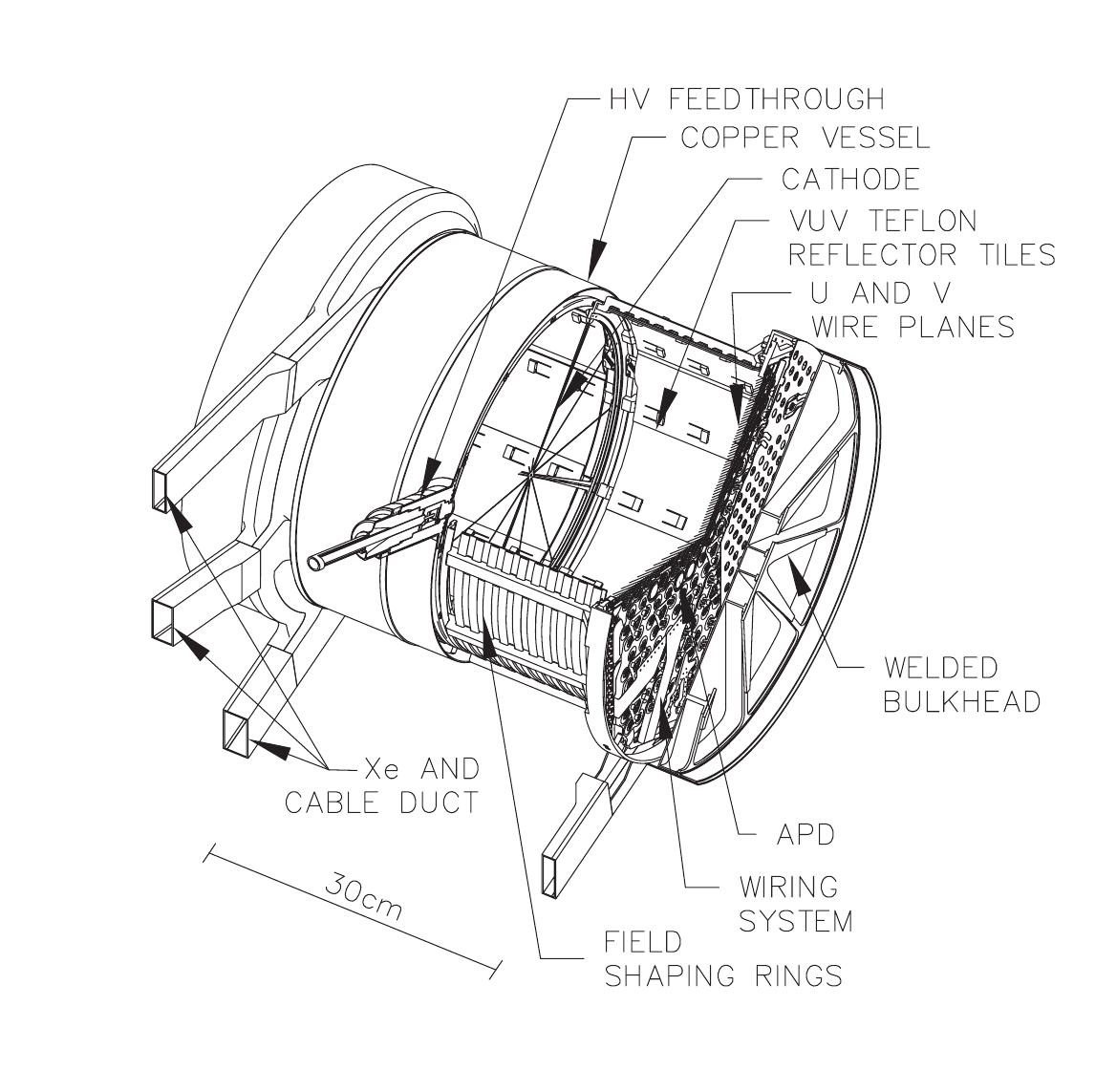}
    \caption{Cutaway view of the EXO-200 TPC.}
	\label{fig:tpc-cutaway}
\end{figure}

The centerpiece of the experiment is a liquid xenon 
(LXe) time projection chamber (TPC), 
as shown in \cref{fig:tpc-cutaway}. 
The EXO-200 LXe is enriched to
$80.672 \pm 0.14$\% in \xeiso{} (\cref{sec:XeRelatedErrs}), 
with the balance being composed mostly of $^{134}$Xe.  
LXe is a good ionizing radiation detection medium because it
produces substantial ionization and scintillation signals. 
The LXe is housed in a cylindrical 
copper vessel of length $\sim44$~cm and diameter $\sim40$~cm, and 
is instrumented by two back-to-back TPCs which share 
a common cathode at the center of the vessel. 
The end caps of the vessel host identical detector packages, 
each of which consists of two crossed and segmented wire grids 
and an array of large 
area avalanche photodiodes (APDs)~\cite{Neilson:2009kf}. 

Energy deposits in the LXe produce free ionization 
charge and scintillation light (at 178~nm). 
The charge drifts along the axis of the 
detector towards the nearest end cap under the action of a uniform electric field,
and the scintillation light is collected and measured 
by the APD arrays. Because the cathode has an optical transparency of 
90\% (at normal incidence), the scintillation light
is detected simultaneously by both APD arrays while the ionization is 
detected only in the TPC in which it was produced.

\begin{figure}
    \includegraphics[width=0.46\textwidth]{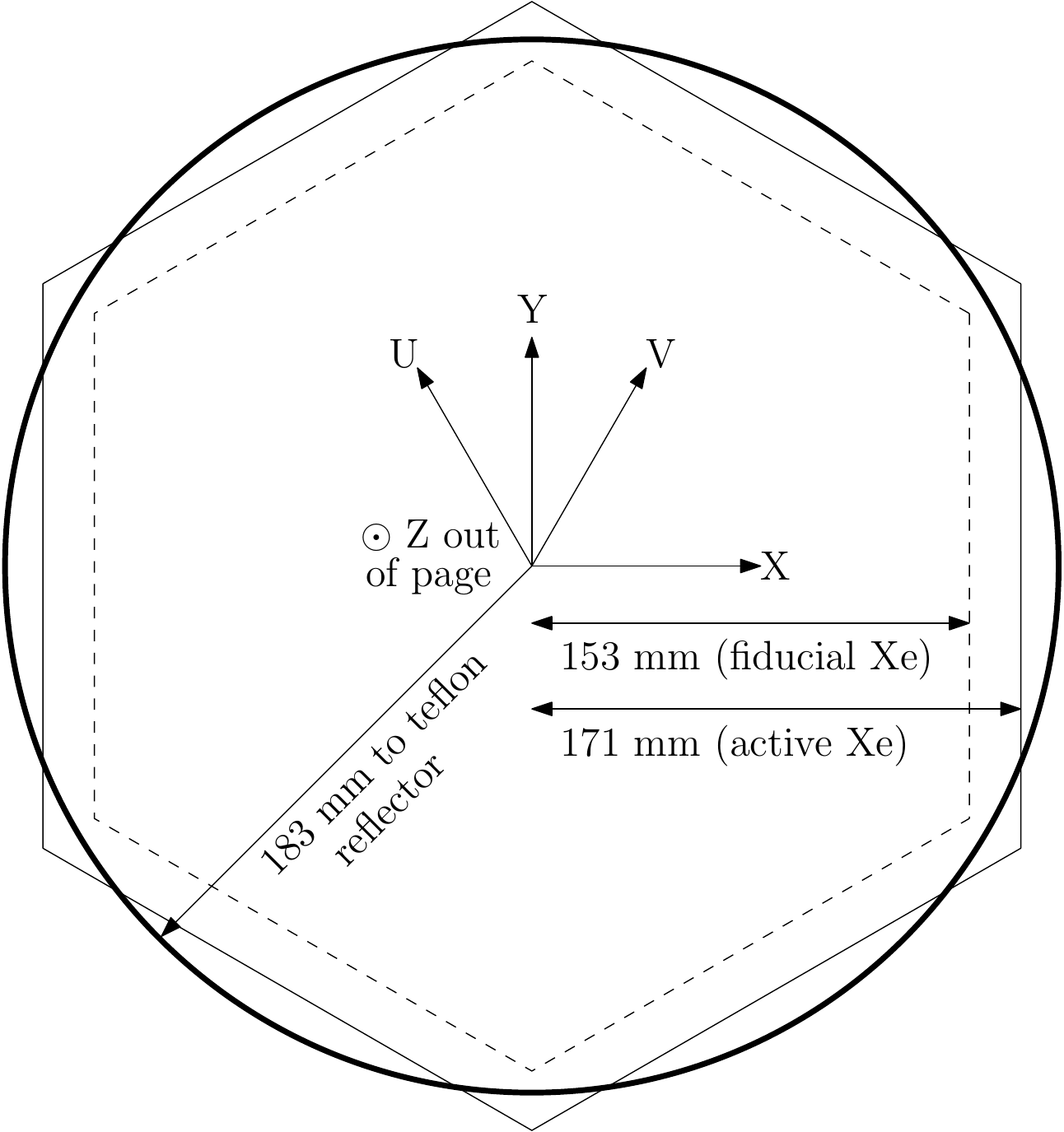}
    \caption{Diagram of the EXO-200 U-V-Z and X-Y-Z coordinate systems. Also shown
    are the hexagonal active region and fiducial region, and
    the circular projection of the Teflon reflector panels.}
	\label{fig:planar-coordinates}
\end{figure}

When drifting ionization reaches the end cap detector package, 
it passes through the first wire grid, known as the shielding grid or induction grid, 
and is collected by the second wire grid which acts as the anode 
(see \cref{fig:WP_Plus_Trajectories}). The shielding grid lies 6~mm in front
of the anode while the APD array lies 6~mm behind it.
Both grids are segmented and read out by charge sensitive preamplifiers. The amplitude of the 
charge collection signal on the anode wires measures the 
ionization energy.  Because the grids are crossed at 
an angle of 60$^\circ$, the anode and the induction signals 
give measurements of two correlated spatial coordinates which we refer to as 
U (anode coordinate) and V (induction coordinate). We transform
these coordinates into an orthogonal X-Y coordinate system, as 
illustrated in 
\cref{fig:planar-coordinates}. We refer to the anode as the 
U-wire system and the induction grid as the V-wire system.

The Z coordinate of the charge deposition (along the axial direction
of the detector) is inferred from the product of 
the drift velocity and the drift time, 
where the start time is given by the prompt scintillation signal. 
The drift velocity is found to be 1.71~mm/\mus{}
(see~\cref{sec:FidVolCut}), which agrees within 10\% 
with previous measurements~\cite{PhysRev.166.871} at this
drift field (374~V/cm from a three dimensional simulation). 
This field is achieved 
by setting the cathode, V-wire, and U-wire voltages to 
-8.0~kV, -780~V, and virtual ground, respectively, while the front face of 
each APD array is biased to $\sim (-1400~\rm{V})$. This arrangement gives an 
electric field of 778~V/cm between the U and V wires, which is sufficient to
ensure 100\% charge transparency of the V-wires.

To reduce the channel count and attendant cabling mass, 
each wire grid is composed of 38 wire-triplets which were 
fabricated by photo-etching phosphor bronze sheet metal. 
Individual wires in each triplet have a roughly diamond-shaped 
cross-section with a full width of about 130~\mum{}. 
The wire pitch is 3~mm, and because the wires are ganged
in groups of three, charge readout channels have a 9~mm pitch, 
for a total grid width of 342~mm. The wire-triplets are mounted
on acrylic beams which are connected to form a hexagonal shape.
The optical transparency of each grid is 96\% at normal incidence.

The distance from the U-wires to the cathode is 198~mm, 
giving a maximum charge drift time of 
115.5~\mus{} (see \cref{sec:FidVolCut}). 
Each TPC has ten field-shaping copper rings mounted on 
acrylic ``combs" which step down the cathode voltage 
to the V-wires and which ensure a uniform drift field in the 
bulk of the LXe. The voltage grading is achieved by a string
of custom low-radioactivity 900~M$\Omega$ resistors. 
The inner side of the field rings and the resistor
package are covered by 1.6~mm thick PTFE (Teflon) tiles capable of 
reflecting the 178~nm scintillation light. 

Each detector package is backed by an APD array 
composed of 234 un-encapsulated silicon devices produced for EXO-200 using 
selected materials.  Each device is circular with a diameter between 
19.6~mm and 21.1~mm and an active diameter of 16~mm.   The APDs are mounted
in two hexagonal-shaped copper platters which provide a common bias
voltage to the APD cathodes of -1400~V and -1380~V on the two detector sides.
The APD anodes are provided a trim voltage, near ground, by their preamplifiers. 
The devices are hexagonally packed such that the sensitive 
area of each platter is 48\% of the total endplate area. The interior side
of each platter is covered by vacuum-deposited aluminum and MgF$_2$
to reflect VUV scintillation photons which do not strike the sensitive areas of the 
APDs.  In each array one APD
device is replaced with a PTFE diffuser which can be illuminated 
by an external laser pulser with a wavelength of 405~nm fed through optical fibers. 
This allows the response of the APD arrays to be periodically monitored.
Ref.~\cite{Neilson:2009kf} describes the testing, performance, and selection
of the APDs. The gain curve, noise, and relative 
quantum efficiency for each APD device was measured at LXe temperature 
as a function of bias voltage, and devices with similar gain characteristics 
were grouped together. The typical grouping is a gang of seven, although
other groupings are also used.
Each APD gang is monitored by a single charge sensitive preamplifier
to reduce cabling material. The anode voltages of the gangs can be 
trimmed by as much as 100~V in groups of six channels.
This allows the gains of the APD channels to be matched to within 2.5\% with
a nominal gain factor of 200. Each APD has a capacitance of 125~pF when biased at 
$\sim (-1400~\rm{V})$, for a channel capacitance of about 1~nF. This leads to substantial
but tolerable electronic noise in the front end (three gangs are disconnected 
due to excessive leakage current).

The charge sensitive preamplifiers for all three systems (U-wires, V-wires, 
and APDs) operate at room temperature 
outside the TPC and are connected to the detector by thin copper-clad polyimide 
flat cables. The flat cables penetrate into 
the xenon volume through custom-made epoxy seals. 

The TPC vessel is fabricated from low-radioactivity copper and sealed
via electron beam and TIG welding. To minimize the mass of the 
vessel, most of the copper is only 1.37~mm thick with 
stiffening features to bolster its mechanical robustness.  
We operate the TPC vessel at
a nominal over-pressure of 8.1~kPa, and changes
of more than 5.3~kPa activate a feedback system that restores
the operating point by either removing or adding xenon.
On the rare occasions when this occurs, the radon level and purity level
of the xenon may be temporarily affected.
To achieve and maintain good xenon
purity with respect to electronegative contaminants such as oxygen 
and water, the xenon is continuously recirculated through two 
heated zirconium getters located outside the detector. Because
the getters must act on the xenon in the gaseous phase, the LXe is evaporated, 
driven through the purifiers with a custom-designed 
xenon gas pump~\cite{Leport:2011hy}, re-condensed outside and above
the detector, and returned to the detector inlet through a vacuum insulated
transfer line. We find that the purity of the LXe is closely correlated with 
the operation of this recirculation loop (see \cref{sec:purity_corrections}).

\begin{figure}
    \includegraphics[width=0.48\textwidth]{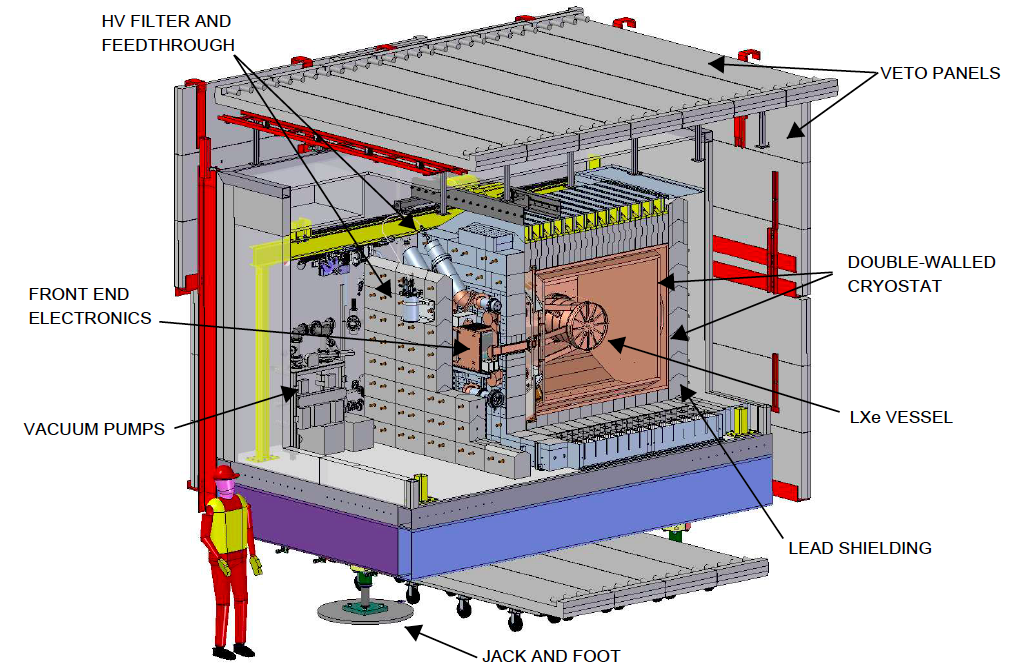}
    \caption{(Color online) The EXO-200 installation at the Waste Isolation Pilot Plant.}
	\label{fig:installation}
\end{figure}

The installation of the detector is shown in \cref{fig:installation}.
The TPC vessel is surrounded a $\ge 50$~cm thick  thermal bath of 
HFE-7000 cryofluid~\cite{3m}, which maintains the temperature
of the TPC and which shields the detector from external gamma radiation. The
HFE-7000 is housed in a double-walled vacuum-insulated cryostat composed
of two nested copper vessels fabricated from low-radioactivity copper plate
of  27~mm thickness.
The outer cryostat is surrounded in all directions by at least 25~cm
of lead. The cryostat features a copper guide tube 
which allows radioactive sources to be inserted past the lead shield and
into the cold HFE volume near the detector. $^{137}$Cs, $^{60}$Co, 
and $^{228}$Th sources of various intensities are available for deployment.

The entire assembly is located in a class 100 clean 
room which is surrounded on four of six sides by a cosmic ray veto 
system. The veto system consists of twenty-nine 5~cm thick Bicron BC-412 plastic 
scintillator panels obtained from the concluded 
KARMEN neutrino experiment~\cite{Gemmeke:1990ix}. 
Each panel is observed by eight photomultiplier tubes (PMTs) and is supported by
4~cm of borated polyethylene. 
The cleanroom laboratory is installed underground
at the Waste Isolation Pilot Plant near Carlsbad, New Mexico, USA
providing 1585~meters water equivalent of overburden~\cite{Esch:2004zj}.

An extensive materials screening and cleanliness campaign, described in 
Refs.~\cite{Leonard:2007uv,Auger:2012gs}, was conducted during 
the design and construction phase 
of the experiment to ensure that the radioactivity of the detector 
would be suitable. Virtually all detector components 
inside the lead shield were custom fabricated for EXO-200.
This effort was informed by a detector Monte Carlo simulation 
which predicted the 
background impact of each detector component.
The enriched xenon itself was screened for noble gas and
electronegative contaminants as described in Ref.~\cite{Dobi:2011zx}.

\subsection{Data acquisition}\label{sec:DAQ} 

The fast data acquisition (DAQ) system for the EXO-200 detector integrates the
readout of 226~hardware channels (76 U-wire signals, 76 V-wire signals, and
74 APD gang signals), muon veto panel output, and a high-voltage (HV) glitch
detector into a single data stream.  
The veto system triggers asynchronously from the TPC when both instrumented ends
of a panel record a hit within a 1~\mus{} coincidence time window. 
The HV glitch detector
monitors high-voltage transients with $\sim\textrm{\mus{}}$ duration.

Each TPC hardware channel is initially coupled to the DAQ via front-end
electronics, consisting of a low-noise charge amplifier with a dual, two-stage
(integration and differentiation) shaper followed by a 12~bit, 1~MS/s
analog-to-digital converter.  The particular values of the shaping times vary
according to the type of channel (e.g.~APD, U, or V).  The digitized data is
fed to a Trigger Electronics Module (TEM), which synchronizes the data from all
hardware channels and forms detector triggers.  The
TEM also incorporates data from the muon veto panels and the HV glitch
detector. 

For the muon veto, the 2 light read-out channels for each panel, each
instrumented with four PMTs, are fed into a discriminator module, which
supplies a bit pattern of the panels above threshold. For a valid muon trigger,
both ends of any one panel are required to be simultaneously above threshold to
reduce the random trigger rate. A secondary, ADC/TDC-based
electronics system is available, allowing the monitoring of panel stability
with detailed semiannual $^{60}$Co source calibration scans of the panels.

In the case of a trigger condition, data is written from the TEM to a control
computer to be stored on disk.  Muon veto or glitch detector events only initiate
transfers of those particular types of records; these are later synchronized
with TPC data by means of their time stamps. If a TPC trigger occurs,
the TEM transfers digitized data for all 226 hardware channels for sample 
times starting 1024~\mus{} before the trigger and ending 1024~\mus{} 
after it. During a normal physics
run, there are four types of TPC triggers used, with rough thresholds
noted in parentheses: (a) individual U-wire trigger for LXe $\gamma$ and
$\beta$ events ($\sim100~\textrm{keV}$), (b) APD individual trigger for activity inside
the respective APDs ($\sim3-4\textrm{keV}$), (c) APD sum trigger for LXe
$\alpha$ events ($\sim25~000$ photons), and (d) solicited (forced) trigger at 0.1~Hz, for
monitoring detector performance. Further information on the DAQ system may be
found in~\cite{Auger:2012gs}.

\subsection{Data processing structure}\label{sec:Processing} 

The processing of data follows a tiered scheme for each run: 

\begin{enumerate}
  \item Tier 0 $\to$ Tier 1, ``Rootification" : conversion of binary data files to
        ROOT~\cite{Bru97} files; low-level verification of data validity.
  \item Tier 1 $\to$ Tier 2, ``Reconstruction" : first two reconstruction stages,
    see \cref{sec:Reconstruction}; noise and muon taggers;
        calculations of waveform characteristics.
  \item Tier 2 $\to$ Tier 3, ``Processing" : final reconstruction stage
    (`clustering'), see \cref{sec:Clustering}; data corrections
    (e.g.~gain, grid, and purity), see
        \cref{sec:CalibrationAndCorrections}.
  \item Tier 3 $\to$ Standard analysis scripts, ``Trending": extraction of
    parameters from a run (e.g. noise, threshold, etc.) useful for tracking
        trends over time.
\end{enumerate}

\subsection{Data analysis strategy}

As expected from the materials screening 
campaign, we find in the data that the primary 
backgrounds to \twonubb{} in EXO-200 are gamma
and beta interactions due to trace quantities of \isot{K}{40}, \isot{Th}{232},
and \isot{U}{238} in the detector materials. 
We separate these backgrounds from \twonubb{} 
candidates by taking advantage of the 
detector's good energy and position resolution
and its ability to perform pattern recognition.
First, we label an event as being ``single-site" (SS) if it is 
consistent with having all charge deposits confined to a single 
volume with a characteristic dimension of $\sim2-3$~mm, 
as expected for most \twonubb{} events in LXe. Otherwise the 
event is labeled as ``multi-site" (MS).
Due to the predominance of the Compton scattering process 
in the energy range of interest (700 keV to 3500 keV), 
gamma events are mostly categorized as MS  
as they  commonly produce two or more localized 
charge deposits separated by at least several centimeters in LXe.
Secondly, we calculate for each event the ``standoff distance", 
or the shortest distance between the various 
charge depositions to an anode wire or reflector suface. 
Gamma events and \twonubb{} events have 
distinguishable standoff probability distributions 
because the latter are uniformly 
distributed in the LXe whereas the former tend
to originate in the passive detector materials 
and exhibit some attenuation in a detector of the size of EXO-200.
Third, we measure the total energy of each event by combining the
charge and scintillation signals in a manner which takes advantage of
the anti-correlation between these channels to improve the energy 
resolution~\cite{Conti2003}.  This last procedure is essential for the 
search for \nonubb{}, where the signal is a resolution-limited feature at the
$Q$ value, but is also utilized in the present measurement of 
the \twonubb{} decay.

We exploit these
three variables by selecting fiducial $\beta$-like events in the 
data, dividing them into the SS and MS categories,
and performing a simultaneous maximum likelihood fit to the 
energy spectra and standoff distance of both event samples. 
The probability distribution functions (PDFs) provided to 
the fit are determined by a Monte Carlo 
simulation of the relevant signal and background sources. This strategy
is validated by comparing data and simulation for calibration
sources of known activity which are periodically inserted near 
the detector. The efficiency of the event selection for \twonubb{} 
events is determined by a combination of data and Monte Carlo 
studies and is cross-checked by the external calibration source
data.

In the following we describe the Monte Carlo simulation (\cref{sec:MC}),
event reconstruction (\cref{sec:Reconstruction}), calibration and energy
measurement (\cref{sec:CalibrationAndCorrections}), simulation-measurement
agreement (\cref{sec:source_agreement}), data quality selection
(\cref{sec:DataQuality}), event selection cuts (\cref{sec:analysis_cuts}), and
likelihood fits and \twonubb{} half-life measurement (\cref{sec:physics}).

\section{Monte Carlo simulations} \label{sec:MC}
The EXO-200 Monte Carlo simulation software is split into two independent components.  The
first component, developed within the GEANT4 simulation package~\cite{GEANT42006},
parameterizes the geometry of the EXO-200 detector and surroundings.  The second
stage uses the output from the first component -- energy depositions within the
simulated detector -- to calculate electronic signals.  The data format produced 
by this process is identical to that of real data and may
be processed through the Tier 1 and Tier 2 reconstruction and analysis chains described in
\cref{sec:Processing,sec:Reconstruction}. 

\subsection{Simulated geometry} 

\begin{figure} 
  \includegraphics[width=0.45\textwidth,clip,trim=70mm 60mm 20mm 60mm]{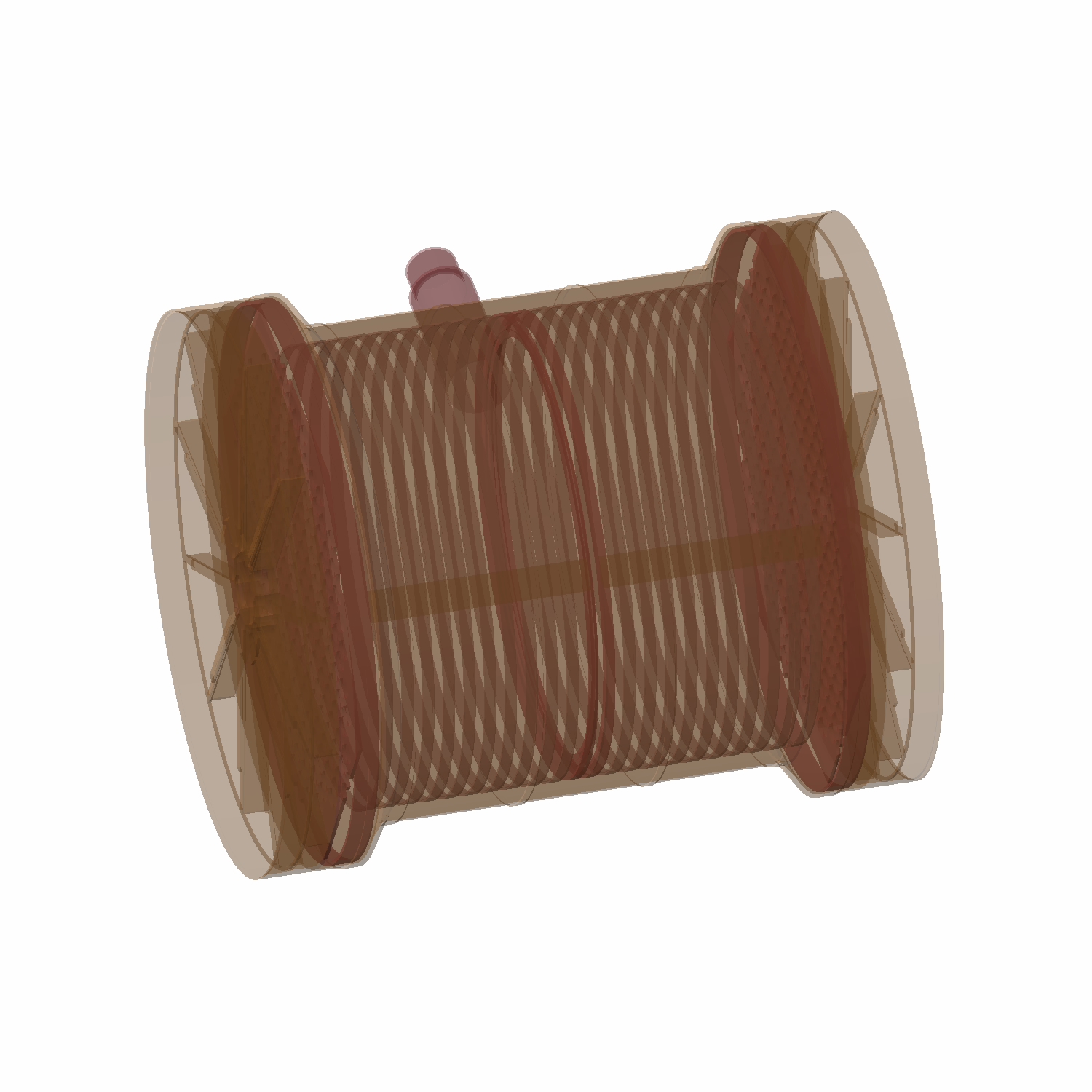}
  \caption[Copper in simulated TPC visualization]{(Color online) Visualization of the
      simulated TPC vessel and internal copper components.}
  \label{fig:vesselraytracer}
\end{figure}

The simulated geometry implements a detailed description of the TPC and its
internal components and includes the surrounding HFE, cryostat and lead shield.  
3-D computer-aided design (CAD) models of the detector are used and coded using GEANT4 shape primitives, 
as illustrated in \cref{fig:vesselraytracer} for the copper components of the 
inner detector.  An approximate geometry is used for the shape of some complex 
components, so a check is made to verify that the mass of materials is accurately
reproduced, as shown in \cref{tab:coppermass} for the materials shown
in \cref{fig:vesselraytracer}.

\begin{table}[ht]
	\renewcommand{\arraystretch}{1.2}
	\begin{ruledtabular}
	\begin{tabular}{l c c c} 
	Component       & Quantity & CAD Mass  & GEANT4 Mass \\
	                &          & Mass (kg) & Mass (kg)   \\
	Outer Cryostat  &     1    &    3453   &    3364     \\
	Inner Cryostat  &     1    &    2843   &    2575     \\ 
	\cline{3-4}
	Total Cryostat  &          &    6296   &    5939     \\
	                &          &           &             \\
  Source Tube     &     1    &    0.533  &    0.207    \\
	                &          &           &             \\
	TPC Leg         &     1    &    6.979  &    6.944    \\
	                &          &           &             \\
	HV Feed         &     1    &    0.491  &    0.303    \\
	LXe Vessel      &     1    &    22.36  &    22.32    \\
	APD Frame       &     2    &    3.144  &    4.440    \\
	Wire Support    &          &           &             \\
	           Ring &     2    &    2.659  &    1.468    \\
	Field Ring      &    20    &    3.055  &    3.240    \\
	Cathode Ring    &     1    &    0.721  &    0.728    \\
	Dummy           &          &           &             \\
	   Cathode Ring &     1    &    0.323  &    0.327    \\
	\cline{3-4}
	Total TPCs      &          &    32.753  &   32.826    \\
	\end{tabular}
	\end{ruledtabular}
    \caption{ Comparison between copper mass in the detector CAD model 
    and GEANT4 model. The two models have independent implementations, and 
    differences are understood to be due to
    simplifications in the GEANT4 geometry. Although these distinctions could 
    lead to minor differences in gamma attenuation between data and simulation, 
    this effect is expected to be small and is cross-checked by external gamma calibration
    source data (see \cref{sec:source_agreement}).}
 	\label{tab:coppermass}
\end{table}

\subsection{Simulated signal generation}\label{sec:MCDigitizer} 

\begin{figure}
  \includegraphics[width=0.48\textwidth]{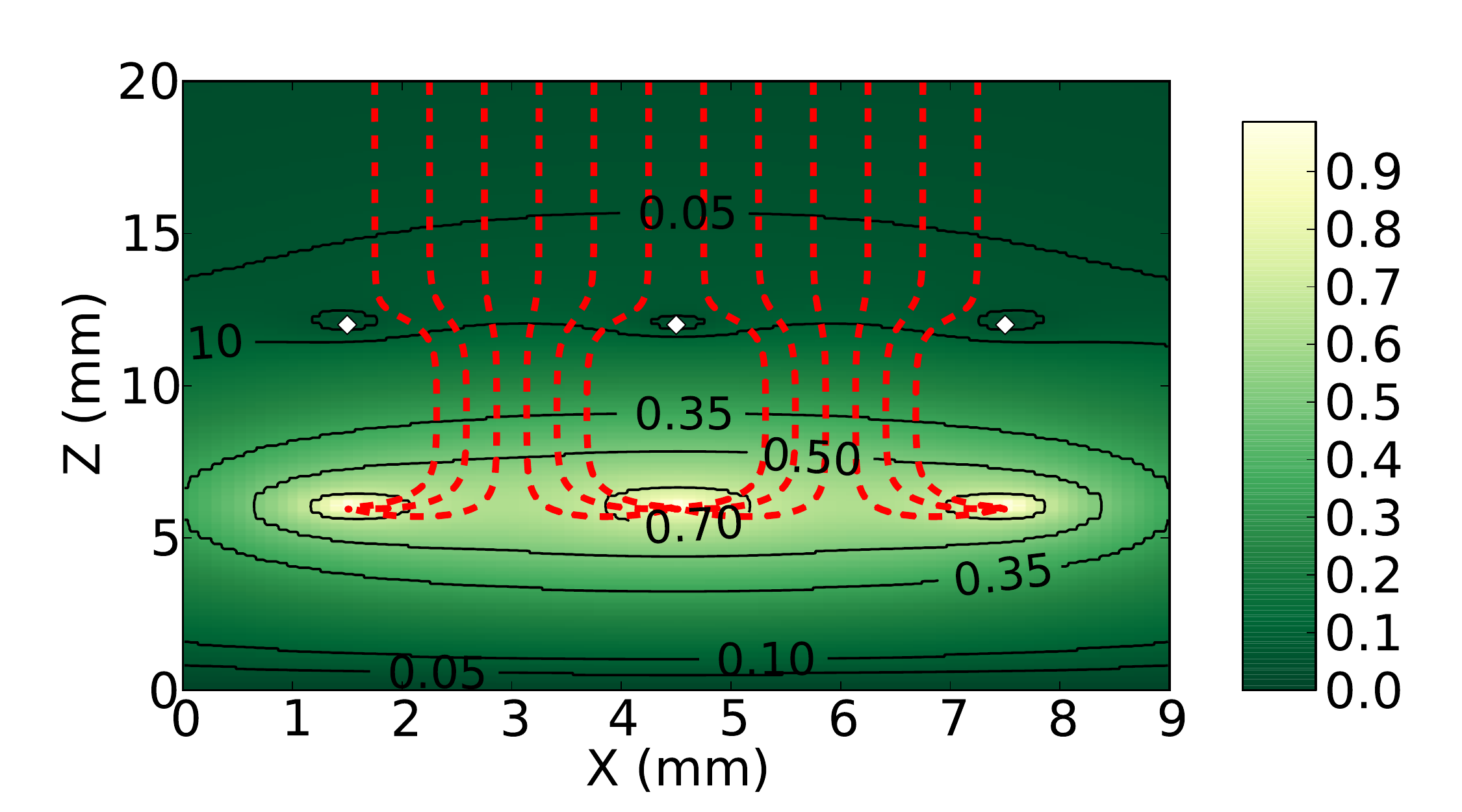}
  \caption[ 2-D weighting potential plus trajectories]{(Color online.) The magnitude 
      of the calculated weighting potential for a single U-wire readout channel. The
      weighting potential peaks to one at the positions of the three wires
      composing a signal channel (X = 1.5, 4.5, 7.5~mm, Z = 6~mm).  Example
      charge drift trajectories are shown by the overlaid red dashed lines; deflections of
      the trajectories arise from the presence of a V-wire channel (diamonds at
      X = 1.5, 4.5, 7.5~mm, Z = 12~mm) situated directly above the U-wire
      channel.  
  } 
  \label{fig:WP_Plus_Trajectories}
\end{figure}

The signal calculation for wire channels employs a two-dimensional (2-D), simplified
geometry which assumes that U-wires and V-wires are parallel to one another,
infinitely long, and perpendicular to the plane of calculation.  In
addition, in this configuration the V-wires lie directly above the U-wires 
in the Z direction.  For this setup the weighting potential,
$\phi(\vec{x})$, and electric field, $\vec{E}(\vec{x})$ have been calculated
using Maxwell 2D~\cite{MAXWELL}, which allows one to model drifting charge
at any location in the detector and calculate the resulting induced
signal using the Shockley-Ramo theorem~\cite{Ramo:1939,*Shockley:1938}.  In
this method, the charge induced on a wire by the movement of charge, $q$, from
$\vec{y}_0$ to $\vec{y}$ is given by $q (\phi(\vec{y}) - \phi(\vec{y}_0))$,
where $\phi$ is the weighting potential for the particular wire.  The weighting
potential for a single U-wire channel along with example charge drift
trajectories are shown in \cref{fig:WP_Plus_Trajectories}.  Charge
diffusion during the drift is not modeled in this calculation.  This does not
significantly affect the quality of the simulation as the larger effect of 
transverse charge diffusion in LXe at the field used here (see e.g.~\cite{Doke198287}), 
produces an RMS spread of $\sim2$~mm to be compared with the 9~mm wire readout pitch. 
The effect of longitudinal diffusion was observed when taking low-electric
field data as it lengthened the U-wire pulse rise-time.  However, this effect
becomes negligible at the nominal electric field.
Once signals are calculated, they are shaped with
the appropriate channel transfer function. For Monte Carlo simulation production runs,
white noise is added to the signal waveforms, though there exists the
possibility to add real noise using event traces
from solicited triggers.  

\begin{figure*}[ht]
  \includegraphics[width=0.97\textwidth]{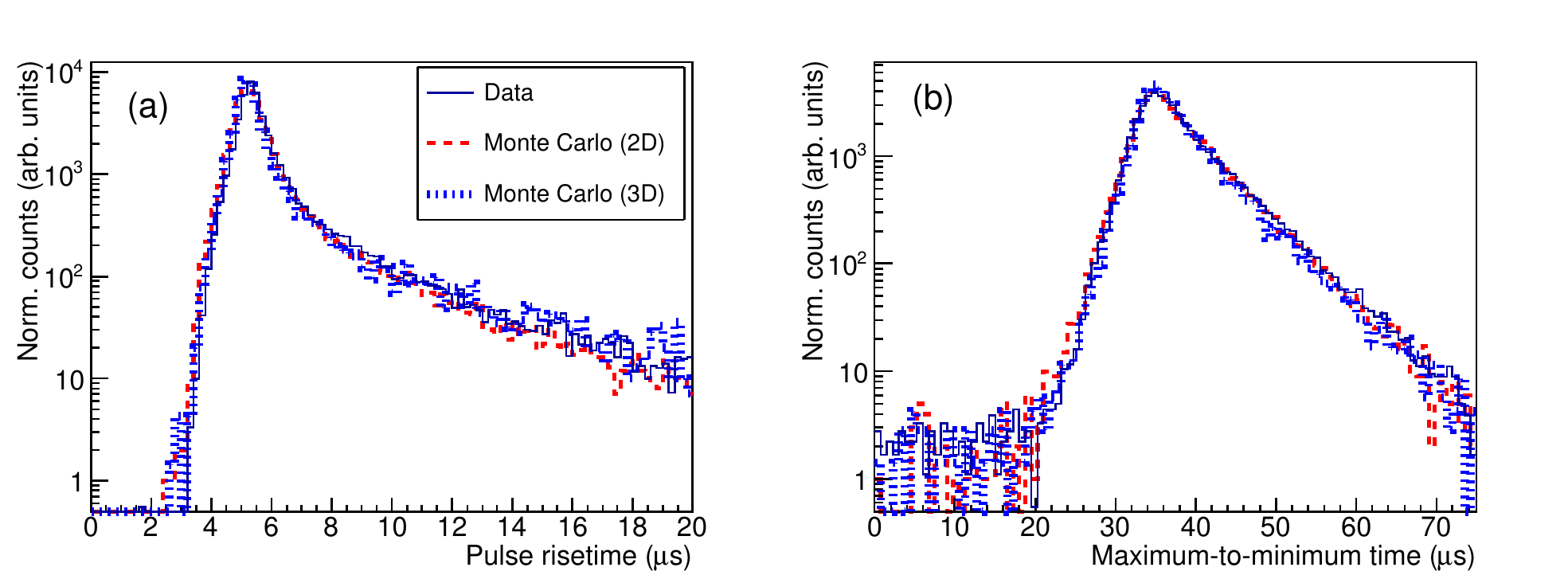}
  \caption{ (Color online) Comparison of the distributions of the pulse
      rise-time (a) and maximum-to-minimum time (b) on U-wires between calibration
      data and simulations using 2-D and 3-D field maps (see
      \cref{sec:InductionIdent}) for \thsrc{} data.  The long tail in the
      rise-time distribution is due to Compton scatters.  Only wire signals
      above 200~keV are compared, as the values of rise-time and
      maximum-to-minimum time for low-amplitude signals depends strongly on the
      waveform noise profile.  The distributions are normalized to the same
      total number of events for comparison.  
  }
  \label{fig:DigiComparison}
\end{figure*}

To determine if the 2-D geometry is a valid approximation to use,
waveform characteristics were calculated and compared for simulated wire
signals and for data from a \thsrc\ source run.  In particular, the rise-time of
the pulses and the time difference between maximum and minimum are calculated,
(see \cref{sec:InductionIdent}).  The results of this study are shown in
\cref{fig:DigiComparison}, where it is clear that signal generation well
reproduces the distributions from data.

In addition, a fully three-dimensional (3-D) COMSOL~\cite{COMSOL} finite element
calculation was used to produce a 3-D electric field and weighting potentials
for the 76~wire channels populating a single TPC at the proper $60^{\circ}$ angle 
between U and V grids.  Although this model is currently too slow to be used in
Monte Carlo simulation production runs, calculations from it have been used
during fiducial volume studies of the detector, in particular to help
understand regions near the edge of the wire grid where the field configuration
is complex.  Results from a similar study of waveform characteristics of traces
generated from this model are included in \cref{fig:DigiComparison}. A
comparison of the two calculations show the 2-D model to provide a valid
approximation of the charge collection.

The signal generation for APD channels uses a parameterized light response 
function which returns the expected amount of light hitting both APD planes given a charge
deposition at a given location in the detector.  The light response function 
was derived from a Monte Carlo study of light collection in the 
TPC which included geometric factors, reflectance, and scattering. 
Effects from the anti-correlation between the ionization and scintillation
(see \cref{sec:AntiCorr}) are not simulated.  The resulting unshaped APD pulse shapes are
assumed to be step functions, a valid assumption given that the integration
time of the APD electronics transfer function ($\sim1$~\mus{}) is much longer
than the intrinsic APD rise-time (10-100~ns).  As with U- and V-wires, these
unshaped pulses are transformed with the appropriate electronics 
transfer functions and white noise is added. 

\section{Event reconstruction}\label{sec:Reconstruction}
Event reconstruction is the process by which waveforms are analyzed to derive
information such as energy content and topology of events.  The reconstruction
of an event has three stages: 

\begin{enumerate}[topsep=1pt]
    \setlength{\itemsep}{-0.1em} 
    \item Signal finding 
    \item Signal parameter estimation 
    \item Clustering, or assembling of found signals to determine event
        topology  
\end{enumerate}

\noindent
The following terminology is used in this section: a ``signal"
refers to a hit which is reconstructed on a particular channel, a ``bundle" is a
group of signals from a single channel type, and a ``cluster" is a set of
bundles which have been associated together to determine the event position. 

\subsection{Signal models}\label{sec:SignalModels}

Signal shape templates are used extensively in
reconstruction in both the signal finding and parameter extraction stages and
are produced for all channels.  In the case of U- and V-wires, unshaped signals
are generated using the signal simulation described in
\cref{sec:MCDigitizer}.  A simple step function is used for APD
channels.  Unshaped waveforms are then filtered using the appropriate transfer
function to create the final signal template shape for a given channel.
Transfer functions are defined by the front-end electronics (see
\cref{sec:DAQ}), resulting in 3~differentiation (one stage
from the preamplifier) and 2~integration stages.  The values for these stages
for each channel type are given in \cref{tab:ShapingTimes}.  It is
important to note that the value of the third differentiation stage of the U-wires
is a measured parameter and varies for each U-wire channel.  Using the measured
value for each channel was found to improve the fits used to determine signal
heights (\cref{sec:rec_par_estimation_signals}) which lead to an improved detector
resolution. 

\begin{table}[ht]
        \begin{center}
        \begin{ruledtabular}
        \renewcommand{\arraystretch}{1.3}
        \begin{tabular}{l@{\hskip 20pt} c c@{\hskip 20pt}  c c C{60pt} } \toprule 

                & \multicolumn{5}{c}{Stage type} \\ \cline{2-6}  
                Channel Type & \multicolumn{2}{m{70pt}}{Integration} & \multicolumn{3}{c}{Differentiation} \\
                \colrule
                APDs & 3 & 3 & 10 & 10 & 300 \\
                U-wires & 1.5 & 1.5 & 40 & 40 & {\centering 51 \textendash{} 85 (nominal 60)} \\
                V-wires & 3 & 3 & 10 & 10 & 60 \\ 
        \end{tabular}
        \end{ruledtabular}
        \caption{Shaping times (all in \mus{}) relevant to the transfer
            functions of different channel types. The third differentiation
            stage for the U-wire signals is measured for every channel by fitting to pulse
            shapes from charge injection data.  }

        \label{tab:ShapingTimes} 
        \end{center}
\end{table}

\subsection{Signal finding}\label{sec:rec_signal_finding}

It is necessary to search for signals on waveform traces because they are not
always guaranteed to arrive at a given time (e.g.~specified by a trigger).  
Two methods are used to find signals on waveform traces: applying a matched
filter, and waveform unshaping.  The second method is used to identify pulses
closely following one another within a signal found by the matched filter.

\subsubsection{Matched filter}\label{sec:matched_filter}

A matched filter is used to find signals due to its simple algorithmic
implementation and because it has been observed to produce stable performance
over time in varying noise conditions.  The filter (see, e.g.~\cite{North1963})
is applied in Fourier space, 

\begin{equation}
    y(t) = \mathfrak{F}^{-1} \left\{X(f) H^{*}(f)\right\} 
    \label{eqn:MF}
\end{equation}

\noindent
where $y(t)$ is the filtered signal, $\mathfrak{F}$ is the discrete Fourier
transform (FT), $X(f)$ is the FT of the original waveform, $x(t)$, and $H^{*}(f)$ is
the complex conjugate of the FT of the transfer function, $h(t)$, for the
particular channel.  For APD channels, \cref{eqn:MF} is further divided by the
noise spectrum of the channels to whiten the spectrum.  This is \emph{not}
performed for U-wire channels as the additional division was found to broaden the
signal due to the longer U-wire shaping times, detrimentally affecting
subsequent peak finding.  V-wires also have no additional noise division. 

\begin{figure}
  \includegraphics[width=0.45\textwidth]{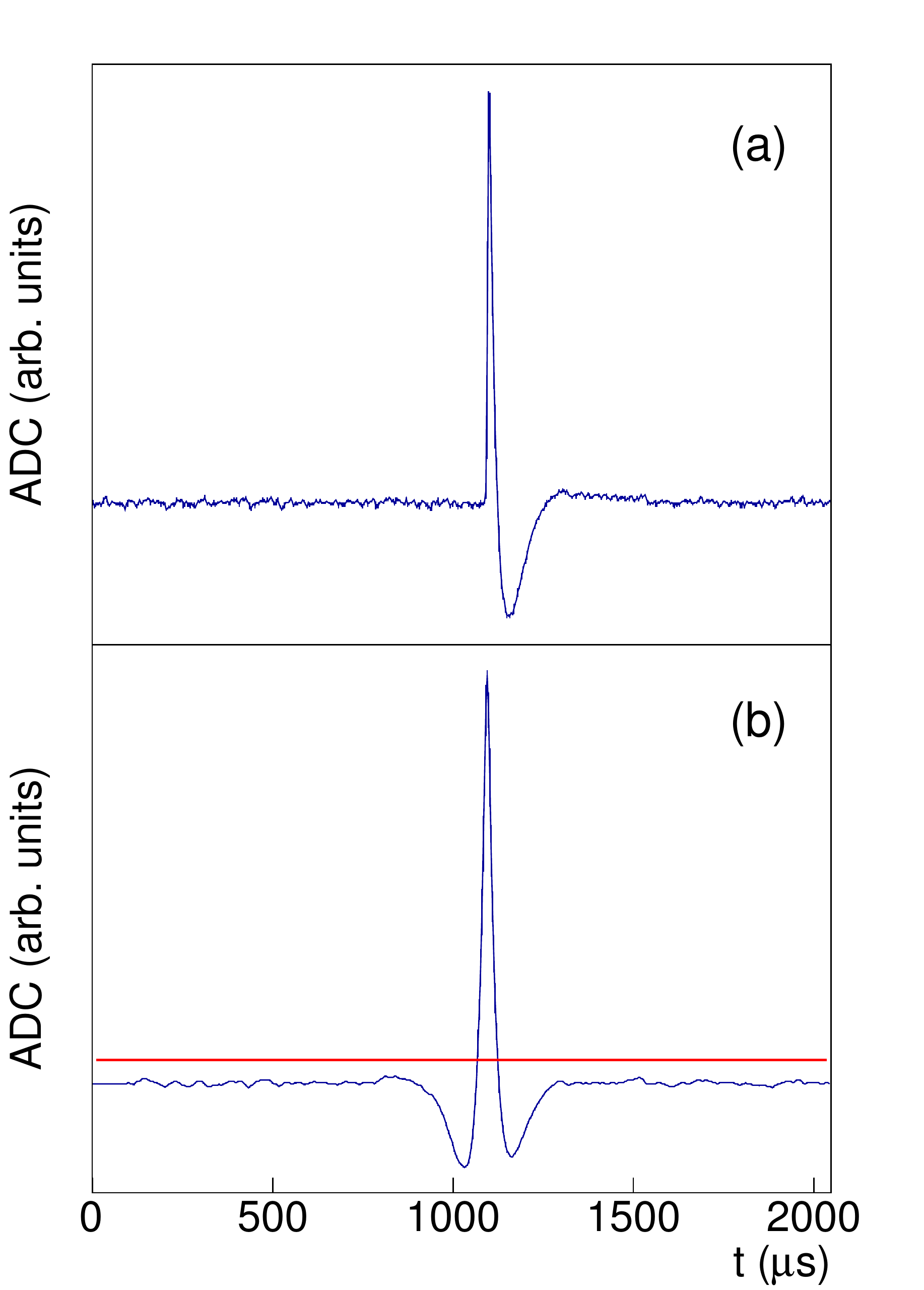}

  \caption{(Color online) A U-wire waveform (a) and the result of the matched filter (b).
  The horizontal line in (b) is the calculated threshold.}

  \label{fig:MFExamp}

\end{figure}

U- and V-wire channels are filtered by applying the template model defined for
each respective channel.  APD channels, in contrast, are first assembled into
two sum waveforms generated by summing together the waveforms from all channels
on each APD plane.  This results in two waveforms, each of which is then
filtered using \cref{eqn:MF} with the APD transfer function.  An example of a
waveform before and after filtering is shown in \cref{fig:MFExamp}.  A
peak-search algorithm is performed on the filtered waveforms, looking for
amplitudes exceeding a certain threshold.  This threshold is calculated for
each waveform on an event-by-event basis to reduce the sensitivity to channel-
and time-based noise variations.  The algorithm proceeds by first determining
the mean absolute deviation (MAD), of the waveform from its baseline, removing
all parts of the waveform exceeding
$\left(3\sqrt{\frac{\pi}{2}}\right)\times{\rm MAD}$, and then recalculating the MAD.
This is equivalent to removing values greater than $3\sigma$ if the deviations
from the baseline are normally distributed.  The threshold is defined as 5 (4)
times this final MAD for wire (APD) signals.

\subsubsection{Waveform unshaping}\label{sec:rec_wf_unshaping}

Since the matched filter is designed to find single pulses, it is ill-suited to 
disentangle multiple signals on a single trace when these signals arrive close to 
one another in time.  Hence a dedicated algorithm is applied to the original 
waveforms after they have been identified by the matched filter.
This algorithm ``unshapes'' the signal, obtaining the original charge deposited
$q(t)$:

\begin{equation}
    q(t) = \mathfrak{F}^{-1} \left\{H^{-1}(f)X(f)\right\} 
    \label{eqn:WUn}
\end{equation}

\noindent
with the same definitions as given for \cref{eqn:MF}, where $H^{-1}(f)$ is the
inverse transfer function.  This process is very sensitive to inaccuracies of
the transfer function used due to inexact pole-zero cancellation.  Detailed
studies show that the procedure works best on a short (265~\mus{}) interval
centered around the pulse time found by the matched filter.    $q(t)$ is
subsequently reshaped with a 2~\mus{} triangular, or moving average, filter
(see e.g.~\cite{Jord94}).  The reshaped waveform is then analyzed with a
peak-search algorithm to determine the presence of any additional signals. 

\subsection{Parameter estimation of signals}\label{sec:rec_par_estimation_signals}

\subsubsection{Amplitude measurement}

\begin{figure}
    \centering
    \includegraphics[width=0.45\textwidth]{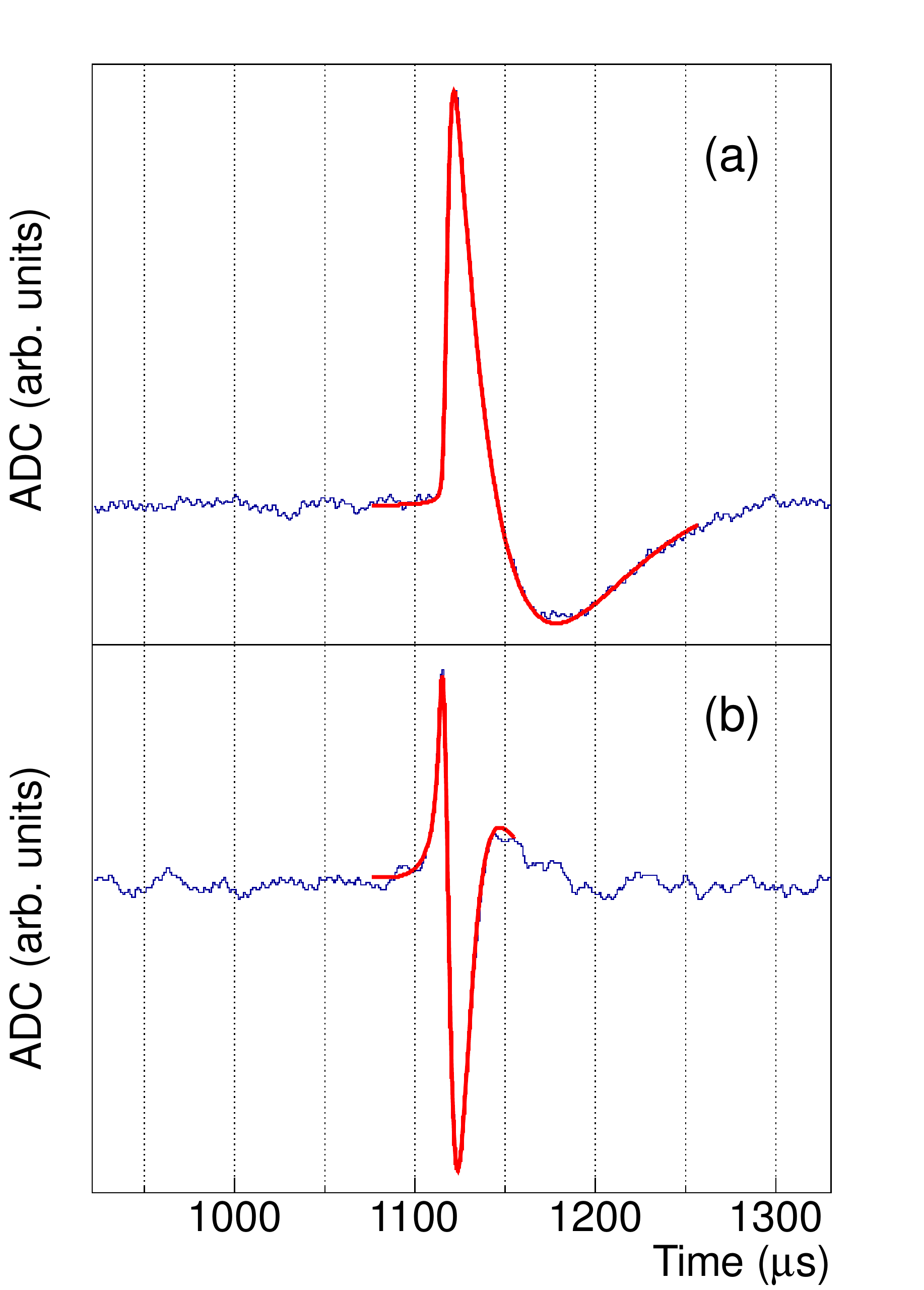}
    \caption{(Color online) Examples of fits to a U-wire (a) and V-wire (b) on an expanded vertical scale.  One can see that
    the V-wire peaks earlier than the U-wire signal.} \label{fig:FitWFs}
\end{figure}

The amplitudes of all U-, V-, and APD sum signals are measured by fitting the
waveforms to their respective signal models (see
\cref{sec:SignalModels}).  A $\chi^2$ function is built using the signal
model, the data, and the output of the previous signal-finding stage:

\begin{equation}
  \chi^2 = \sum_{l=0}^{L} \frac{\left[x_l - b - \left(\sum_{i=0}^{N} \left\{ A_i
      f_{\rm SM}(x_{l}, t_i)\right\}\right)\right]^2}{\sigma_{\rm noise}^2}
  \label{eqn:ChiSquareFit}
\end{equation}

\noindent
where $x_l$ is the data sample at time $l$, $b$ is the measured baseline,
$i$ is the index of the $N$ signals on the waveform, $A_i$ and $t_i$ are the
amplitude and time of the $i^{\rm th}$ signal, and $f_{\rm SM}$ is the signal model.  The
baseline and the RMS noise $\sigma_{\rm noise}$ are calculated and fixed for each
waveform.  $A_i$ and $t_i$ are the only floating parameters, and the values
estimated during the signal fitting stage are used as initial input.  The size
of the fit window, $L$, extends $\pm 40$~\mus{} around the signal (defined in
the finding stage).  In the case of U-wires, the
upper fit window limit is extended to 140~\mus{} to include the longer undershoot
induced by the larger differentiation times in the U-wire transfer functions.
When multiple signals are found on the waveform traces, the fit windows are
determined for each signal and combined.  In the case of signals further apart
than the 40 (or 140)~\mus{}, several separated fit windows are produced.  
An example of fits to U and V waveforms is given in \cref{fig:FitWFs}. 

The $\chi^2$ function is minimized using MIGRAD~\cite{Jam75}, resulting in 
signal amplitudes, times and errors, in addition to the overall value of the
minimized $\chi^2$ function.  Before the fitting stage is completed, the results of the
fits to the two APD sum signals are used to then fit individual APD gang
channels separately and extract the amplitudes of each gang channel.  The $t_i$
parameters from the fits to the two APD sum signals are used as input to
\cref{eqn:ChiSquareFit}, but these values are fixed during the
subsequent fit; only the amplitude parameter(s), $A_i$, are allowed to float
when fitting the gang signals.

\subsubsection{Waveform characteristics}\label{sec:InductionIdent}

In addition to the pulse characteristics derived from the waveform fits, several
metrics are calculated directly from the U-wire waveforms for use in
pulse-shape-based discrimination between ``collection" and ``induction" signals.
In this sense, we define induction signals on U-wires as signals that occur
when charge drifts close enough to a channel to induce a signal, but does not
deposit on that channel.  These induction signals must be corrected for 
so that they are not mistakenly reconstructed as a low-energy charge cluster
(see \cref{sec:Clustering}), causing a single-site topology event to be
interpreted as a multi-site event.  This has particularly important
implications for event classes that are predominantly SS (e.g.~\nonubb{} and
\twonubb{}) and would be mistakenly classified as MS with a loss of
efficiency.

The following discriminants are calculated:
\begin{enumerate}[topsep=1pt]
    \setlength{\itemsep}{-0.1em} 
    \item Pulse timing: 
        \begin{enumerate}[topsep=-0.6em] 
            \setlength{\itemsep}{-0.2em} 
            \item The rise time of the pulse, measured from the last time the
                pulse crosses a minimum threshold (set to 10\% of the pulse
                height above the baseline) to the first time the pulse crosses
                a maximum threshold (at 95\% of the pulse height above the
                baseline) 
            \item The time from the pulse maximum to the following pulse
                minimum, defined similarly as the time between the last
                crossing of a maximum threshold (90\% of the pulse height,
                measured from the pulse minimum) and the first crossing of a
                minimum threshold (10\% of the pulse height above the pulse
                minimum) 
        \end{enumerate}
    \item Pulse integral:  The pulses are unshaped using inverse transfer functions
        (see \cref{sec:rec_wf_unshaping}), and the integral of the pulse
        is calculated in a window within 10~\mus{} before and 40~\mus{} after
        the pulse maximum.
    \item Fit $\chi^2$: Following the standard fitting procedure for U-wire
        signals described in \cref{sec:rec_par_estimation_signals}, the
        same signal finding and fitting procedure is repeated for each U-wire
        waveform using a signal template describing a U-wire induction signal
        instead of the standard collection template.  The fit $\chi^2$ value is
        then calculated for fits to both the collection and induction signal
        templates in a time window restricted to 20~\mus{} before and 30~\mus{}
        after the signal. 
    \item Nearest-neighbor amplitude: For each U-wire signal, the total energy
        on neighboring channels within 50~\mus{} is calculated (U-wire induction
        signals reconstructed as collection signals have poor time estimations
        due to the template mismatch.) 
\end{enumerate}

\begin{figure*}[ht]
  \centering
  \includegraphics[width=0.95\textwidth]{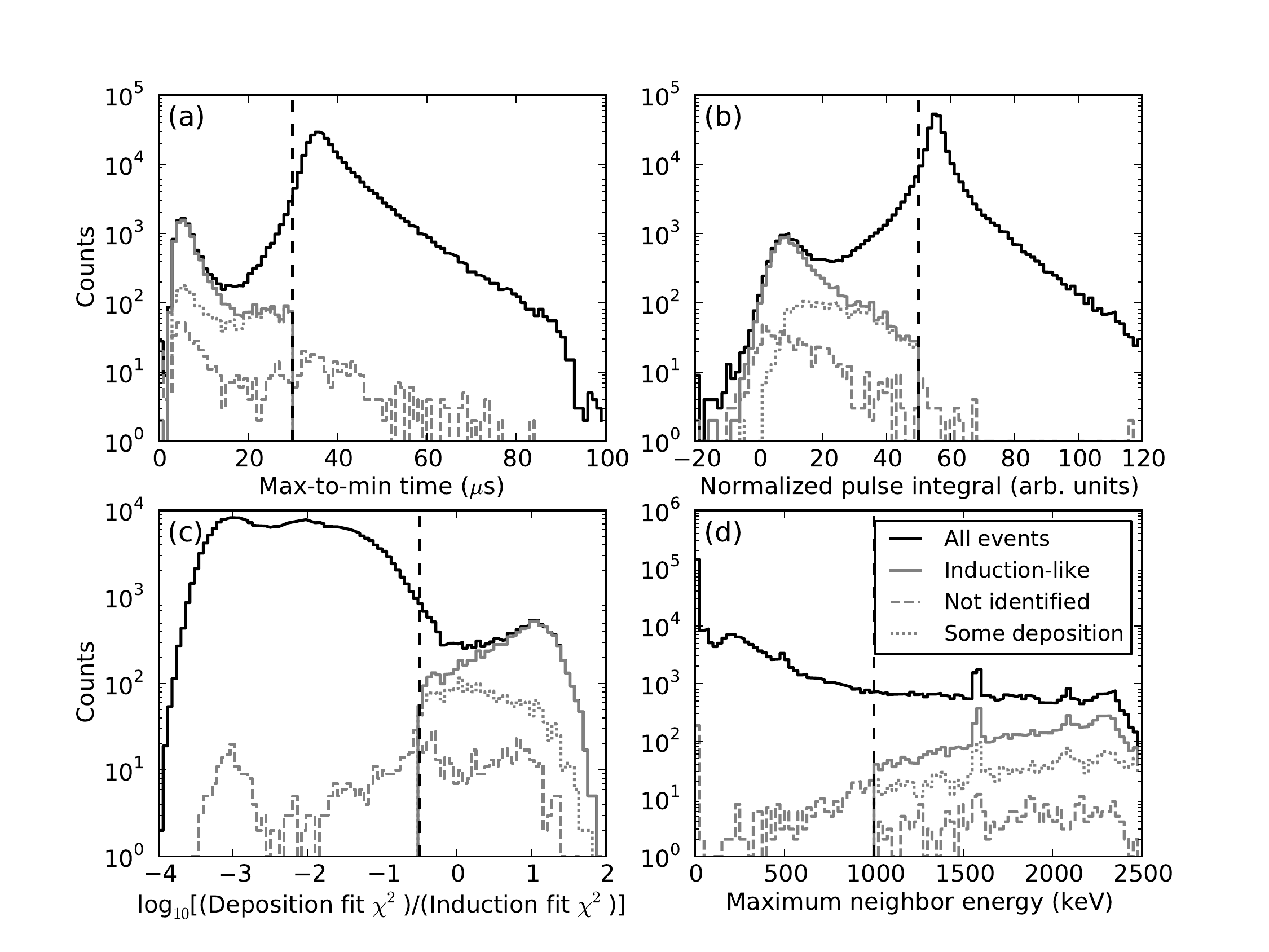}
  \caption{Distributions of the discriminants described in the text
      used to identify induction signals, calculated for waveforms generated
      from $^{228}$Th Monte Carlo simulations.  The distribution of these values
      for all U-wire signals is shown as the solid histogram, with the cuts
      used for each value denoted by the vertical dashed lines.  U-wire signals
      must satisfy \emph{all} cuts to be identified as ``induction-like"
      signals.  The cuts are chosen to ensure minimum impact on collection
      signals (acceptance efficiency $>99.9\%$), corresponding to a rejection
      efficiency of induction signals of 77\%, integrated over the \twonubb{}
      energy spectrum.   Also shown are the distributions of simulated
      induction pulses which were \emph{not} identified as induction-like by
      the cuts (``Not identified") and simulated pulses containing at least
      some deposition, but which were identified as induction-like (``Some
  deposition").  }
  \label{fig:InductionDists} 
\end{figure*}

\noindent
Distributions of these discriminants are given in
\cref{fig:InductionDists}, comparing the values for induction signals and
collection signals.  A full discriminator is built from a combination of these
values to ensure the collection efficiency for collection signals with at least
250~keV of deposited energy is $>99.9\%$.  With the chosen cut, the rejection
efficiency for U-wire induction signals is 77\%, integrated over the \twonubb{}
energy spectrum. Since signals identified as induction by these selection
criteria are required to have an energy deposit on a neighboring channel of
$>1000$~keV, no events can be forced below the 700~keV analysis threshold due
to removal of misidentified collection signals.  These efficiencies were
calculated using Monte Carlo studies. 

\subsection{Signal clustering}\label{sec:Clustering}

Once the time and amplitude of signals on U-, V-, and APD channels have been
found and properly gain corrected (see \cref{sec:ChannelCorrections}),
these signals are grouped together to form 3-D ``clusters''.  
Since signals from different channels arising from the same charge deposit 
will have correlated characteristics (e.g.~amplitudes and times), these 
characteristics can be used to guide the clustering process that 
proceeds in a step-wise fashion.  First, signals of like 
channels are associated together (U-wires with other U-wires, etc.) in a 
process called bundling.  U-wire signals
identified as due to induction, using the criteria described in
\cref{sec:InductionIdent}, are ignored when constructing bundles.    
The Z positions of these bundles are then 
determined by associating them with APD bundles.  Finally these wire 
bundles are grouped together to form fully 3-D reconstructed clusters.

\subsubsection{Signal bundling}
U-wire signals on adjacent channels arriving close in time (within 3.5~\mus{})
are bundled together.  The time of each bundle is defined by the
amplitude-weighted average of the associated signals.   V-wire signals are also bundled
according to time, using to the relationship:

\begin{equation}
    \left| t_{i} - t_{0} - (2.97~\text{\textmu{}s/chan}) \Delta \V \right| \leq 4.5~\text{\textmu{}s}  
\end{equation}

\noindent
where $t_0$ is defined by the V channel with the largest amplitude, $t_i$ is
determined from the V channel of interest, and $\Delta \V$ is defined 
as the absolute channel number difference of the two signals.  For
example, for two V-wire signals occurring on channel 39 and 37, $\Delta \V$ would
be 2.  This is because ``outer'' V signals are
reconstructed earlier in time than the ``central" (largest-amplitude) V-signal
and the arrival time difference grows roughly linearly
with the number of channels between the signal and the central signal. 
This occurs because the V-wire channels further from the drifting charge become
shielded by the nearby wires as the charge nears the V-wire plane.  
In contrast to U-wire bundles, only the time of the \emph{largest} V-wire signal
is chosen as the time of the whole V-wire bundle; no weighted average is
performed.  Using the weighted time average in the V-wire bundle has been found
to dilute the time correlation between U-wire and V-wire bundles. 

The bundling of APD signals is performed solely based on time,
associating signals if they arrive within 6~\mus{} of one another, with the
timing defined by the sum of the integration times in the APD electronics
(see \cref{tab:ShapingTimes}).  When multiple APD signals are
grouped to form a scintillation bundle, the time of the scintillation bundle
is calculated by performing the weighted average over the time and energy of
the component signals. 

\subsubsection{Determining two-dimensional (2-D) position}
\label{sec:ClusOverview}
A determination of the 2-D event position is achieved by grouping
together U- and V-wire bundles in their most likely configurations to generate
charge ``clusters"  The clustering stage employs probability density functions
(PDFs) that describe how likely it is that a particular U-bundle is associated
with a given V-bundle.  There are three PDFs to: (1) describe the time
difference between U- and V-bundles, (2) describe the consistency between
the sum of signal amplitudes in the U- and V-bundles, and (3) ensure the
resulting (U,V) coordinate is physically allowed (not all U- and V-wires
intersect one another because of the hexagonal shape of the grid frame).  These
PDFs are described in more detail in \cref{sec:ClusPDFs}.  

The PDFs used to cluster U- and V-wire bundles together are dependent on the Z
position of the U-bundle.  To calculate them, an associated scintillation
bundle is found in the following way: from all scintillation
bundles that occurred between 3~\mus{} \emph{after} and 3~\mus{} plus the maximum
drift time \emph{before} the U-wire bundle, the one scintillation bundle with
the \emph{smallest} absolute time difference from the U-wire bundle is chosen.
The expansion of the search window on both sides by 3~\mus{} is again given by the
integration times of the APD transfer functions.  If no scintillation
bundle lies in that time range, the Z position is set as undetermined and the
particular U-/V-bundle will not be further clustered.  

The negative log of the product of the three PDFs described above defines a
test metric, the ``cost", which is used as a measure of how well a U-bundle
matches a V-bundle.  This may be translated into the following expression: 
\begin{equation}
   {\rm cost} = \sum_{i=1,2,3} -\ln P_i
\end{equation}
where the sum is over the three PDFs.  A lower cost indicates a better match,
or a higher likelihood for a given configuration.  The matching algorithm
rigorously tests all
combinations of the V and U-bundles with one another, including whether
or not multiple bundles of one type may actually correspond to a single bundle
of the other type.  The best matching configuration is the one for which the
sum of the cost divided by the number of connections is minimal, where the
number of connections is defined as the smaller of the number of U-bundles or
V-bundles.  Once this has been determined, a charge cluster is created for each
of the connections within the matching configuration and each charge cluster is
linked with its associated scintillation bundle.  

It is possible that clustering fails to associate one or more of the U-, V-, or APD
bundle types together, resulting in a cluster without full 3-D position.  This may occur due to a
clustering error or because signals are not found because they are below
threshold.  This produces
an associated error on the \twonubb{} measurement, which is quantified in
\cref{sec:part_rec_cut}.  It is also possible for clustering to skip an event
completely if too many signals are found.  This may introduce an error on the
\twonubb{} efficiency if reconstruction mistakenly finds too many signals and
these events are then ignored.  However, the total number of skipped events is
$<0.18$\% of the final \twonubb{} counts.  We choose to assign an additional
0.18\% error on the \twonubb{} measurement to account for the possibility that
events are mistakenly skipped by the clustering stage in reconstruction.

To determine the analysis threshold, the efficiency to find a signal on the
three types of channels (U-wire, V-wire, and APD) was studied versus deposition
energy.  The three types of channels exhibit efficiencies with different
dependencies on deposition energy as well on position.  An analysis threshold
(700~keV) was chosen where all channel types demonstrated 100\% signal-finding
efficiency independent of the event deposition position.

\section{Calibrations and corrections}\label{sec:CalibrationAndCorrections}
\subsection{TPC source calibrations}

\begin{figure}[ht]
    \includegraphics[width=\columnwidth]{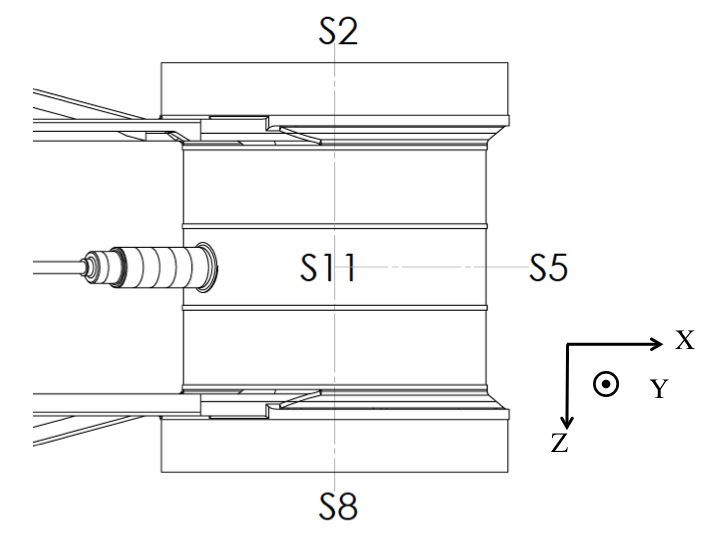}
  \caption{The calibration source locations around the TPC vessel 
  as viewed from above. The XYZ coordinates of the source 
  locations are: S2 = (0.0, 0.0, -29.5~cm), S5 = (25.5~cm, 0.0, 0.0 ), 
  S8 = (0.0, 0.0, 29.5~cm), and S11 = (0.0, 25.5~cm, 0.0). Not shown:
  S17 = (0.0, -25.5~cm, 0.0).  }
  \label{fig:locations}
\end{figure}

\isot{Cs}{137}, \isot{Co}{60}, and \isot{Th}{228} sources are 
utilized to calibrate the TPC response to gamma radiation. 
The sources have been 
selected to span the energy range of interest.
The source activities, listed in \cref{tab:sourceRates},
were chosen to collect calibration data quickly while 
not saturating the DAQ system. The \isot{Th}{228} 
source is deployed every few days to a position 
near the cathode to monitor the electron lifetime and 
measure the energy response. Occasionally all 
three sources are deployed in series to other positions 
around the TPC vessel for comprehensive calibration studies.
These positions are shown in \cref{fig:locations}.

\begin{table}[ht]
\centering
\begin{ruledtabular}
    \renewcommand{\arraystretch}{1.3}
\begin{tabular}[c]{ccc}
\toprule
Source & Activity (Bq) & Half-life (years) \\\hline
$^{\phantom{0}60}$Co & $\phantom{0}530\pm\phantom{0}6$ & $\phantom{0}5.27$\\
$^{\phantom{}137}$Cs & $\phantom{}2820\pm\phantom{}33$ & $\phantom{}30.1$\\
$^{\phantom{}228}$Th & $\phantom{}1417\pm\phantom{}17$ & $\phantom{0}1.91$\\
\bottomrule
\end{tabular}
\end{ruledtabular}
\caption{EXO-200 calibration sources. The activities shown here are 
referenced to September 1, 2009 and were verified by gamma-ray spectroscopy
within the collaboration. Other sources with greater activity are also
available for deployment.}
\label{tab:sourceRates}
\end{table}
		
\subsection{TPC channel-based corrections}\label{sec:ChannelCorrections}

The absolute U-wire channel gains were measured prior to Run 2a using a 
pulser coupled to the front-end  electronics through a precision (1\%) capacitor.  The gain
value, measured in units of electrons per ADC count, is calculated by a
linear fit to the measured amplitude versus the injected charge.
The stability of the gain values is monitored daily using a charge injection circuit
which is integrated into the front-end card. This daily charge injection run is also
used to monitor the time constant of the third differentiation stage of the preamplifier 
(see \cref{tab:ShapingTimes}).

The relative U-wire gain values are also determined with
\isot{Th}{228} source calibration data using the pair-production peak of
2615 keV gammas.
The resulting channel-to-channel gain values are strongly
correlated with those determined from the charge injection runs, as expected.
The mean gain is $\sim380$~electrons/ADC unit,
with 30\% variation over all channels.   The observed drift of the U-wire gains 
is $< 1$\% over the Run 2a dataset, with $<0.1$\% relative channel-to-channel drift.
For the \twonubb{} half-life measurement
the gain values determined from the pair-production peak are used.

The V-wires are not directly used in the energy measurement, but variations in their
relative gain can affect position reconstruction and clustering. We
correct for these variations using the precision pulser charge injection data.
The V-wire gains as determined from these calibrations are found to vary
from 300 to 360~electrons per ADC count. These gain values are accurate within
2\textendash3\%.

The APD channel gains are monitored periodically using the external laser pulser. 
The observed channel-to-channel gains vary by 12\%,
and the time variation is 1\% over the Run 2a dataset.
However, for the \twonubb{} half-life measurement,
the APD signals are not explicitly gain corrected on a channel-by-channel basis, as
gain variations are absorbed in the light map correction described below.

\subsection{TPC position-based and time-based corrections}
\label{sec:PosBasedCorrections}

\subsubsection{APD light map}

The amount of scintillation light collected by the APDs 
depends on the location of the energy deposition.
This variation is caused by differences in the solid 
angle covered by the APDs and by their gain differences.
3-D correction
functions are used to account for this position dependence. Such correction functions are 
generated from \isot{Th}{228} calibration runs with the source placed at the two anodes 
and three positions around the cathode plane.
The detector volume is divided into 1352 spatial voxels (13 radial bins, 8  azimuthal bins, 
and 13 Z~bins).
The bin widths are chosen to ensure adequate statistics, and to optimally map
the response in the regions with a high light collection gradient.

The light map is normalized such that the mean response is 1. A continuous correction
function, \(f(r,\phi,z)\), is created with a trilinear interpolation between the centers of the voxels
in the light map. 

\begin{figure}[!tbhp]
\includegraphics[width=\columnwidth]{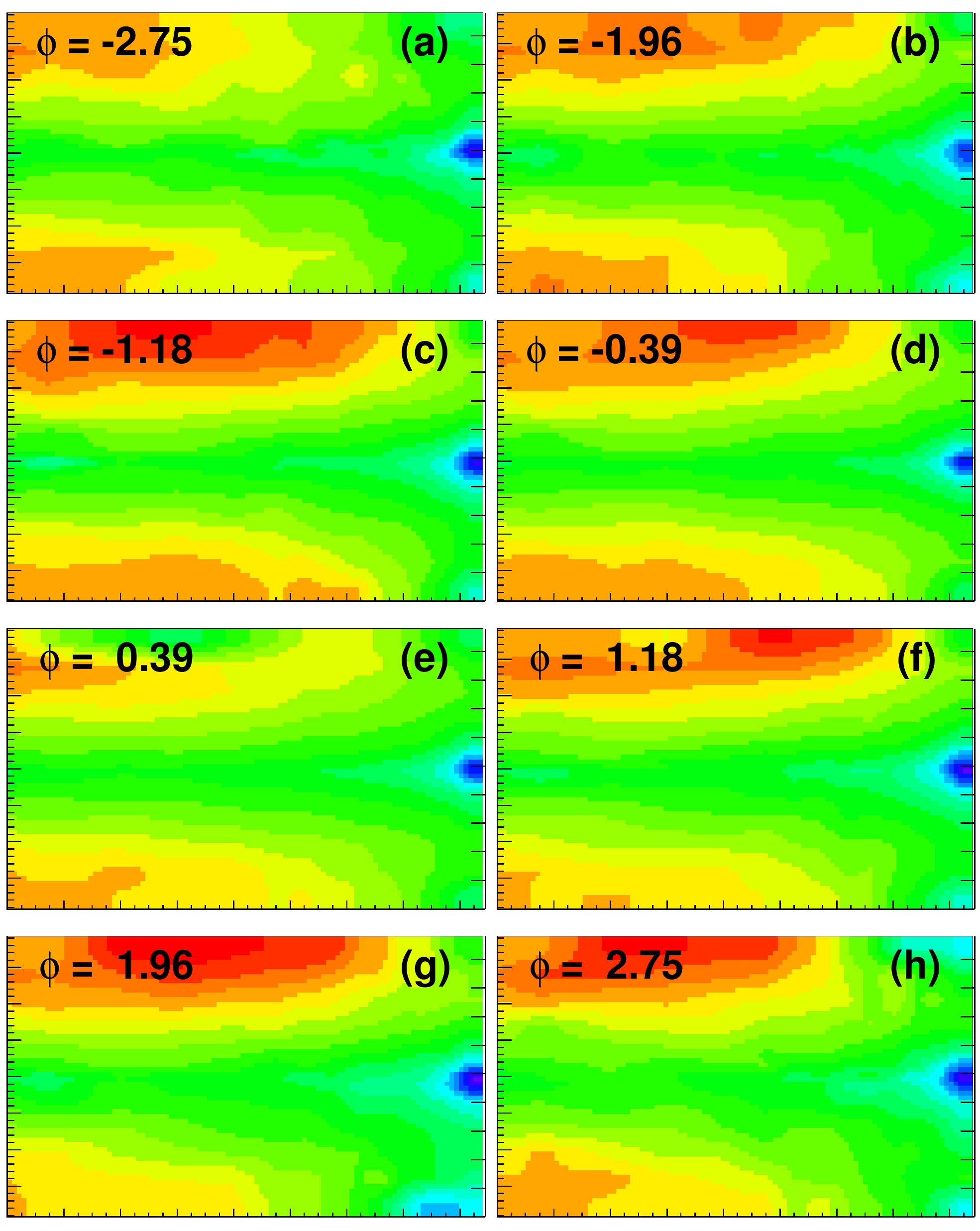}
\includegraphics[width=\columnwidth]{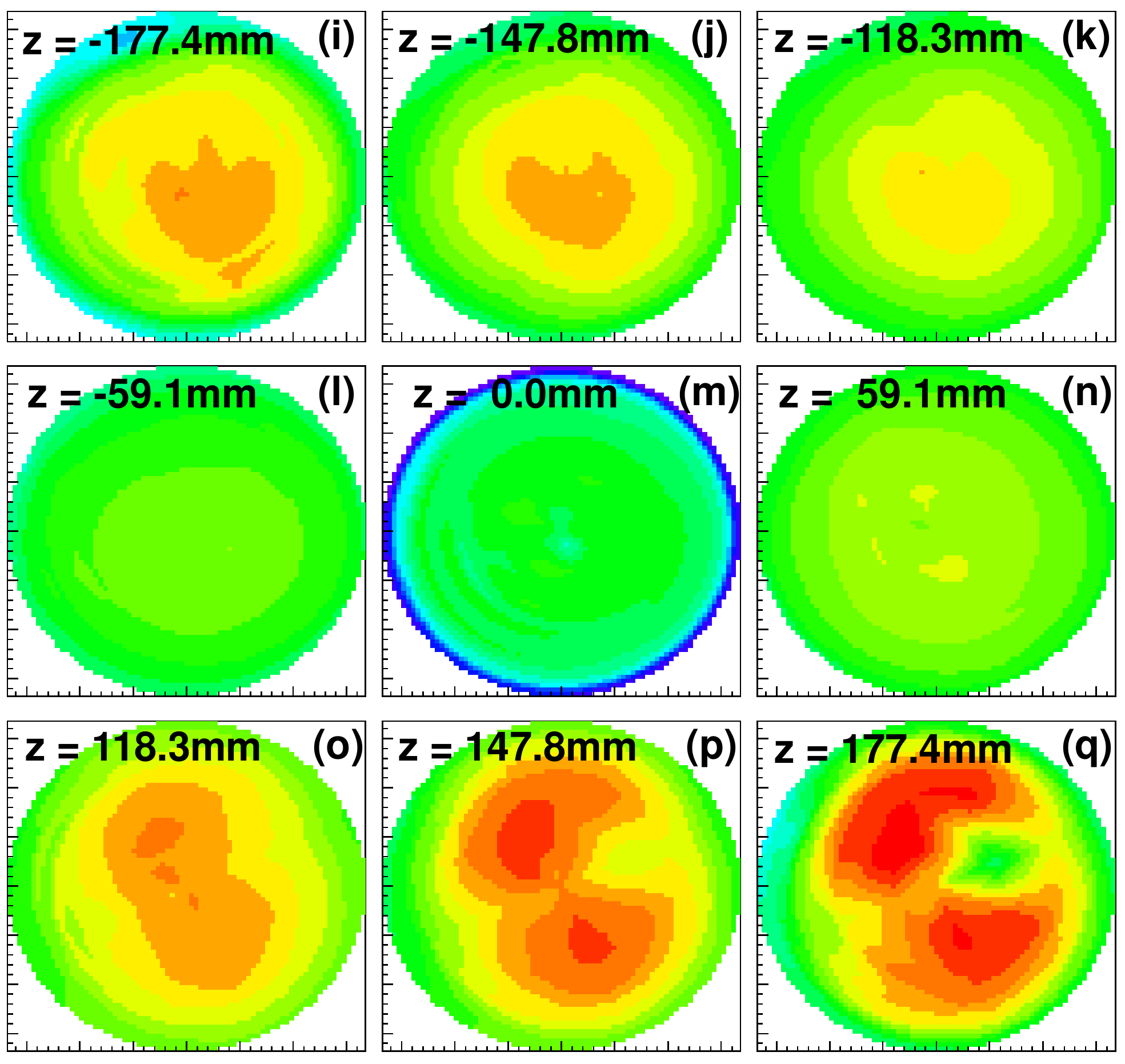}
  \caption{(Color) The light response function \(f(r,\phi,z)\) from September 22, 2011 until February
      20, 2012 is shown here indicated by the color coding.  For both plots the 
      correction factor ranges from 0.70 (blue) to 1.19 (red). 
      Panels (a) \textendash{} (h): the function evaluated at 8 discrete values of the
      azimuthal angle $\phi$ over the full range of \(r\) (horizontal axis,
      $0~\mathrm{mm} < r < 168~\mathrm{mm}$) and Z (vertical axis,
      $-192~\mathrm{mm} < \mathrm{Z} < 192~\mathrm{mm}$). Panels (i)
      \textendash{} (q): the function evaluated at 9 discrete values of Z over
      the full range of X (horizontal axis, $-168~\mathrm{mm} < \mathrm{X} <
      168~\mathrm{mm}$) and Y (vertical axis, $-168~\mathrm{mm} < \mathrm{Y} <
      168~\mathrm{mm}$). A second response function (not shown here) is used
      from February 20, 2012 until
      April 15, 2012.  }
\label{fig:correction_slices}
\end{figure}

Because maintenance was performed on the APD front-end boards in the middle of Run 2a, two 
light maps are used for this data. The first light map covers the period from October 1, 2011 
until February 20, 2012, and the second covers the period from February 20, 2012 until 
April 15, 2012. Some representative sections of the first light map are shown 
in \cref{fig:correction_slices}.

For a SS event, the correction function is applied by multiplying
the sum of the two APD plane signals by \(1/f(r,\phi,z)\), while
for a MS event, a correction factor is deduced
by taking the appropriate charge-cluster energy-weighted sum.
The light map correction function improves the scintillation-only energy resolution
at 2615~keV from 7.9\% to 6.0\% for SS events and from 8.1\% to 6.3\% for MS events, respectively.
The largest correction factor within the fiducial volume is $\sim15\%$ (the fiducial cut 
near the cathode eliminates the region of the detector that sees the largest
gradient in the correction function).
	
\subsubsection{Electron lifetime correction}\label{sec:purity_corrections}
\begin{figure}[htbp]
\centering
\includegraphics[width=\columnwidth]{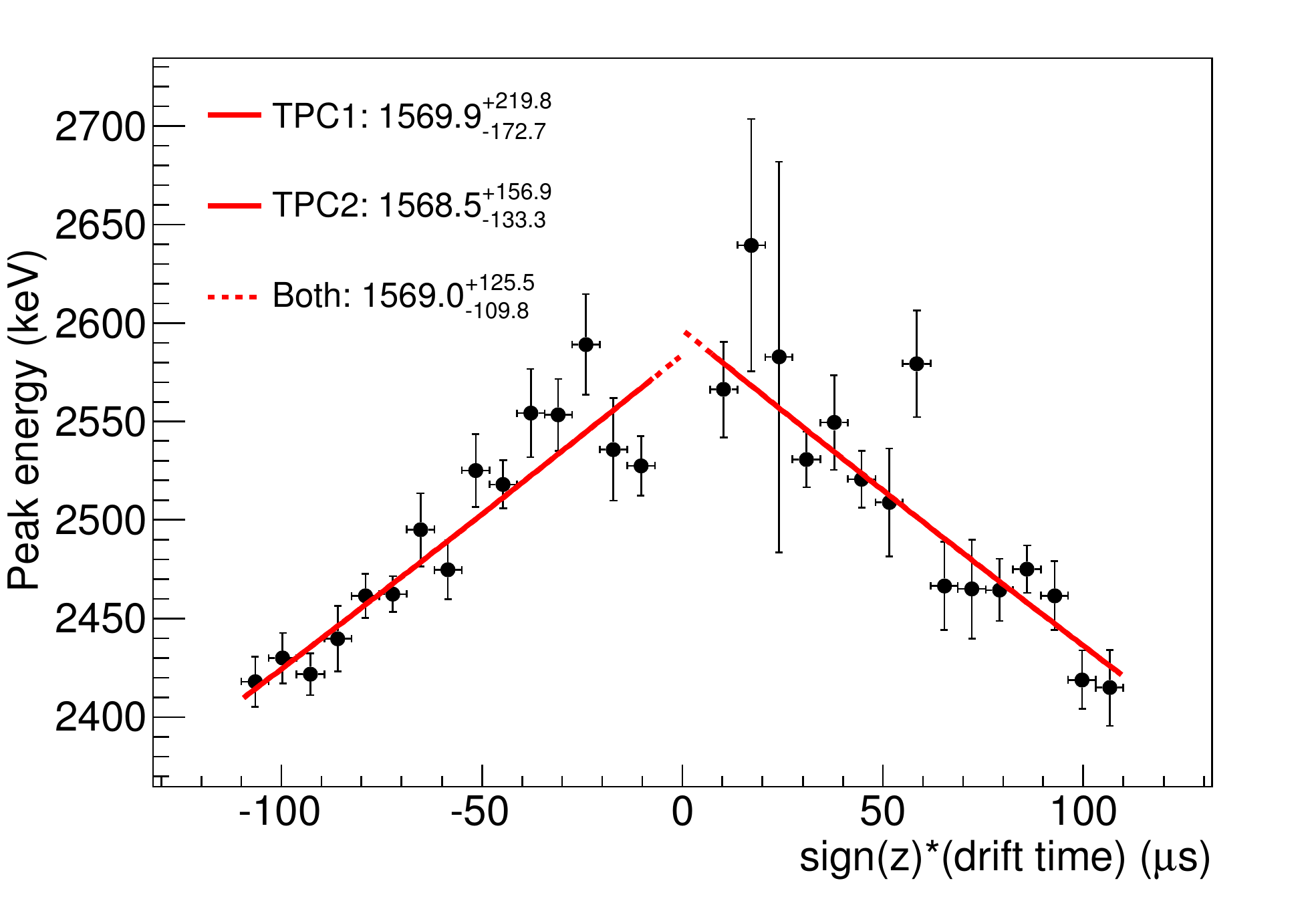}
\caption{(Color online) An example electron lifetime $\tau_{\rm e}$ measurement  
obtained by fitting a decaying exponential to the $^{228}$Th full-absorption 
peak energies binned by drift time. TPC~2 (negative Z) is assigned a 
negative drift time for convenience in visualization. Fits to 
separate $\tau_{\rm e}$ for each TPC and the combined fit value are shown. 
This data is from a single source calibration run.}

\label{fig:elfit}
\end{figure}

Electrons drifting in LXe can be captured on electronegative impurities leading to an 
exponential decrease with time. This attenuation is described by

\begin{equation}
N_{\rm e}(t) = N_0 \exp (-t/\tau_{\rm e})
\label{eq:exponentialtaue}
\end{equation}

\noindent
where \(N_0\) is the original number of electrons,  \(\tau_{\rm e}\) is the
electron lifetime, and \(t\) is the drift time. 
We correct for this attenuation by measuring the electron lifetime
every few days using the \isot{Th}{228} source deployed near the cathode.
We divide the data in each TPC into 16 drift time bins and fit a Gaussian plus
error function model to the full absorption peak in each bin. The central value
of the peak is plotted versus drift time as shown in \cref{fig:elfit}, and the
electron lifetime is extracted from a fit to \cref{eq:exponentialtaue}.
It is found that the
goodness-of-fit function is asymmetric around the minimum and
larger electron lifetime values have larger uncertainties.

\begin{figure}[htbp]
\includegraphics[width=0.97\columnwidth]{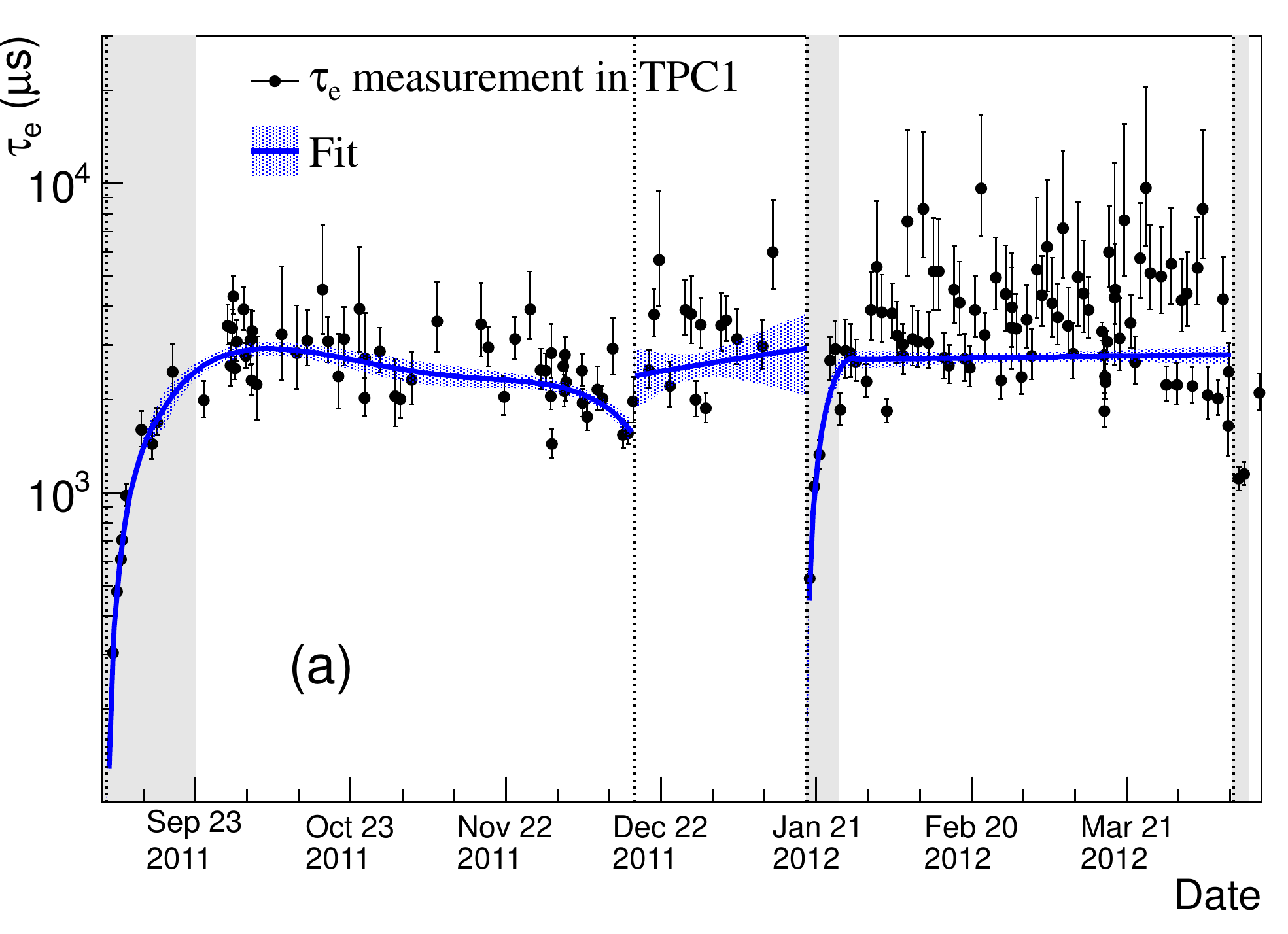}
\includegraphics[width=0.97\columnwidth]{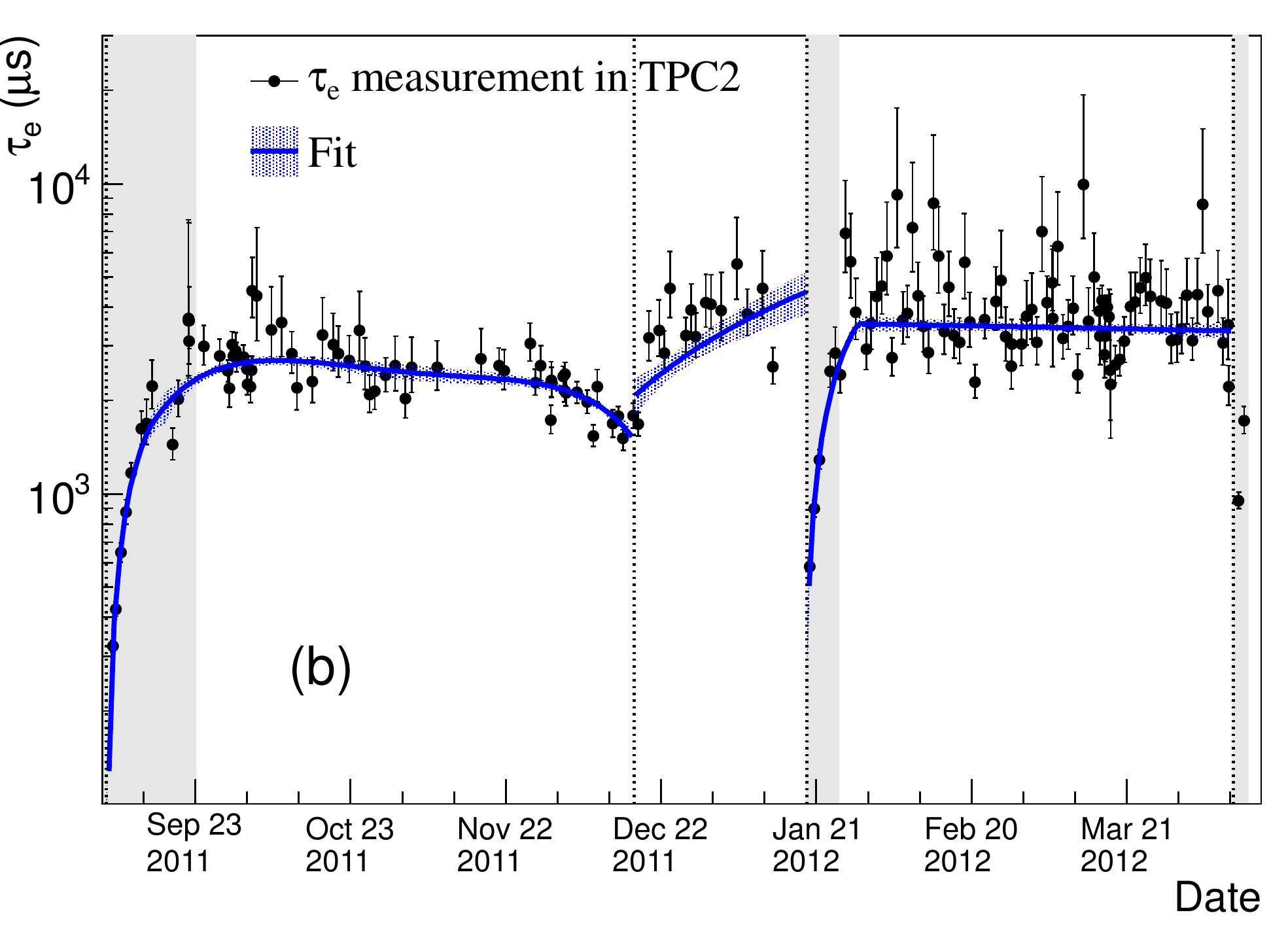}
\label{fig:el_time_variation:TPC2}
\caption{(Color online) The fit of a piecewise polynomial to electron lifetime
in TPC 1 (positive Z, (a)) and TPC 2 (negative Z, 
(b)). The colored bands show the 68\% confidence
interval on the fit.  The vertical dashed lines indicate discontinuities in the
electron lifetime due to interruptions in the xenon recirculation or xenon gas
feed events. The vertical shaded regions indicate time periods which were
excluded from the final dataset due to poor electron lifetime.} \label{fig:el_time_variation}
\end{figure}

The electron lifetime varies over time due to small changes in the xenon recirculation
rate, occasional interruptions due to xenon pump maintenance or failure and power outages, and
events where xenon gas is added to the detector by the detector monitoring system. 
To account for this, a piecewise polynomial is
fit to the measured \(\tau_{\rm e}\)  history, as shown in \cref{fig:el_time_variation}. 
Separate \(\tau_{\rm e}\) are used for each TPC to allow for spatial variation in the 
LXe impurity content, although the two TPCs track each other quite well.
A correction factor of $\exp(t/\tau_e)$ is applied to all ionization signals
in the data by evaluating the \(\tau_{\rm e}\)
history function at the time of each event.

\subsubsection{Shielding inefficiency corrections}
\label{sec:grid_eff_correction}

The shielding inefficiencies of the V grids (see e.g.~\cite{Gook12}) 
produce a small residual dependence of the U-wire pulse amplitude on the Z
position of the charge deposition.    We measure this effect in the data by
fitting the purity-corrected peak-position of the $^{208}$Tl gamma line at
2615 keV as a function of Z.
We fit this data to the function:

\begin{equation}
    E_{\rm meas} = \frac{E_0}{1+p_0e^{(|\Z|-\Z_{\rm max})/p_1}}
\end{equation}

\noindent
where $E_{\rm meas}$ is the purity corrected peak energy, $E_0$ is the true
peak energy, Z is the Z coordinate of the energy deposit in units of mm and 
$\Z_{\rm max}$ is the maximum drift distance (192.5~mm).
We find that the best fit parameters
are ($p_0$, $p_1$) = (0.043, 7.02~mm) and (0.064, 8.09~mm) for one-wire and
two-wire charge deposits, respectively. The function is then inverted to correct
the measured charge deposit energy to the true energy. For events in the fiducial region
this correction is much less than 1\%.
	
\subsection{Rotated energy measurement}\label{sec:AntiCorr}

\begin{figure}[ht]
    \includegraphics[width=\columnwidth]{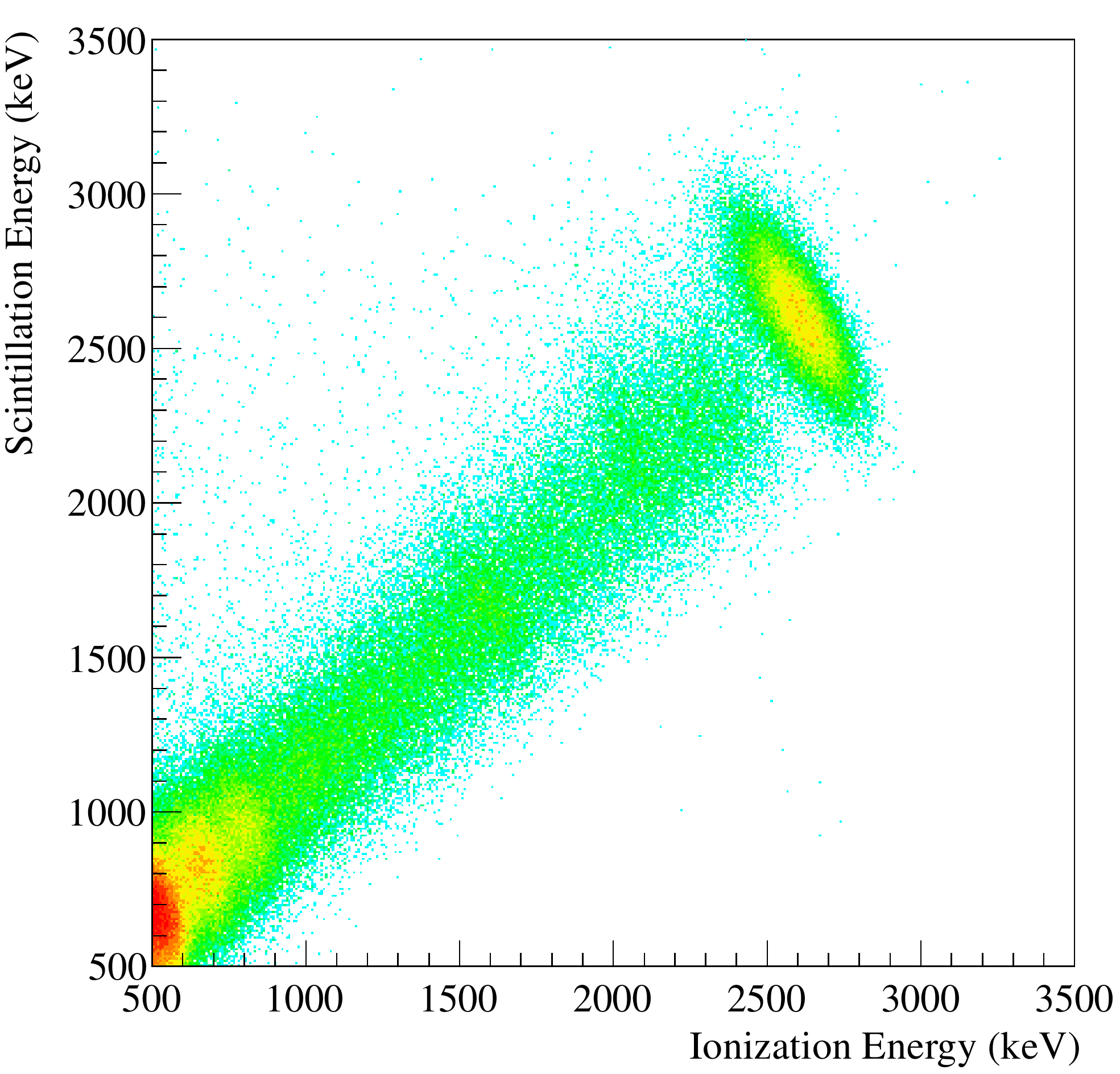}
  \caption{(Color online) Anti-correlation between ionization and scintillation
for SS events from a $^{228}$Th source. The prominent island at the upper end 
of the distribution is the 2615 keV gamma line of $^{208}$Tl.}
  \label{fig:antiCorr}
\end{figure}

To optimize the energy resolution of the detector we take advantage of the
microscopic anti-correlation between ionization and scintillation in LXe~\cite{Conti2003}.
This effect is illustrated for \isot{Th}{228} source calibration data
in \cref{fig:antiCorr}.
The optimal energy variable is
calculated by combining the charge, $E_I$, and scintillation, $E_S$, measurements
with appropriate weights according to
\begin{equation}
E_R = E_{S}\cdot\sin(\theta^R) + E_{I}\cdot\cos(\theta^R)
\label{eqn:rotE}
\end{equation}
where $E_R$ is the ``rotated'' energy and $\theta^R$ is the rotation angle.

\begin{figure}[ht]
  \includegraphics[width=\columnwidth]{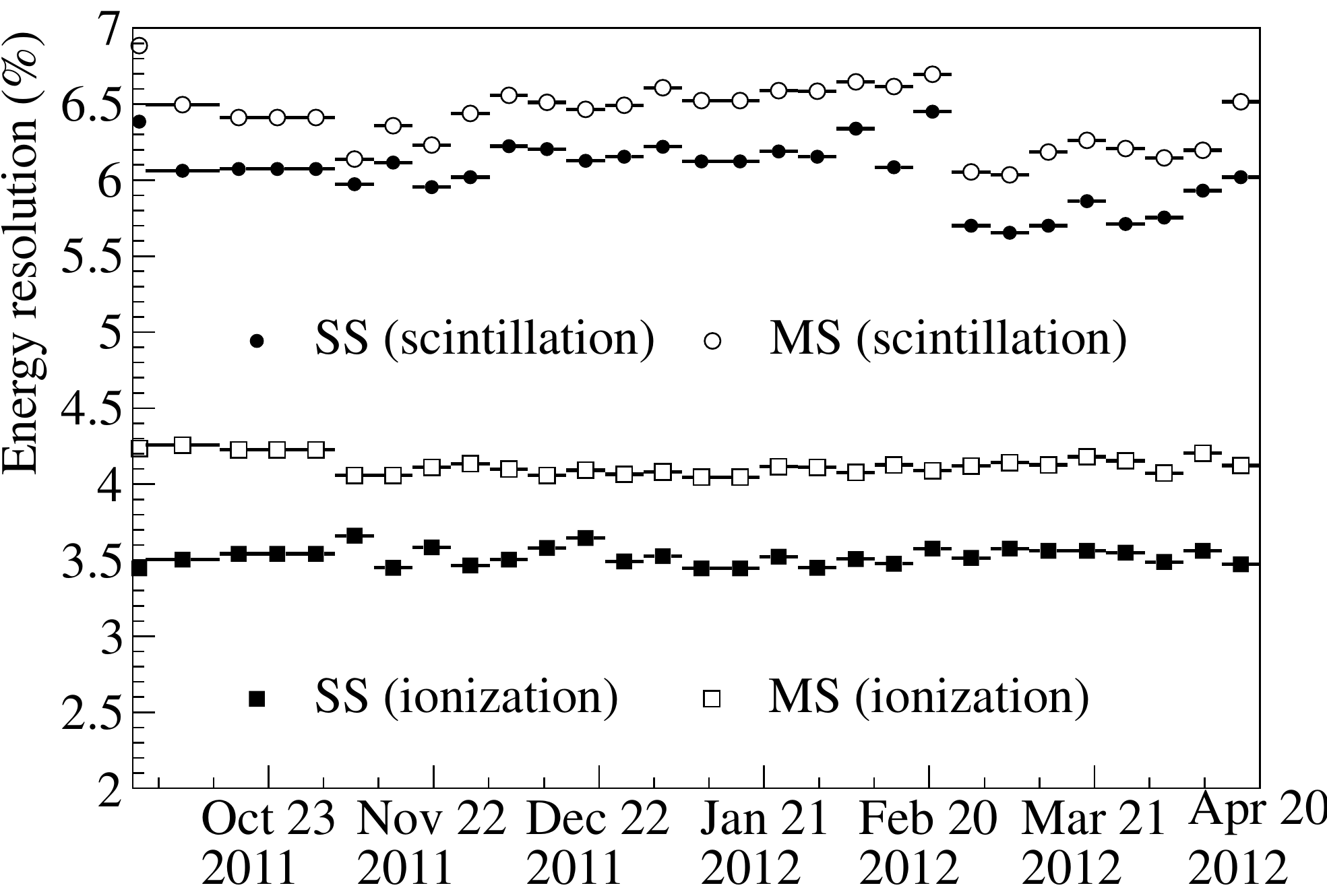}
  \caption{Energy resolutions of ionization and scintillation channels during \RunTwoA{}.}
  \label{fig:Thresolutions}
\end{figure}

We find that the energy resolution of the ionization channel is nearly constant in
time with a value of $\sigma / E =$ 3.5\% and 4\% at 2615 keV for
SS and MS events, respectively. The scintillation-only energy resolution,
however, does show significant time variation as shown in \cref{fig:Thresolutions}.
The exact cause of this variation is under investigation, but it is likely
related to noise in the front-end electronics of the APD channels.

Due to the variation in the scintillation energy resolution, the optimal
$\theta^R$  to use in \cref{eqn:rotE} also varies with time. We
measure $\theta^R$ weekly with \isot{Th}{228} source calibration data
and apply it to calculate the appropriate rotated energy for events
in the ``low-background data" (physics data) as a function of their date. This is done
separately for SS and MS events. We apply the same rotation angle to 
all events regardless of the event energy. 

\begin{figure}[ht]
  \includegraphics[width=\columnwidth]{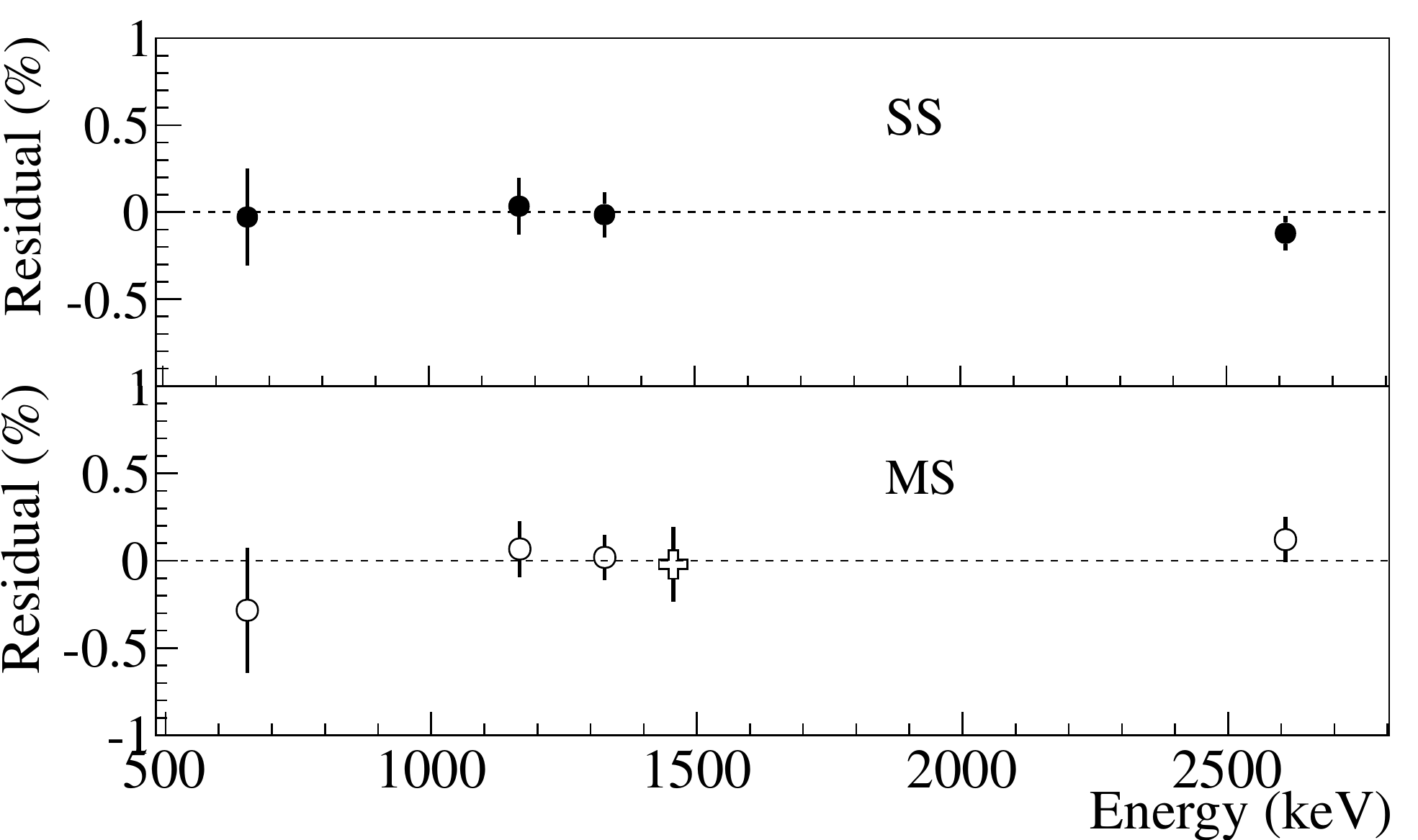}
  \caption{Residuals between the calibrated source energies and the true energies for SS and MS events.
  Also shown is the residual for the \isot{K}{40} peak (cross, MS only) which is obtained from 
  the low-background data set.}
  \label{fig:residual}
\end{figure}

After applying all corrections, the residuals defined as
$(E-E_{\rm true})/E_{\rm true}$, are 0.36\% for \isot{Cs}{137}, and 0.17\% for
\isot{Co}{60} and \isot{Th}{228} sources, as shown in \cref{fig:residual}. The
energy measured for the $^{40}$K peak in the low-background data
has a residual
of 0.21\% and is also shown in \cref{fig:residual}.

\begin{figure}[ht]
  \includegraphics[width=\columnwidth]{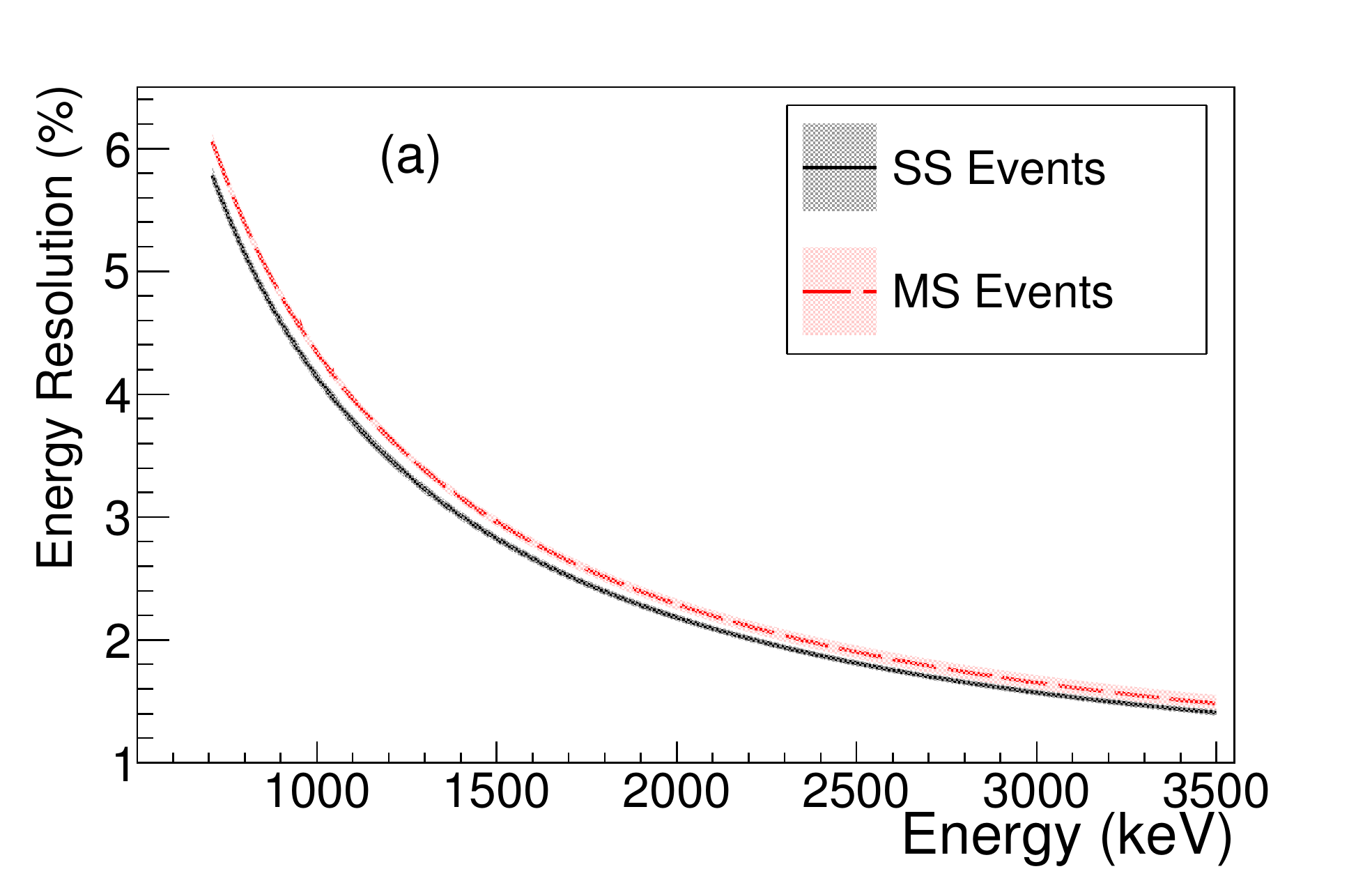}
  \includegraphics[width=\columnwidth]{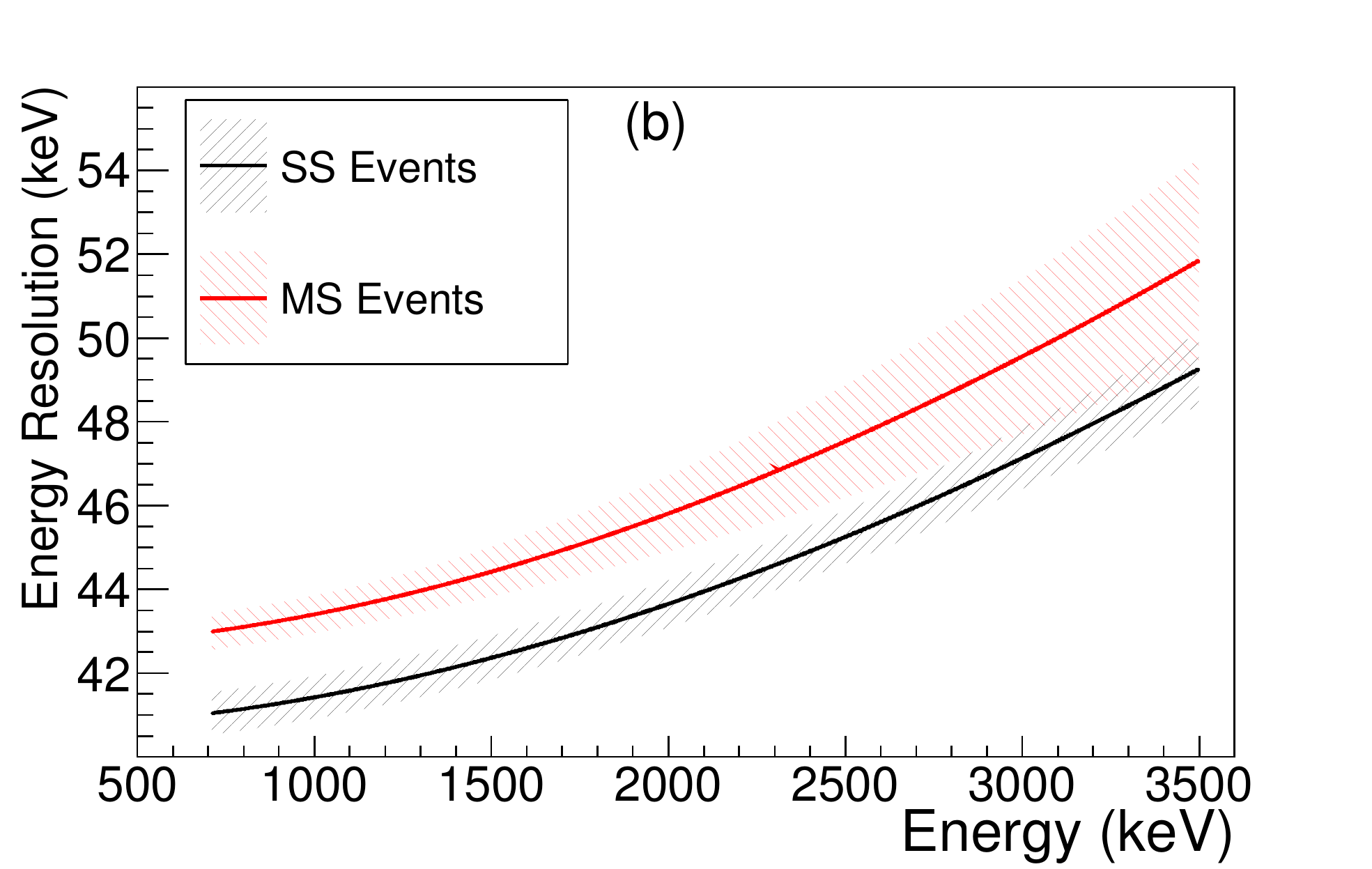}
  \caption{(Color online) (a) The fractional energy resolution ($\sigma / E$)
  for SS and MS events. The width of the two lines indicates the uncertainty.
  (b) The absolute energy resolution for SS and MS events. The cross-hatched
  areas indicate the uncertainty.} \label{fig:resol-curve}
\end{figure}

The energy resolution $\sigma (E)$ is parameterized as a function of energy,
$\sigma^{2}(E)=\sigma^2_{\rm elec}+bE+cE^{2}$, where $\sigma_{\rm elec}$ is the
electronic noise contribution, $bE$ represents statistical fluctuations in the
ionization and scintillation, and $cE^2$ is regarded as a position- and
time-dependent broadening. This parameterization is used to smear the Monte
Carlo data set to produce the energy PDFs. To estimate the covariance
matrix of the resolution parameters, an iterative approach (inspired
by~\cite{Patterson07}) was developed to fit the smeared \isot{Th}{228} Monte Carlo
spectrum to the calibration data set, with an underlying assumption that all
the parameters compose a multivariate Gaussian distribution. We calculate the
resolution curve $\sigma(E)$ weekly for source agreement studies (see
\cref{sec:source_agreement}). In order to take into account the time
variation of the energy resolution, the \isot{Th}{228} Monte Carlo data set is initially
smeared by the weekly resolution parameters, then weighted by the low
background live-time fraction in each week. In the end the Monte Carlo \isot{Th}{228}
data sets are combined and fitted by using the iterative approach to get the
averaged resolution parameters. The time-averaged energy resolution is plotted
versus energy in \cref{fig:resol-curve}, and its values at the
$\beta\beta$ decay $Q$ value are $1.84 \pm 0.03$ \% ($1.93 \pm 0.05$ \%) SS
(MS) events, respectively. These values are slightly different than those reported
in Ref.~\cite{Auger:2012ar} because we now average over time.

\section{Measurement and Monte Carlo simulation agreement}\label{sec:source_agreement}
The maximum-likelihood fit requires, as an input, probability density
functions (PDFs) in energy and standoff distance space for both the signal and
backgrounds. Separate PDFs, one containing SS events and the other containing
MS events, are generated using the EXO-200 simulation package. The energy
observable in these simulated PDFs is derived from a convolution of the
measured energy resolution function (\cref{sec:AntiCorr}) and the deposited
energy given by the simulation.

An important mechanism for determining the accuracy of simulation-generated
PDFs is the direct comparison of the detector response to certified calibration
sources with the full detector simulation of the same source.  Calibration
source data, however, differs from the \twonubb{} signal in several significant
ways.  The \twonubb{} events are predominantly single-site ($\sim95\%$ SS)
whereas the interactions of photons from the external sources are mostly
multi-site ($\sim20\%$ \textendash{} 50\% SS, see e.g.~\cref{fig:ss_frac}).  In
addition, \twonubb{} events occur uniformly throughout the LXe volume while the
source of calibration events lies in a single point external to the LXe volume.
The photons emitted by the calibration source are strongly attenuated by the
LXe, concentrating the interactions in the active xenon nearest to the
deployment position of the calibration source.

Because of the differences outlined above, the verification of source agreement
is augmented by direct analysis of the low-background data wherever feasible
(see \cref{sec:part_rec_cut}).  The calibration source/simulation agreement
studies are partitioned as follows: (A) Source shape agreement, which compares
the shapes of distributions derived from both; (B) Source rate agreement, which
considers the ability of the simulation to predict the activity of the source
that is observed; and (C) SS fraction agreement, which verifies the fidelity of
SS/MS discrimination predicted by the simulation.

\begin{figure*}
 \includegraphics[width=0.97\textwidth]{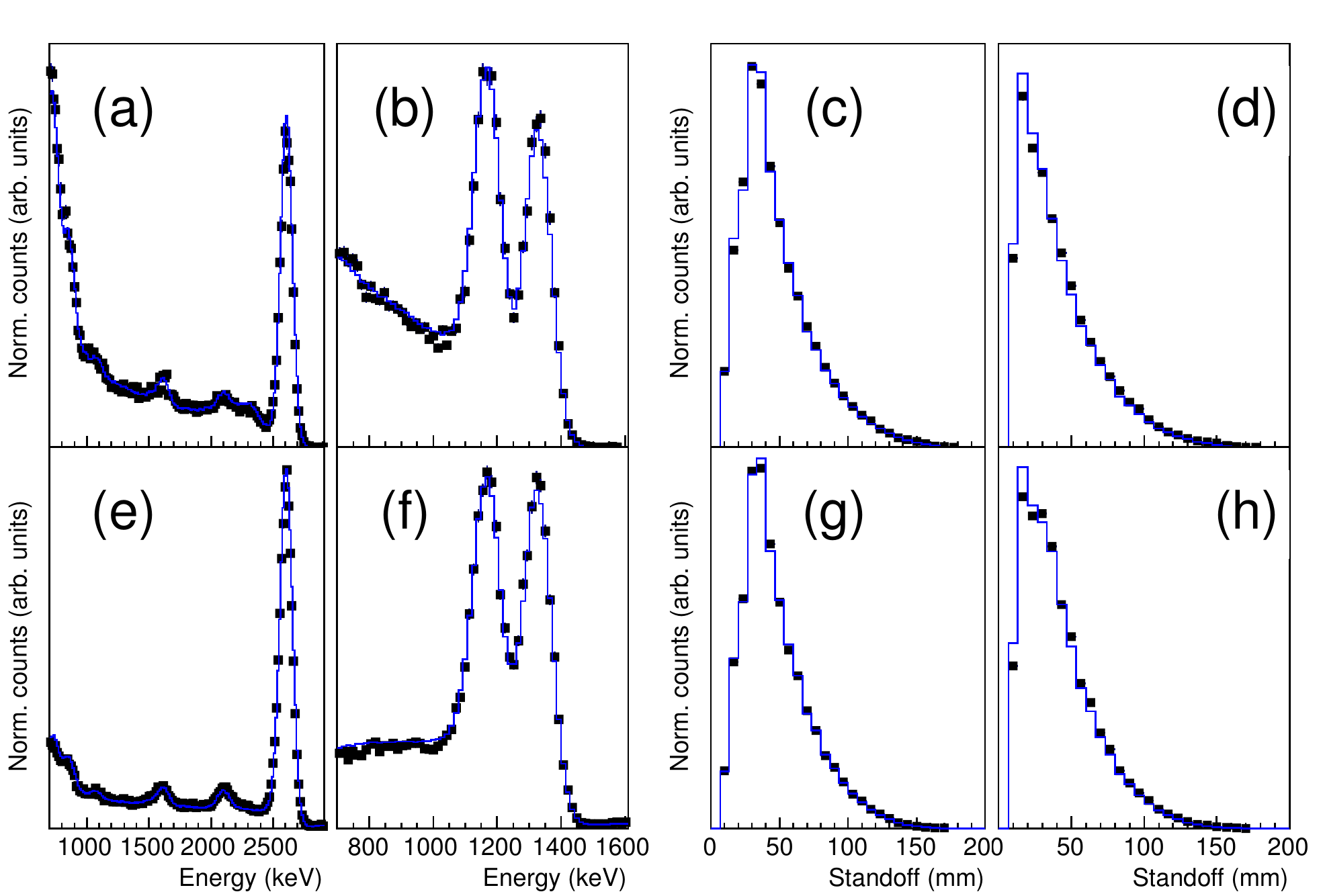}

  \caption{(Color online) Comparison of shape between data (points) and
      simulation (histograms) for energy and standoff distance distributions.  All
      distributions have been normalized to 1 and in all cases, the top
      (bottom) plots are single-site (multi-site) distributions. (a), (c), (e),
      and (g) are for a \thsrc{} source near the cathode.  (b), (d), (f), and (h) are
      for a \cosrc{} source located near the anode.  
}

 \label{fig:SourceCompare}
\end{figure*}

        \subsection{Source shape agreement}
        \label{sec:shape_agreement} 

The shape of \thsrc{} and \cosrc{} calibration source energy spectra and
standoff distributions were compared to the those from simulation at various
calibration source deployment positions.  The \isot{Cs}{137} source was not
considered as its main feature (the 662~keV gamma-ray full-energy absorption
peak) lies below the analysis energy threshold (700~keV). The primary source
deployment positions near the cathode (S5 and S11) illuminate the bulk of the
active LXe in both TPCs. Source calibrations near either anode (S2 and S8) are
additionally used to confirm adequate modeling at each end of the TPC.

\begin{figure}

 \includegraphics[width=0.45\textwidth]{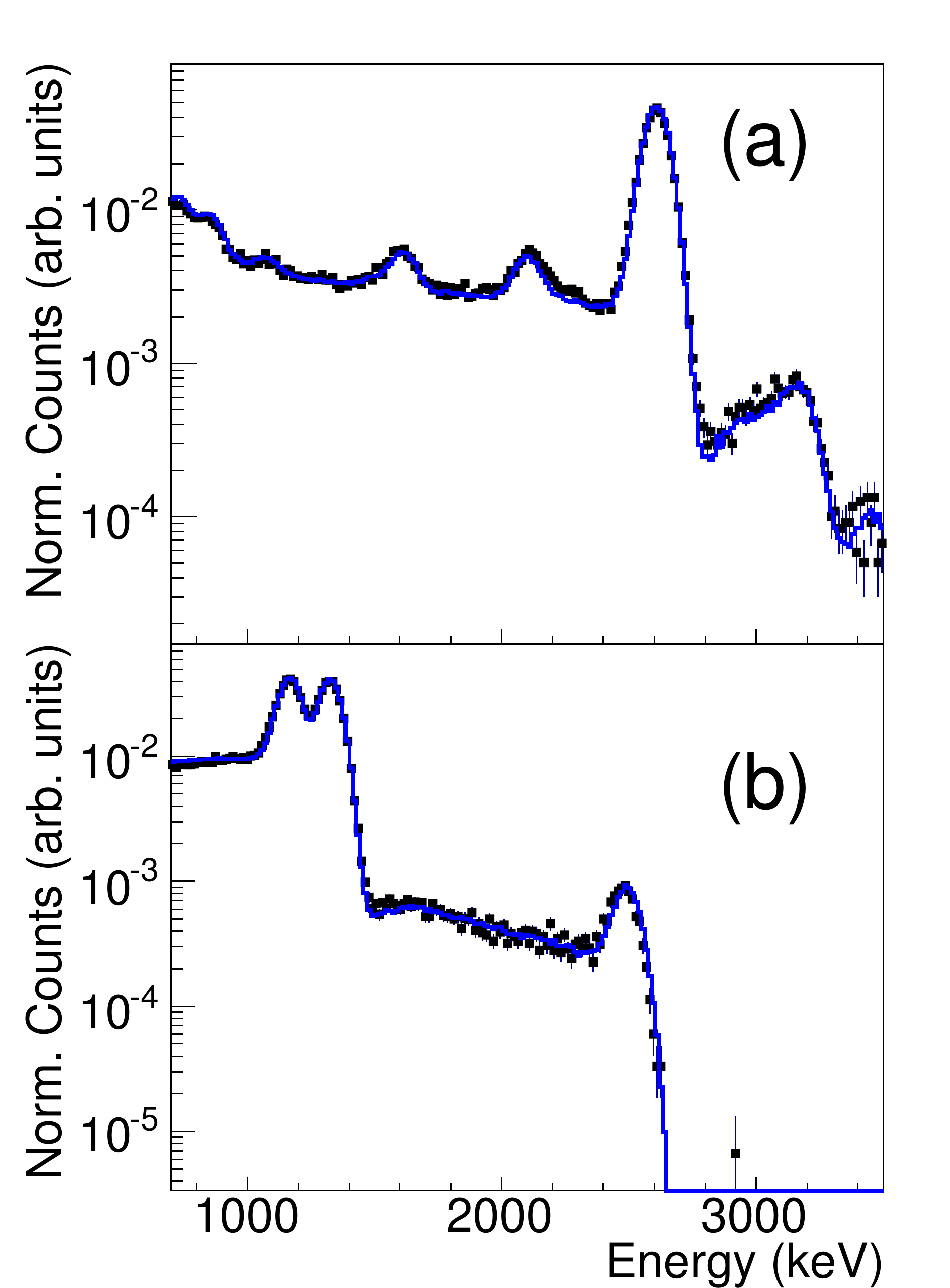}
 \caption{(Color online) Comparison of multi-site energy distribution between
     data (points) and simulation (histograms) using \thsrc{} (a) and \cosrc{}
     (b) sources near the cathode (S5).  Note the excellent agreement of the
     full-energy-deposition gamma-ray peaks all the way up to the summation
     peaks.  } \label{fig:sum_peaks} 
 \end{figure}

\Cref{fig:SourceCompare} demonstrates, for the shape of the energy and standoff
distance distributions, the agreement between \thsrc{} calibration source (at
S5) data and simulation and the agreement between \cosrc{} calibration source
(at S2) data and simulation. \Cref{fig:sum_peaks} highlights the Gaussianity of
the \thsrc{} and \cosrc{} (at S5) full-energy deposition peaks along with the
excellent agreement achieved between both sets of calibration data and the
simulation in the vicinity of the respective summation peaks. All of these
distributions are normalized to unity to study only the shape agreement.

\begin{figure}
      \includegraphics[width=0.47\textwidth]{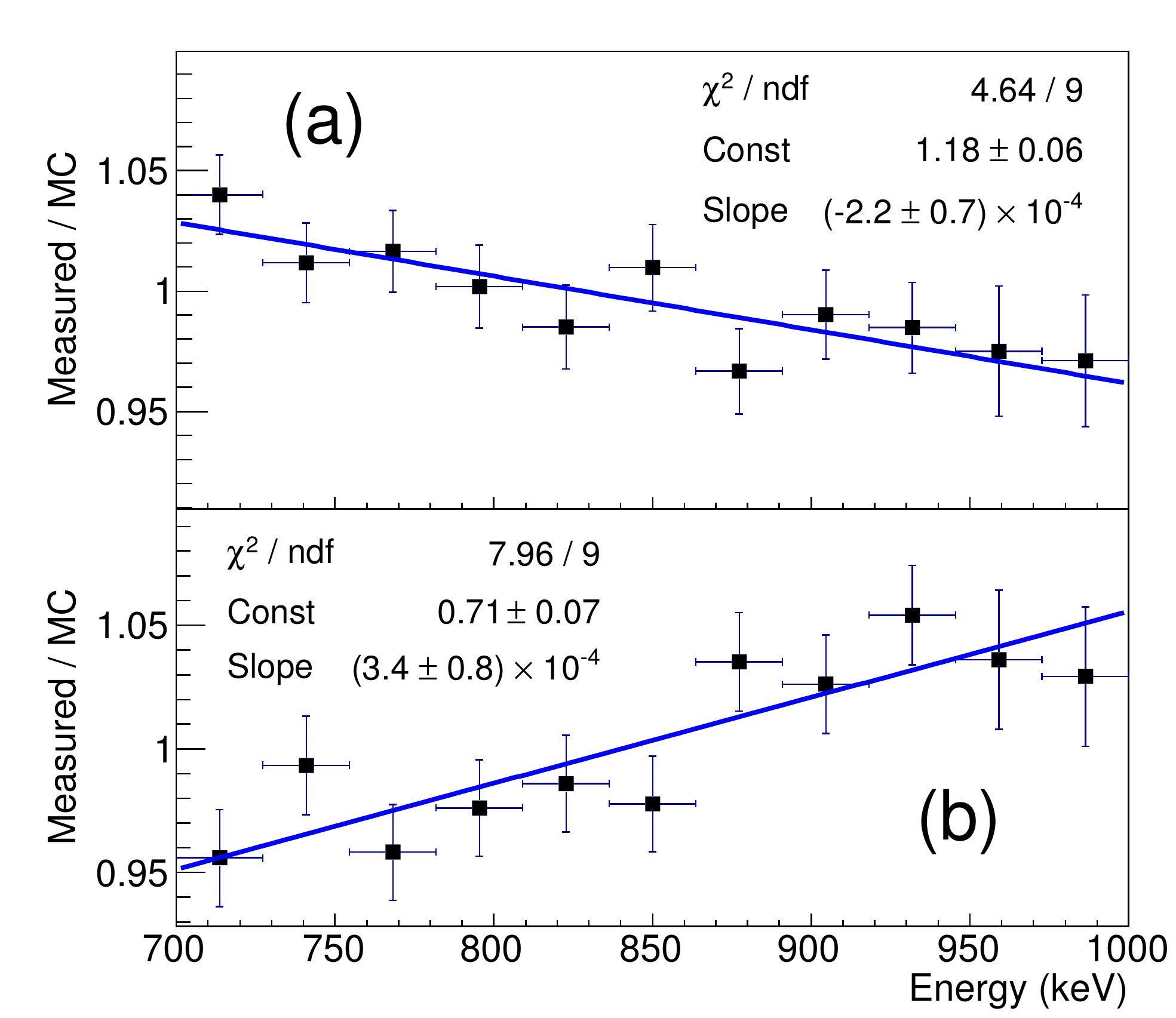}
      \caption{(Color online) Ratio of \cosrc{} source (at the cathode, S5)
      calibration spectra to that generated from simulation.  Single- and
      multi-site events are shown in (a) and (b), respectively, and linear fits
      to the data are shown.  Ratio plots for \thsrc{} source data and
  simulation demonstrate the same behavior.  } 
  
  \label{fig:shape_ratio} 
      
\end{figure}

The ratio of \cosrc{} calibration source spectra, below 1000~keV, to that
predicted by the simulation is shown in \cref{fig:shape_ratio}. A linear
parameterization as a function of energy is also shown over the data points.
Both the \cosrc{} (shown) and the \thsrc{} calibration source data-simulation
discrepancies can be parameterized by a single linear function of energy.  This
suggests a single mechanism is responsible for the common disparate shapes.
These linear skewing parameterizations (one for each the SS and MS energy
distributions) are used in \cref{sec:SimModelInadeq} to estimate the systematic
error induced in the \twonubb\ rate measurement.

        \subsection{Source rate agreement}
        \label{sec:sourceRateAgreement}

To test the ability of the simulation to predict the observed activity of the
\thsrc{} and \cosrc{} calibration sources, the total number of events passing
all selection criteria (see \cref{sec:analysis_cuts}) was compared between
simulation and data.  The results from simulations were normalized to the
NIST-traceable activity of the respective source and the counting times of the
corresponding data sets.  The agreement is quantified by the fractional
difference $(\text{Data}-(\text{MC~Sim}))/\text{Data}$ in number of events.
The results of these studies are listed in \cref{tab:source_agreement} for both
calibration sources, deployed at four distinct locations, and indicate that the
simulation can predict the activity of external sources within $\pm 4\%$.

There are several important contributions to this result, which make it not
directly reflective of the ability of the simulation to predict the detection
efficiency of \twonubb{}.  In particular, the observed rate of a source depends
strongly on its exact position due to solid angle changes.  To quantify how the
uncertainty on the source position propagates to the uncertainty on the rate,
simulations were generated varying the source locations around their nominal
positions.  Maximum-likelihood fits between simulation and data were performed
using the spatial distributions of events within the TPC at each simulated
position.  The error on the rates presented in \cref{tab:source_agreement} were
taken from the results of these fits. 

An additional important consideration is that the simulation predicts fewer MS
events to be fully position reconstructed, which would result in the removal of
more events in simulation. For external sources of gamma-rays, of which more
than half are observed to be MS, this leads to a more significant
under-prediction of the activity than would be expected for \twonubb{} decay
events, of which only about 5\% are observed to be MS.  The effect of this
under-prediction in simulation on the \twonubb{} rate is estimated in
\cref{sec:part_rec_cut} using low-background data. 

\begin{table}[ht]
\centering
\renewcommand{\arraystretch}{1.2}
\begin{ruledtabular}
\begin{tabular}{lcD{.}{.}{-1}} \toprule
    Source location & Source type & \multicolumn{1}{p{0.2\textwidth}}{\centering Absolute rate agreement $({\rm Data} - ({\rm MC~Sim}))/{\rm Data}$ [\%]} \\
    \colrule\\[0.1ex] 
    \multirow{2}{*}{S2 (anode)}    & \thsrc{} &  \pmasy{3.5}{0.8}{1.3}  \\
                                   & \cosrc{} &  \pmasy{2.4}{0.4}{1.6}     \\[1.5ex]
    \multirow{2}{*}{S5 (cathode)}  & \thsrc{} &  \pmasy{1.1}{1.0}{0.9}     \\
                                   & \cosrc{} &  \pmasy{-3.7}{1.5}{1.2}     \\[1.5ex]
    \multirow{2}{*}{S8 (anode)}    & \thsrc{} &  \pmasy{-3.2}{0.8}{0.9}    \\
                                   & \cosrc{} &  \pmasy{1.8}{0.8}{1.1}     \\[1.5ex]
    \multirow{2}{*}{S11 (cathode)} & \thsrc{} &  \pmasy{3.1}{2.3}{2.7}     \\
                                   & \cosrc{} &  \pmasy{1.3}{3.1}{4.0}     \\

\end{tabular}
\end{ruledtabular}
  \caption{Constraints on the absolute rate agreement at each calibration
      location, including uncertainties due to the source location.  These
      numbers, which \emph{do not} include the source activity uncertainty may
      be compared to the uncertainty on the absolute activity of the sources,
      known to be 1.2\% from the NIST-traceable certificates issued by the
      vendor and confirmed through independent collaboration measurements.} 
  
  \label{tab:source_agreement}
\end{table}

\subsection{Single-site fraction agreement}\label{sec:ss_agreement}

\begin{figure}
 \includegraphics[width=0.45\textwidth]{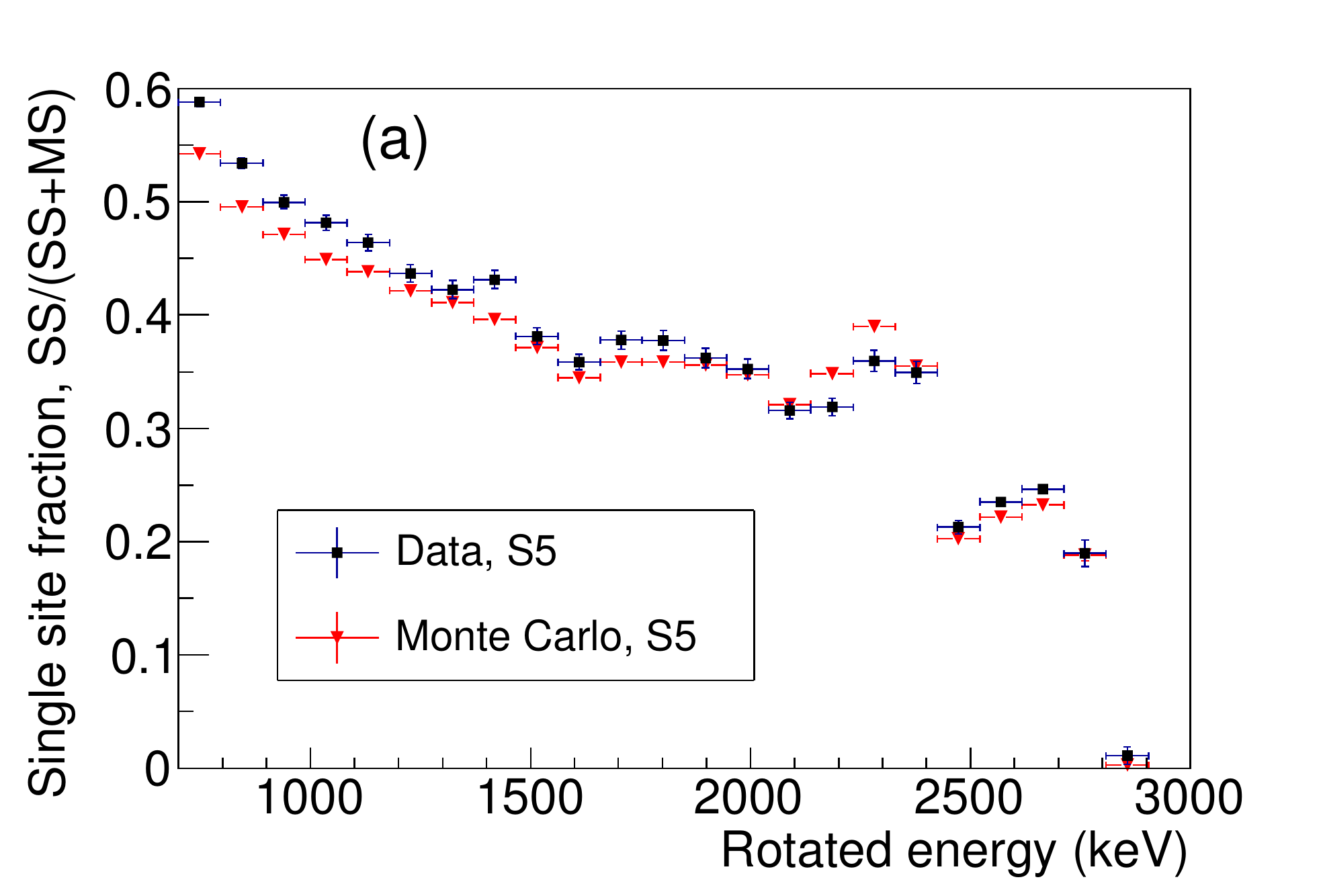}
 \includegraphics[width=0.45\textwidth]{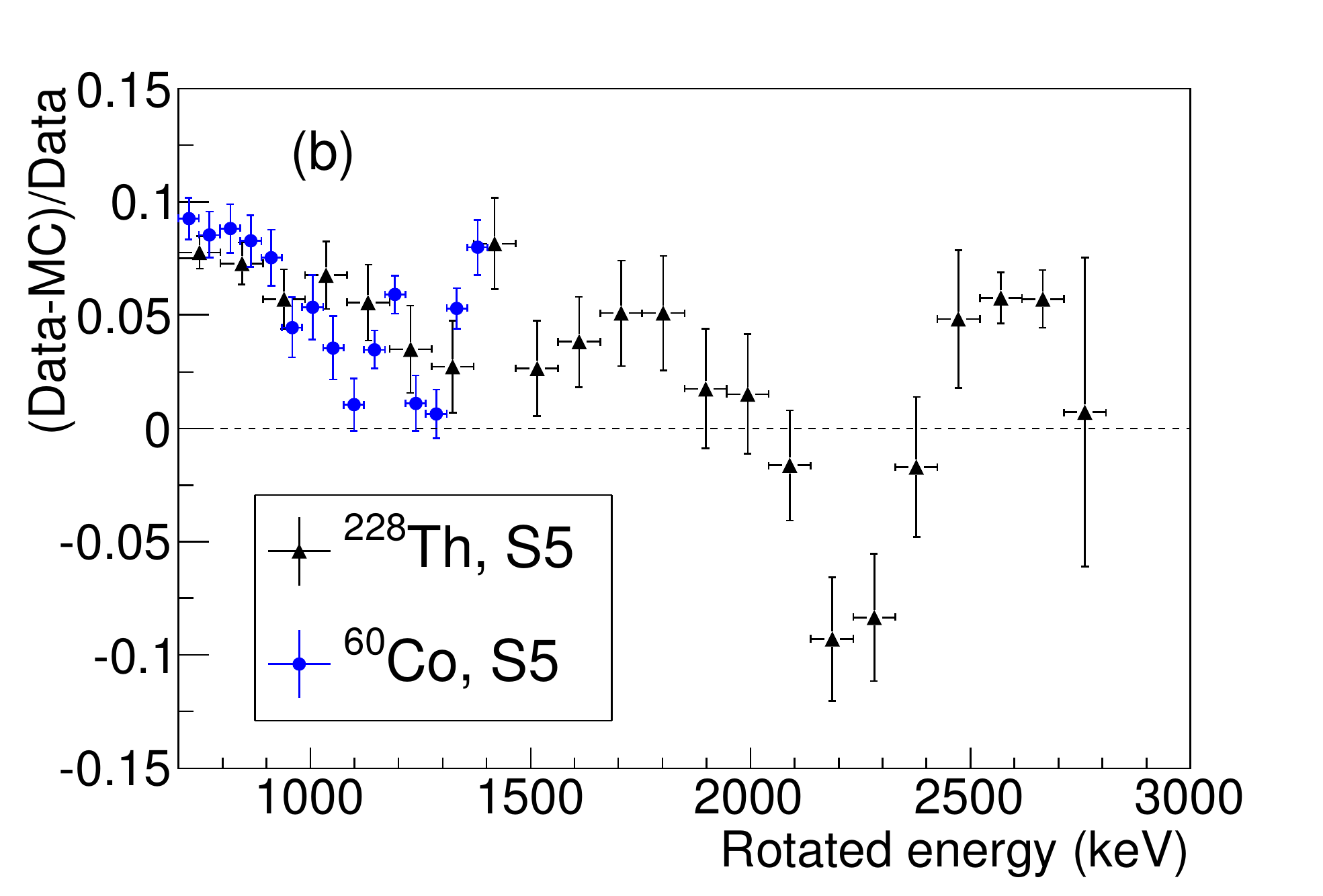}
 \caption{(Color online) (a) Single-site fraction for \thsrc{} calibration
 data and simulation.  (b) Fractional difference between single-site
 fraction for data and simulation as a function of energy for \thsrc{} and
 \cosrc{} sources.} \label{fig:ss_frac}
\end{figure}

The fraction of SS events, defined as the event number ratio $\text{SS}/(\text{SS}+\text{MS})$, is
calculated from both calibration source data and source simulations. As the SS
fraction is observed to be energy dependent, the discrepancy between the SS
fraction of calibration source data and the SS fraction of the simulations is
determined versus energy. \Cref{fig:ss_frac} shows that the maximum fractional
difference between the SS fraction of calibration source data and the SS
fraction of source simulations is $\pm10\%$.

From these observations, a spectral-weighted-average SS fraction error is
calculated for the \twonubb{} spectrum. This error, determined to be 5.9\%, is
applied as a constraint in the fit to the low-background data set. Any
remaining error on the \twonubb{} rate resulting from the residual energy
dependence observed in the SS fraction has been studied using the
low-background data and is presented in \cref{sec:SimModelInadeq}.

\section{Data Quality and Detector Monitoring} \label{sec:DataQuality}

To ensure operational consistency of all relevant detector parameters over the data-taking
period, a number of environmental and analytic parameters are monitored.  This
monitoring enables a quantitative assessment of the quality of data taken within
a particular time period, which we use when selecting runs for the physics
analysis.  Checks include muon veto system monitoring, low-level checks, 
and high-level checks, the last of
which requires data to be processed with the analysis framework described in
\cref{sec:Processing,sec:Reconstruction}. 
Observations made by monitoring shifters during the run are additionally taken
into account to ensure that external activities not monitored by the DAQ
system (e.g.~construction or mine activity) do not have an adverse affect on
the data.  The runs selected for the physics analysis total 135.2~days,
representing $\sim77\%$ of the total low-background data taken during
\RunTwoA{} (23\% loss due to data quality requirements).  

\subsection{Muon veto system source campaigns}
\label{sec:veto_source_campaigns}

The health and response of the muon veto system is 
evaluated twice per year
with a $^{60}$Co source and a stand-alone DAQ system
employing a charge sensitive ADC. This monitoring program
is necessary due to the large environmental 
temperature swings to which 
the system is exposed outside of the EXO-200 cleanrooms. 
This may lead to yellowing of the scintillator, PMT failure, 
or cracks in the PMT optical coupling.
During these campaigns the $^{60}$Co source is
placed at four different distances to each instrumented end of each 
of the 29 veto panels.
The average signal from of backscattered $^{60}$Co gammas is
determined by fitting the Compton edge, $p(x_j,t_i)$, in the spectral 
distributions of each instrumented detector end, where
$t_i$ denotes the time of the measurement and $x_j$ the distance of
the source from the PMTs. 
$i$ identifies the calibration campaign, $j$ numbers the different 
distances. It is assumed that the scintillator light
yield $L$ is time and location independent. This gives
the following expression for the Compton edge:
\begin{displaymath}
p(x_j,t_i)\; =\; g(t_i)\cdot L\cdot e^{-x_j/\lambda(t_i)},
\end{displaymath}
where $g(t_i)$ denotes the time-dependent product of 
PMT gain and quantum efficiency and $\lambda(t_i)$
is the light attenuation length. 
Forming two ratios of the measured Compton edges allows both detector
parameters to be tracked independently:
\begin{equation}
\frac{p(x_j,t_i)}{p(x_1,t_i)}\; =\; e^{-(x_j-x_1)/\lambda(t_i)} \label{eq:att_length}
\end{equation}
A fit to the distance dependence of Compton edge ratios
(normalized to one at the closest distance to the PMTs in a particular
calibration campaign), as shown in \cref{eq:att_length}, determines the light attenuation length of
each panel at time $t_i$.
\begin{equation}
\frac{g(t_i)}{g(t_2)}\; =\; \frac{p(x_1,t_i)}{p(x_1,t_2)} \cdot 
e^{x_1\cdot \left[\lambda(t_2)-\lambda(t_i)\right]/\left[\lambda(t_2)\cdot\lambda(t_i)\right]} \label{eq:gain_ratio}
\end{equation}
Knowledge of attenuation lengths at time $t_i$ and for the reference measurement at time $t_2$
allows the computation of the relative gain change of the PMTs from the ratio of Compton edges given
in \cref{eq:gain_ratio}. The ratio is corrected for the change in light attenuation
length. The second calibration campaign serves as the reference dataset as thresholds
settings were not finalized during the first calibration. 

The attenuation length and gain-ratio data exhibits
linear trends for all panels. 
The data indicates an average rate of 
change per year of $\Delta \lambda/\lambda_2=-2.9\%$ with an RMS of 5.9\% and $\Delta g/g_2=-5.2\%$ with
an RMS of 8.5\%, showing adequate stability in the system's
response to ionizing radiation.

\subsection{Low-level checks}

A typical low-background run has a duration of $\sim 24$~hr, although, occasionally, 
shorter runs are taken.  A number of baseline requirements must be met for a 
low-background run to be used in the physics analysis: 

\begin{enumerate}
    \item Run length greater than 1800~seconds.
    \item Average solicited trigger rate measured within 0.5\% of its nominal value (0.1~Hz) . 
    \item Calculated live-time no more than 30~s different than the run duration.
\end{enumerate}

The run length requirement ensures enough statistics for the calculation of
data quality indicators.  The second and third requirements ensure that the DAQ
system is operating nominally throughout the run and that no significant
reductions in live-time caused by noise bursts or DAQ interruptions have been
seen. 

\subsection{High-level checks}

The higher level checks involve monitoring:

\begin{enumerate}
    \item the electron lifetime $\tau_e$
    \item the efficiency of muon veto system
    \item rates of certain classes of events. 
\end{enumerate}

Runs satisfying all checks may be automatically approved to be used in the
physics analysis.  Those runs that do not satisfy all checks are not
immediately rejected, but are instead subjected to additional scrutiny to determine
their final status. 

\subsubsection{Electron lifetime}

Even though the charge data is corrected for the effects of finite electronegative 
purity, as described in \cref{sec:purity_corrections}, 
uncertainty in the electron lifetime correction or rapid
changes in the purity can 
degrade the detector energy resolution. We calculate that this contribution
is acceptably small when the electron lifetime is greater than  1~ms.
We also calculate that for an electron lifetime of 1~ms, a rate 
of change of $500$~\mus{}/day can contribute 1.04\% to the 
energy resolution. Hence the following requirements on the
detector purity are applied:

\begin{enumerate}
    \item $\tau_{\rm e} > 1000$~\mus{}.
  \item Four or more consecutive measurements of similar electron lifetimes 
  over several days with constant xenon recirculation.  This ensures stability of the purity.
  \item $\tau_{\rm e}$ must not be increasing at a rate $ > 50\%$ 
 or decreasing at a rate $>  25\%$ of the previous measurement per day. 
\end{enumerate}

The variation of the electron lifetime during Run 2a is shown in
\cref{fig:el_time_variation}.

\subsubsection{Muon veto panels}\label{sec:MuonVetoPanels} 

The absolute efficiency of the individual veto panels is monitored bi-annually 
as described in \cref{sec:veto_source_campaigns}.
The muon veto system is also monitored semi-continuously with 
low-background data by counting the fraction of events
identified as muons in the TPC that are also tagged with a veto trigger
within 2~\mus{}.  For the Run 2a dataset this fraction is $0.9618\pm0.0021$
on average.

The individual veto panels are monitored on a run-by-run basis as follows.
Because of geometrical effects, the TPC-correlated muon rate varies among veto panels.
If one or more panels do not register a single trigger during a low-background 
run, and those panels account for more than 5\% of the average veto-TPC 
coincidence rate, then 
the run is rejected. This threshold was chosen so that the global veto system 
efficiency remains $>90\%$, which is a
requirement for achieving the background goals of the \nonubb{} search. 

\subsubsection{Event rates}

\begin{figure}
     \centering
          \includegraphics[width=.45 \textwidth]{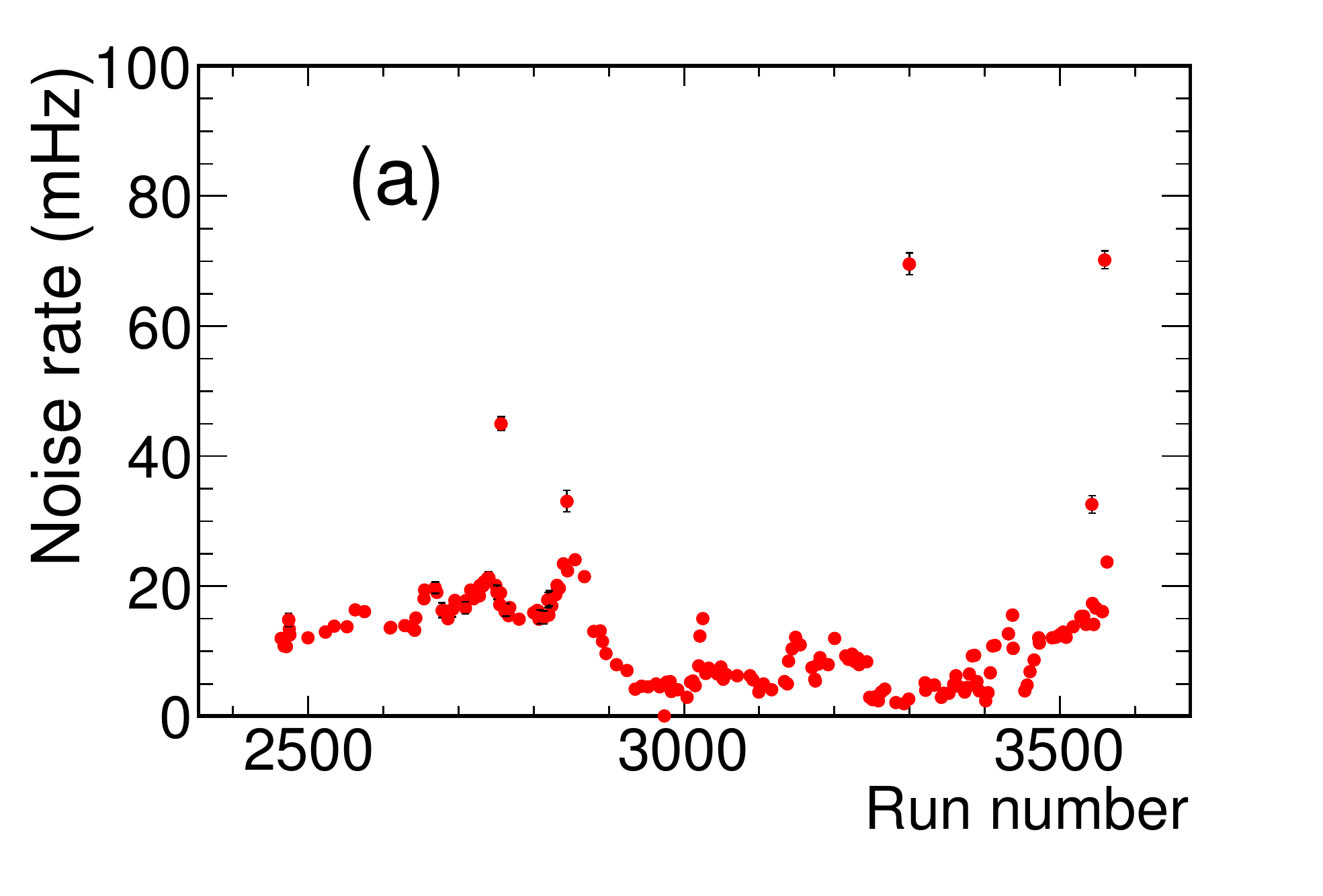}
      \includegraphics[width=.45 \textwidth]{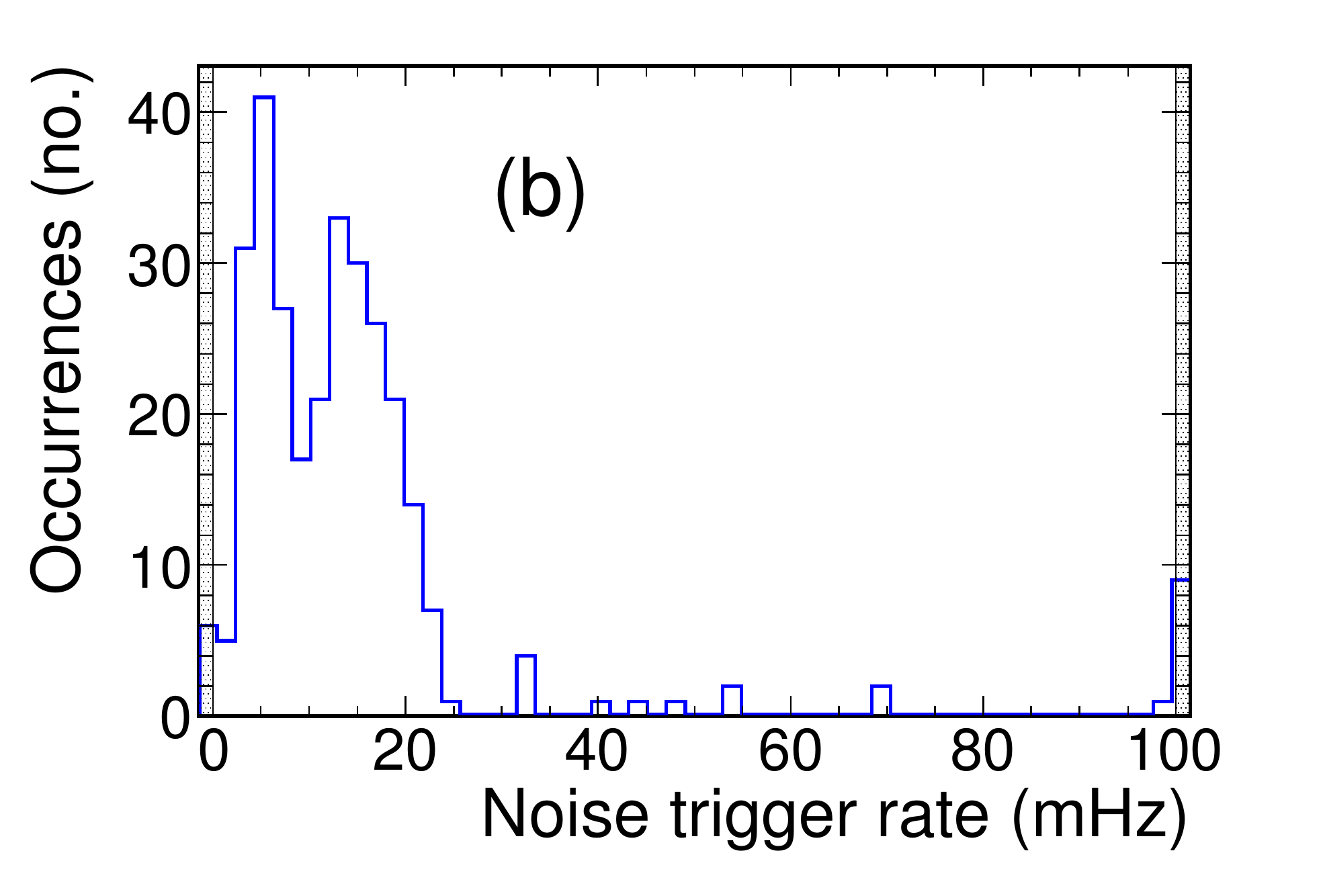}
        \caption[Noise rates]{(Color online) History (a) and distribution (b) of rate of events
            tagged as noise.  The drop in noise starting with run 2855 is due
            to a cooling fan change in the electronics box.  The unshaded
            region in (b) indicates the accepted parameter values ranges. 
        }
\label{fig:noise}
\end{figure}

\begin{figure}[ht]
      \centering
         \includegraphics[width=.46 \textwidth]{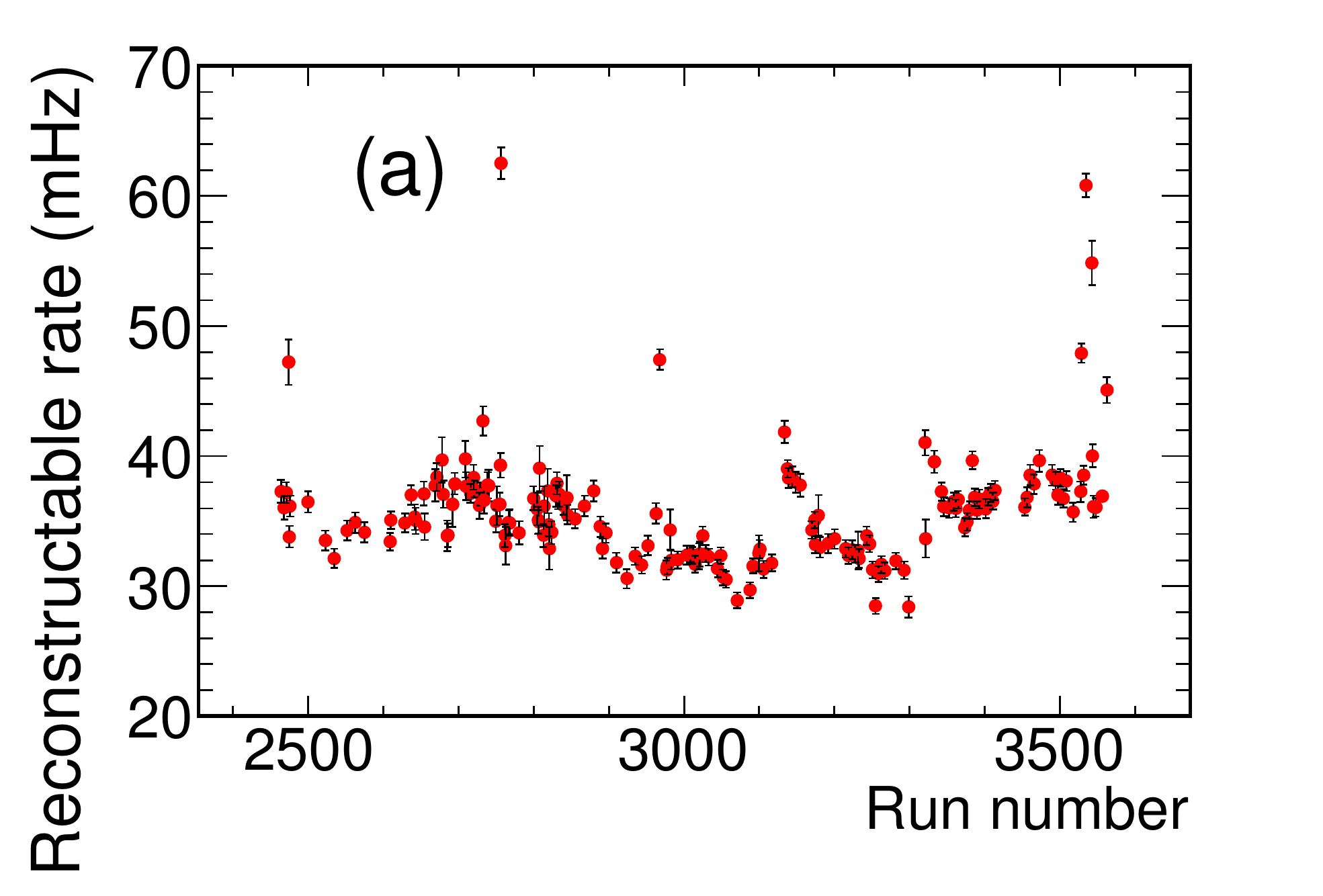}
         \includegraphics[width=.46 \textwidth]{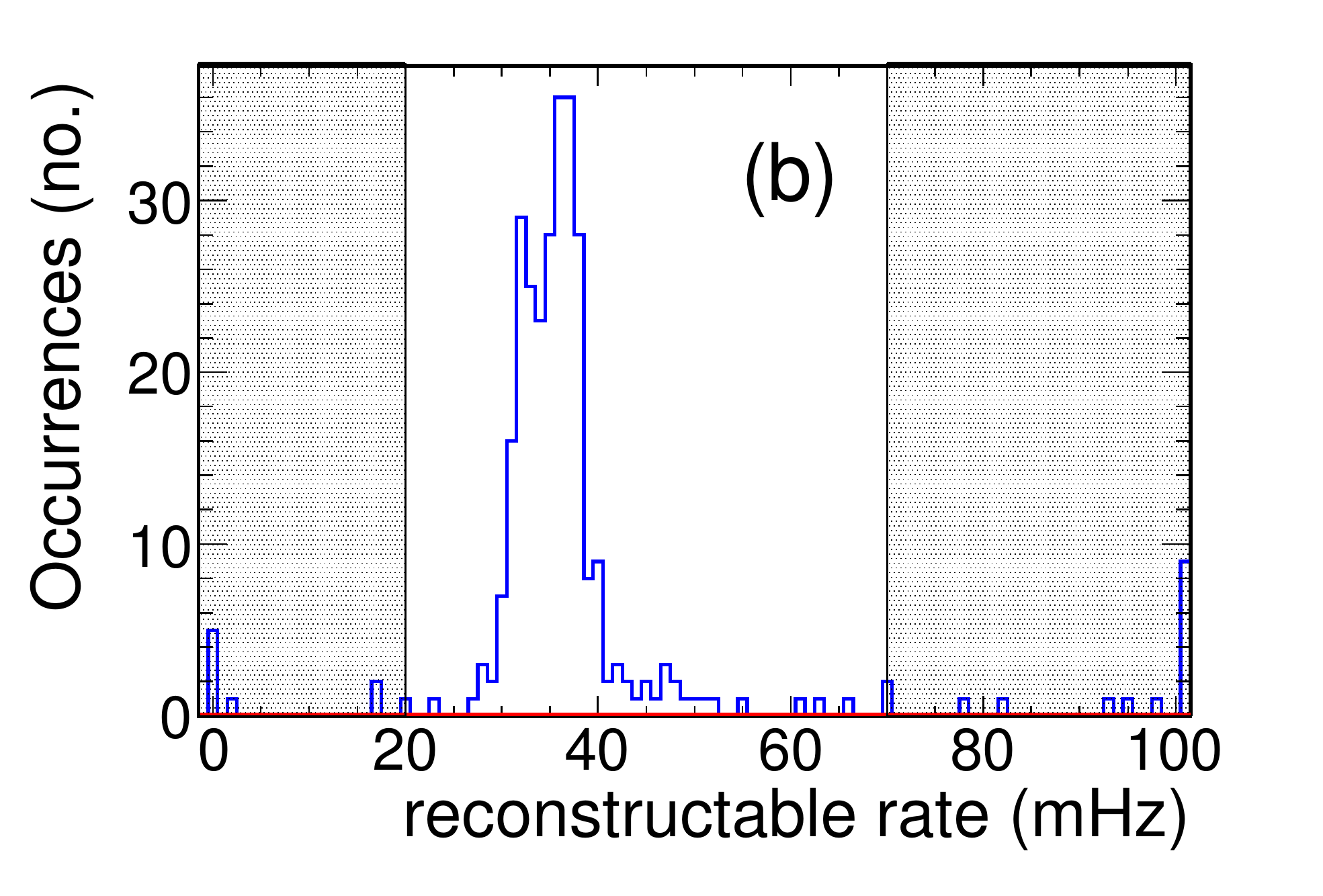}
        \caption[Reconstructable event rate]{(Color online) History (a) and distribution (b)
        of the rate of events that have at least a scintillation and charge
        cluster.  The increase in rate seen after January 13, 2013 (run 3121)
        came from increased Rn in the system after a Xe feed event.  The
        discontinuity on February 22, 2013 (run 3333) was from an APD
        electronics change.  The unshaded region in (b) indicates the
        accepted parameter values. 
        }
\label{fig:recon}
\end{figure}

Rates of different classes of events are monitored over time, as deviations
from mean rates can provide an indication of a detector problem or an
important environmental change. In particular, rates of the following seven
classes of events are monitored, with rough, nominal rates of each class given
parenthetically: (a) events tagged as noise by the analysis processing
(10~mHz); (b) reconstructable events, defined as having at least one
scintillation cluster and not tagged as a muon, noise, or a solicited trigger
(35~mHz); (c) non-reconstructable events, those failing criteria (b) and not
tagged as a muon, noise, or a solicited trigger (25~mHz); (d) events with
$>0$~keV (25~mHz); (e) events with $>300$~keV (20~mHz); (f) events with
$>1000$~keV (3~mHz); and (g) events with $>2000$~keV (1.5~mHz).  The energy
ranges in the last four classes of events are defined for ionization energy.
For each of these seven rates, an acceptance region is defined so that a value
falling within these limits confirms the data as high quality.  An example set
of plots is given for event classes (a) and (b) in \cref{fig:noise,fig:recon},
respectively, showing the rate versus time as well as the distribution of rates
of the two parameters over the data-taking period.

\section{Event selection requirements}\label{sec:analysis_cuts}
The live-time of the data set differs from the data-taking time due to the
following rejection criteria, with the percentage live-time lost due to each in
parentheses:

\begin{enumerate}
    \item \label{itm:MuonTrigger} 1~ms before to 25~ms after the muon veto system triggers (0.6\%) 
    \item \label{itm:TPCMuon} 1~\mus{} before to 60~s after TPC events tagged as muons (4.5\%)
    \item \label{itm:PoorDataTaking} Portions of runs flagged as coincident with poor data-taking conditions, 
        including times during a run where one or more of the selection criteria
        described in \cref{sec:DataQuality} are not satisfied. (0.6\%)
    \item \label{itm:RunBuffer} 60~s after the beginning and 1~s before the end of any run ($<0.1$\%)
\end{enumerate}

\noindent
These result in a total live-time reduction of 5.6\% over the total data-taking
time of all selected runs in the \RunTwoA{} data set.  This does not exactly
correspond to the sum of the above percentages because some of the criteria
overlap. 

Selection criterion~\ref{itm:MuonTrigger} is motivated by the observed
elevated event rate in the TPC following a muon veto system trigger.  This rate is
found to return to its nominal value after 25~ms. In addition, the analysis of
the time behavior of 2.2~MeV full-energy deposition events within 25~ms
following a muon trigger yields a neutron capture time in hydrogen in the HFE
of $740\pm120$~\mus{}, well within the 25~ms time window.  
Selection requirement~\ref{itm:TPCMuon} is motivated by the expectation of 
the cosmogenic activation of short-lived isotopes in and near the LXe following
a muon event in the TPC.  The 60~s coincidence window is determined by limiting 
the impact of this requirement on the live-time of the data set to be $<5\%$.

Data-taking times may be flagged as unacceptable
(criterion~\ref{itm:PoorDataTaking}) due to data quality concerns arising
from e.g.~bursts of electronic noise or mining activity.
Criterion~\ref{itm:RunBuffer} ensures that neither a muon TPC event or an
event within 1~s coincidence could have been missed immediately before the
initiation of or immediately after the termination of a run.  The application of
these live-time selection criteria results in \pmasy{127.6}{0.012}{0.000}~days
live-time out of 135.2~days of cumulative data taking during all selected runs
in the \RunTwoA{} data set.

The following data selection criteria are evaluated on an event-by-event basis.
For each data reduction, the resulting loss of \twonubb{} signal efficiency is
estimated along with an associated error. These efficiencies are used to correct
the measured \twonubb{} and the associated errors on the \twonubb{} rate are
included via a normalization term (see \cref{sec:FitConstraints}). The
following data reductions are applied in the order that they are listed (note
that event frames are 2048~\mus{} in length). In order for events to be selected
for analysis they must

\begin{enumerate}
    \item \label{itm:SolTrg} not be in coincidence with 0.1~Hz solicited triggers (explicitly tagged by the DAQ system),
    \item \label{itm:Noise} not be in coincidence with events tagged as noise, 
    \item \label{itm:one_sec_veto} not occur within 1~s of another TPC
        event (to address e.g.~Bismuth-Polonium (Bi-Po) fast $\beta$-$\alpha$ coincidences in
        the \isot{Rn}{222} and \isot{Rn}{220} decay chains and other correlated decays),
    \item \label{itm:nsc_gt_one} not contain more than 1 reconstructed scintillation signal, 
    \item \label{itm:late_sc} not begin within 120~\mus{} of the end of the
        waveform trace, 
    \item \label{itm:partial_rec} not contain any partially reconstructed signals (must be fully 3-D reconstructed),
    \item \label{itm:fid_vol} be reconstructed inside the fiducial volume,
    \item \label{itm:diag_cut} not exhibit a high light-to-charge ratio (e.g.~appear to be $\alpha$-like),
    \item \label{itm:energy_cut} contain total rotated energy $>700$~keV 
\end{enumerate}

The estimation of systematic errors related to items~\ref{itm:one_sec_veto} \textendash
\ref{itm:fid_vol} are described in more detail in the following sections.  The
removal of solicited triggers results in a $10^{-3}$\% reduction of signal
efficiency, estimated from the probability of a solicited trigger occurring
coincidentally with a \twonubb{} event.  A comparable reduction in signal
efficiency ($10^{-3}$\%) has been calculated for events tagged as noise.
Whereas the majority of noise events are tagged because of grossly unphysical
characteristics (e.g.~channel range saturation), there is a possibility that
some physics events are falsely tagged as a particular class of noise.  The
probability of this occurring has been estimated by randomly selecting a set of
events failing this cut and hand-scanning them.  No false positives were found,
leading to a conservative $6\times10^{-2}$\% systematic error on the signal
efficiency.  

Events with a high light-to-charge ratio are indicative of an event including
$\alpha$-particles or of events with incomplete charge collection, which may occur if the
charge cloud drifts to the edges of the wire grid.  This cut is highly degenerate with
some of the other selection cuts and, after the previous application of all other cuts, only
removes $\sim35$~events from the 23082-event dataset.  We conservatively take
this ratio (0.15\%) as a systematic error. 

The removal of events below the threshold of 700~keV is subject to errors in
the measurement of the detector calibration and resolution.  The contribution
of this effect to the \twonubb{} measurement was found to be 0.4\%, estimated
by the residual calibration error of the \isot{Cs}{137} 662~keV peak (see
\cref{sec:AntiCorr}).

\subsection{Scintillation-signal-driven data selection}

Two cuts are performed, which are related exclusively to scintillation signals.
Events which have a scintillation signal within 120~\mus{} of the end of a
waveform are removed because they cannot be fully reconstructed.  This cut
removes only 1 event from the final dataset therefore contributing a
negligible amount to the systematic error on the efficiency. 

In addition, events are required to have no more than one scintillation signal
in order to remove correlated decays (e.g.~Bi-Po correlated decays).  The false positive rate
(the probability to incorrectly reconstruct more than one scintillation signal
when only one \emph{real} signal exists) has been estimated by running the
reconstruction software on noise trace (solicited trigger) events from data with an added
simulated signal of known amplitude.  This study estimates a false positive
rate integrated above 700~keV of $<0.1$\%.  This study was cross-checked by
analyzing the 613~low-background events removed by this cut after
applying all other analysis cuts, and was determined to remove
0.7\% of all events because of an incorrectly found scintillation signal.  We choose
to take the more conservative value and so a normalization error of 0.7\% is
added to account for this effect.

\subsection{1~s event-to-event coincidence}
A one-second coincidence cut on TPC events is used to reduce the presence of
backgrounds arising from fast, correlated decays in the TPC that may arise, for
example, after through-going muons or from the decays of radioactive species in
the LXe (Bi-Po decays) not removed by the scintillation event cut.  For the
purpose of this cut, two TPC events are considered to be in coincidence if
their DAQ triggers were not solicited, the events are not tagged as correlated
noise, and both events have some scintillation signal or saturate a scintillation channel.  The
\twonubb{} signal efficiency was calculated by considering the rate of random
coincidences of TPC events and was determined to be~$0.931\pm0.002$. 

\subsection{Fiducial volume selection}
\label{sec:FidVolCut}

The fiducial volume is chosen to include regions of the detector that are
well understood and properly modeled in simulation.  Since the measurement of
the \twonubb{} half-life is not constrained by statistics, a smaller fiducial
volume is chosen than was presented in~\cite{Auger:2012ar} to minimize the
impact of fiducial-volume-related systematics on the result.  As
in~\cite{Auger:2012ar}, since signals are fundamentally reconstructed at U- and
V-coordinates parallel to the detector wire channels, the fiducial volume is
chosen to be hexagonal in shape, thus avoiding effects arising from radial
coordinate transformations.  However, this still requires using a third
coordinate (X) derived from U and V to define the hexagon.  This is because the
U- and V-coordinate axes are each perpendicular to two sides of the hexagon,
whereas the last two sides are perpendicular to the derived X axis (see
\cref{fig:planar-coordinates}). 

The Z range of the fiducial cut is chosen to include events satisfying 182~mm
$>|\Z|>$ 15~mm in both TPC halves, or 10.2~mm from the V-wire planes and 15~mm
from the cathode.  The distance from the V-wire plane ensures the grid
efficiency correction to the charge energy (see
\cref{sec:grid_eff_correction}) is $\ll1$\%.  The distance from the
cathode was determined from studies using the 3-D electric field model (see
\cref{sec:MCDigitizer}), which demonstrated slightly distorted (not
parallel in Z) charge drift trajectories within 15~mm of the cathode. 

\begin{figure}
        \includegraphics[width=0.47\textwidth]{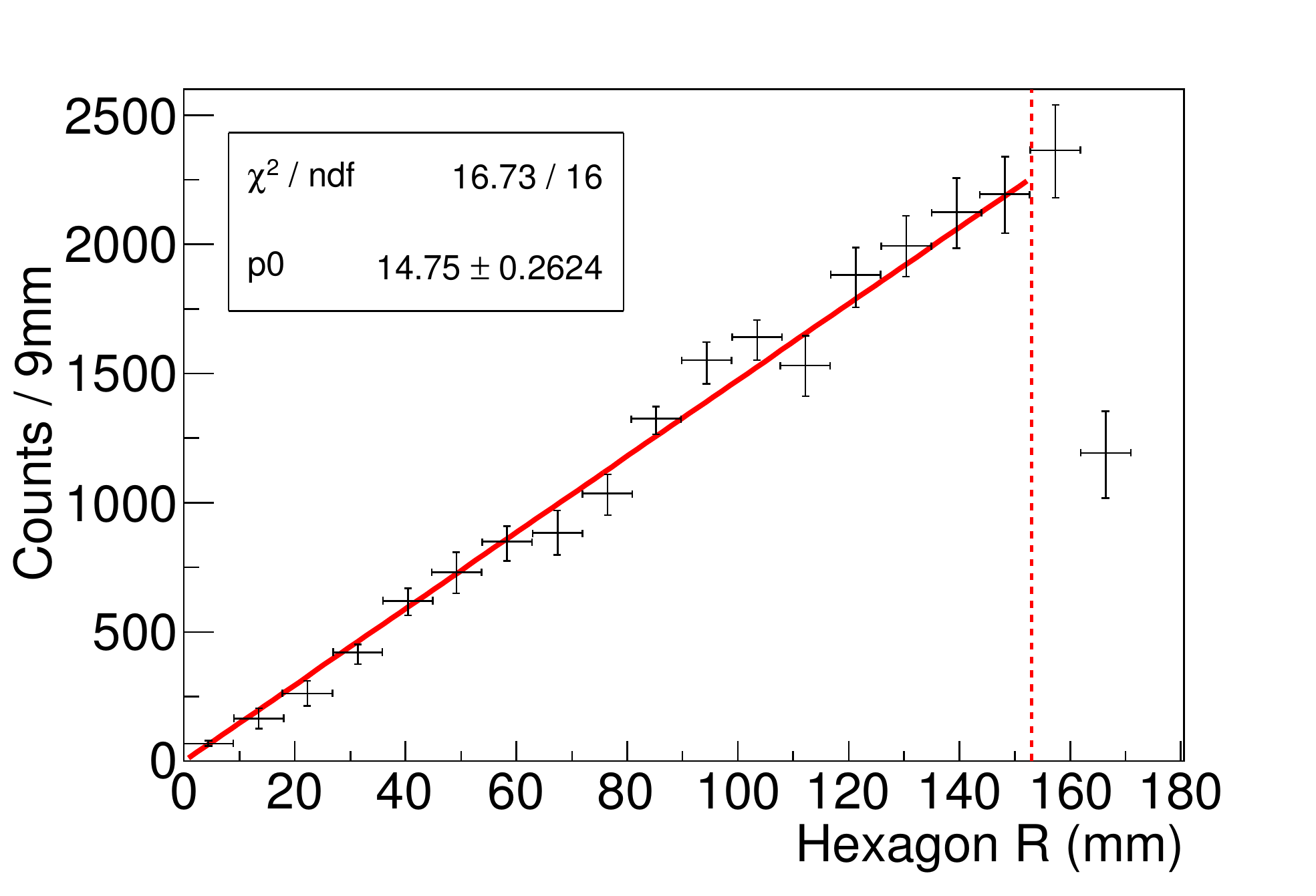}
        \caption{(Color online) Rate of \twonubb{} versus hexagonal apothem.
        The vertical dashed line indicates the fiducial cut we have chosen to
        use, at an apothem of 153~mm.  The data within that region is
        consistent with a rate that is proportional to volume.} 
        \label{fig:bb2nRateVsHex}
\end{figure}

To determine the cut in the U- and V-coordinate system, the single-site
\twonubb{} rate was studied as a function of detector position by fitting the
low-background spectrum in separate hexagonal position bins.  The results of
this study are shown in \cref{fig:bb2nRateVsHex}. We choose
the apothem of the hexagonal fiducial volume to be 153~mm, as indicated
in \cref{fig:bb2nRateVsHex} and illustrated in  
\cref{fig:planar-coordinates}. This cut removes events
on the boundaries of the sensitive area of the hexagonal wire plane.  
The rate of \twonubb{} decay was also studied in Z position bins and found to be
independent of position within the chosen fiducial volume. The fiducial volume
selected in this way corresponds to 27.08~liters of liquid Xe, corresponding to
66.20~kg of \isot{Xe}{136}.  See \cref{sec:XeRelatedErrs} for a discussion on
the error on this number. 

To understand the error on the chosen fiducial volume we first study the
cluster position reconstruction accuracy in the two original coordinates - U,
V.  Given that the detector has a discrete wire spacing, the position
resolution is finite and the expected uncertainty is given by the standard
deviation of the uniform distribution.  For a 9~mm wire pitch in the U- and
V-directions this gives $\pm$2.6~mm for uniformly distributed events.  

To study the position reconstruction uncertainty in simulation we generate the
distribution of differences between reconstructed cluster coordinates and true
location of simulated charge deposits of a \twonubb{} source.  The average RMS
is 1.2~mm in the V direction and 2.4~mm in the U direction, and both
distributions are centered on zero with negligible bias.  The small RMS in the
V direction arises due to the larger number of multi-wire events, which allow
better position determination through weighted averaging.  The third coordinate
that defines the hexagonal volume - X - is different than the fundamental
coordinates U and V because it is a combination of the two and so may be
subject to an additional uncertainty due to errors in the clustering algorithm
(e.g.  incorrect wire association).  The uncertainty on the X coordinate was
determined to be 2.6~mm, consistent with the individual U and V errors added in
quadrature, indicating that the clustering error is negligible.

We can cross-check these numbers with the data for specific cases, in
particular by using $\alpha$- and $\beta$-decays from the cathode that deposit
energy in both TPC halves.  Such events typically originate from a common
position on the cathode and can be used to check how accurately both TPCs
reconstruct a shared coordinate. Because the U- (V-)wires in TPC1 are
parallel to the V- (U-)wires in TPC2, these events may be used to cross-check the U
and V position reconstruction.  The
distributions of $\U_{\rm TPC1} - \V_{\rm TPC2}$ and $\V_{\rm TPC1} - \U_{\rm TPC2}$ both have a
width of 3.07~mm, consistent with the uncertainties obtained in the Monte Carlo
simulation study considering that the measurement of $\U - \V$ corresponds to a measurement
of the sum of $\Delta \U$ and $\Delta \V$.  The mean values are 0.4~mm for
$\U_{\rm TPC1} - \V_{\rm TPC2}$ and 1.5~mm for $\V_{\rm TPC1} - \U_{\rm TPC2}$, which could be
explained if the wireplanes do not exactly mirror one another.  Since we do not
have an independent measurement of the wireplane alignment after cooling the
TPC with sufficient precision to confirm this possibility, a systematic
uncertainty of 1.5~mm on the reconstructed U and V positions is included in
the overall fiducial volume uncertainty to account for any residual bias.

\begin{figure}
 \includegraphics[width=0.45\textwidth]{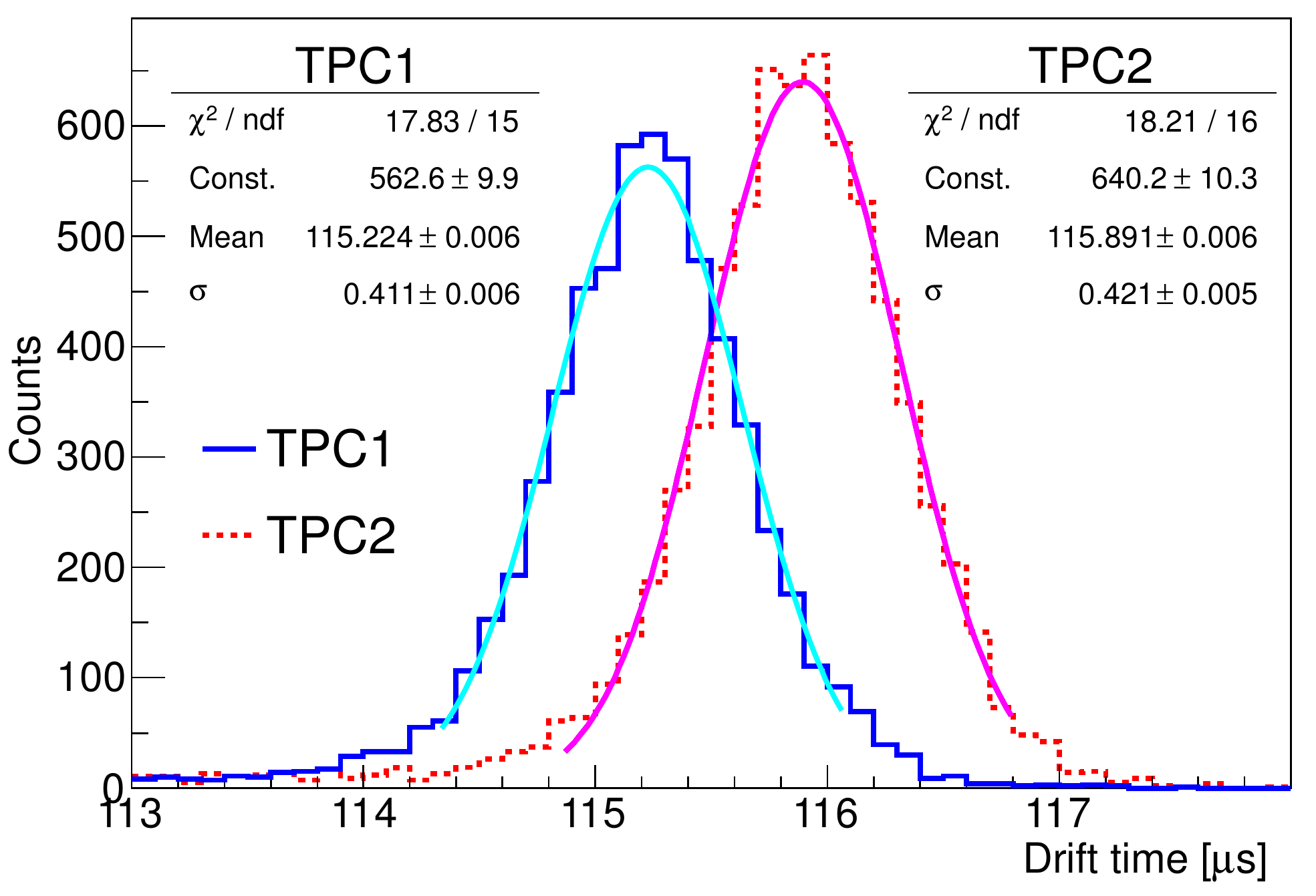}
            \caption{(Color online) Drift time distribution for $\alpha$ events
              originating from the cathode for events collected in
              TPC1 (solid histogram) and TPC2 (dashed histogram). Gaussian fits
          to the distributions and their results are also shown.}
            \label{fig:CathodeAlphasFit}
\end{figure}

The overall error on the reconstructed Z position for events drifting the full
length of each TPC can be determined from the drift time distributions for
$\alpha$ particles emitted from Rn-daughter contamination on the cathode, as shown in
\cref{fig:CathodeAlphasFit}.  The observed drift time distributions for
$\alpha$ events originating on the cathode have means of $115.224 \pm 0.006$~\mus{}
and $115.891 \pm 0.006$~\mus{} for TPC1 and TPC2, respectively.  The widths of
the drift time distributions, $\sigma_{t,\rm{TPC1}} = 0.411 \pm 0.006$~\mus{} and
$\sigma_{t,\rm{TPC2}} = 0.421 \pm 0.005$~\mus{} correspond to the expected error on
the determination of Z for a single cluster for a given charge drift velocity.  Assuming the same drift
velocity in both TPCs and using the expected spacing between the wire planes
gives an error on Z of $\sigma_{\Z} = 0.42$~mm for an individual cluster.
The 0.6~\mus{} difference in the means between the TPC halves was verified
using an independent analysis that determined the maximum drift time for the
distribution of events observed in \thsrc{} calibration data with the source
positioned at the cathode.  This offset may be accounted for if the cathode
were offset by $0.5$~mm from center in the direction of TPC1, which is within
the expected tolerance after vessel cool down.  As with the
observed U, V position bias, this offset is included as a systematic error on
the Z position to account for the possibility that this bias instead arises from
an error in the position reconstruction.

The respective errors on the component coordinate reconstruction are propagated
to the fiducial volume error, including any possible biases introduced by the
cut.  A 0.42~mm spread in Z direction would only result in a systematic bias on
the chosen volume if the cut position was in the region where the density of
reconstructed events is different inside and outside the cut (e.g.\ if the cut
was close to the cathode or wire planes). As this is not the case for the
chosen Z cut location, there is no systematic bias in the number of selected
events. 

In the case of the U, V, and X directions, varying the fiducial cut by
$1\sigma$ above (below) the default cut position does \emph{not} result in the
addition (subtraction) of an equal amount of volume.  This means that for a
constant density of events the cut will be biased to accept \emph{more}
events.  We estimate this effect to be +0.36\% by varying the fiducial cut and
comparing the selected number of events in the hexagonal layer \(1\sigma\)
above and below the default cut position.  

The uncertainty on the Z position of the cathode translates to the \(\pm\)0.3\%
systematic error on the selected volume, given the cut position. The 1.5~mm
uncertainty on U and V positions corresponds to \(\pm\)1.7\% error on the
chosen volume. Combining the uncertainty in the Z and U/V directions in
quadrature, the total error on the chosen fiducial volume is \(\pm\)1.73\%.  
We choose to add the above bias of +0.36\% symmetrically (i.e.~$\pm$0.36\%) in
quadrature with $\pm$1.73\%, which yields a total systematic error on the
normalization due to the fiducial volume cut of 1.77\%. 

\subsection{Rejection of partially reconstructed events}\label{sec:part_rec_cut}

The requirement that all events are fully reconstructed ensures that the
fiducial volume cut and position dependent corrections (see
\cref{sec:PosBasedCorrections}) have been properly applied.  The Monte Carlo simulation is
used to calculate the efficiency of this cut, which is defined as how often,
given an
event of a certain energy, all clusters in the event have a reconstructed U, V, and Z position.  We
estimate the resultant systematic error on the \twonubb{} measurement by
looking at the discrepancy between the simulation model and data.  As
noted in \cref{sec:sourceRateAgreement}, the simulation models fewer
events with full position reconstruction in the MS spectra for external sources
than observed in data.  To estimate the systematic difference between
simulation and data for the
\twonubb{} signal, the result of the fit to the low-background data is taken
and background subtracted to obtain a more pure \twonubb{} sample. 
This is done both before and after applying the partial reconstruction cut, and
the ``efficiency" is calculated, defined as the integrated event count above
700~keV of the spectrum \emph{with} the cut applied, divided by the number
of events \emph{without} the cut applied.  This value (93.9~$\pm$~0.2 (stat)
\%) can be compared directly to the value calculated for simulation for the
\twonubb{} PDF (94.2\%).  There are two primary sources of error that come from
using this background subtraction method: uncertainty from the background
normalizations and the systematic discrepancy (4\%) seen between simulation and data
for this cut for external $\gamma$ sources.  The propagation of these errors
was performed by generating a toy Monte Carlo dataset, drawing the
normalization of the different background components from the correlation
matrix measured from the low-background fit.  The 4\% systematic difference
observed for source data was incorporated as a corresponding error on the cut
efficiency calculated for each background component.  The systematic error on
the \twonubb{} result is then taken from the $1\sigma$ width of the efficiency
distribution, which translates to a 1.6\% error. 

\Cref{tab:event_selection_eff} provides a summary of the efficiency
reduction and associated systematic errors introduced by the event selection
cuts presented in this section. 

\begin{table}
\begin{ruledtabular}
    \renewcommand{\arraystretch}{1.3}
    \begin{tabular}{lrr}
        \toprule
        Cut type & \multicolumn{1}{p{0.14\textwidth}}{Signal efficiency (\%)} & Error (\%) \\
        \colrule
        Solicited triggers         & 99.99 & -    \\
        Noise                      & 100   & $<0.06$ \\
        1~s coincidence            & 93.1  & 0.2 \\
        $>$ 1 scintillation signal & 100   & \pmasy{}{0.7}{0.0} \\
        Partial reconstruction     & 93.9  & 1.6 \\
        Fiducial Volume            & -     & 1.77 \\
        Light-to-charge ratio      & 100   & 0.15 \\
        Energy $>$ 700~keV         & -     & 0.4 \\
        \hline
        Total    & 87.4 & 2.53 \\
    \end{tabular}
\end{ruledtabular}
    \caption{Signal efficiency and associated systematic errors for events
        occurring in the fiducial volume and above threshold introduced by
        event selection requirements.  Signal efficiency reductions are
        corrected for in the final \twonubb{} measurement.  Listed errors
        contribute directly to the final systematic error on the \twonubb{}
        measurement.  } 
\label{tab:event_selection_eff} 

\end{table}

\section{Fit to the low-background data}\label{sec:physics}
\subsection{Fit model}
\label{sec:FitModel}
To derive the number of \twonubb{} events from the \RunTwoA{} dataset we fit
both single- and multi-site event sets with their corresponding probability
density functions using a binned maximum likelihood method.  The data are
binned in energy (200 bins, $700 < E < 3500$~keV) and standoff distance (SD) (20
bins, $0 < r_{\rm SD} < 200$~mm) and the negative log likelihood function is
defined as: 

\begin{widetext}
\begin{equation} \label{eqn:negLL}
    -\ln L = 
    \sum_i \left[ \left( \mu_i^{\rm SS} + \mu_i^{\rm MS} \right) - 
         \left( k_{obs,i}^{\rm SS} \ln \mu_i^{\rm SS} + k_{obs,i}^{\rm MS} \ln \mu_i^{\rm MS} 
         \right) \right] + G_{\rm const}
\end{equation}
\end{widetext}

\noindent
where $k_{obs,i}^{\rm SS(MS)}$ are the number of SS (MS) counts observed in a
given bin $i$, $\mu_i^{\rm SS(MS)}$ defines the expected number of SS (MS)
events from the fit model in the $i^{\rm th}$ bin, and the sum proceeds over all
4000~bins in energy and standoff distance space.  $G_{\rm const}$ are Gaussian
constraints applied to the fit, which are described in more detail in
\cref{sec:FitConstraints}.  $\mu_i$ may be written as: 

\begin{equation}
    \mu_i^{\rm SS(MS)}(\bm{s}, \bm{n}, N) = 
    \int_{i^{\rm th}~{\rm bin}} F^{\rm SS(MS)} (\bm{s}, \bm{n}, N, \bm{y})~d\bm{y}
\end{equation}

\noindent
which is a function of SS/(SS+MS) fractions for each PDF, $\bm{s}$,
the relative number of events in different PDFs, $\bm{n}$, and an overall
normalization parameter, $N$, used to include the error on detector efficiency
(see \cref{sec:FitConstraints}).  The integral is performed over the $i^{\rm th}$
bin for the observables energy and standoff distance ($\bm{y} = (E, r_{\rm SD})$).  $F^{\rm
SS(MS)}$ are defined as:

\begin{eqnarray}
    F^{\rm SS} (\bm{s}, \bm{n}, N, \bm{y}) & = & N \sum_j n_j s_j f_j^{\rm SS}(\bm{y}) \\ 
    F^{\rm MS} (\bm{s}, \bm{n}, N, \bm{y}) & = & N \sum_j n_j (1-s_j) f_j^{\rm MS}(\bm{y}) 
         \label{eqn:allPDFs}
\end{eqnarray}

\noindent
where the sums are performed across the total number of PDFs included in the
fit ($N_{\rm PDF}$), $s_j$ is the relative fraction of SS events in PDF $j$
($\bm{s} = \{s_0, ..., s_{N_{\rm PDF}}\}$), $n_j$ is the total number of
events in PDF $j$ ($\bm{n} = \{n_0, ..., n_{N_{\rm PDF}}\}$), and $f_j^{\rm
SS(MS)}(\bm{y})$ is the $j^{\rm th}$ PDF for SS (MS) events and a function of
energy and standoff distance.  

The list of components comprising the total model is: \twonubb{},
\nonubb{}, backgrounds in the copper vessel (\isot{U}{238}, \isot{Th}{232},
\isot{K}{40}, \cosrc{}, \isot{Zn}{65}, and \isot{Mn}{54}), backgrounds in the
liquid Xe (\isot{Xe}{135} and \isot{Rn}{222}), and backgrounds in the air gap between
cryostat and lead wall (\isot{Rn}{222}).  The PDFs for the background isotopes were
produced using the standard GEANT4 Radioactive Decay Module (RDM)
generator~\cite{GEANT42006}.   For \isot{Th}{232} and \isot{U}{238} the entire
chains of decays, assumed in secular equilibrium, are used.  For
\isot{Rn}{222}, only decays between \isot{Rn}{222} and \isot{Pb}{210} are used.
The \twonubb{} event generator uses the Fermi function suggested in~\cite{Schenter:1983uq}. 

Additional sources of radioactive background were also considered to be added
to the background model, in particular \isot{U}{238} and \isot{Th}{232} in the
surrounding HFE fluid.  However, studies comparing the shapes of these
backgrounds in both the energy and standoff distance distributions found no significant differences
with the corresponding distributions from simulated backgrounds in the copper
vessel and so these additional PDFs were not included in the background model
of the fit.  This implies that values returned from the fit (e.g.~$n_{{\rm
Th-232},{\rm copper}}$) also include contributions from backgrounds located
further away from the detector.   

   \subsection{Constraints}
   \label{sec:FitConstraints}

Using the results of additional studies, several Gaussian constraints are added to
the negative log-likelihood function ($G_{\rm const}$ in \cref{eqn:negLL}).  In
particular, these constraints contribute the following function to
\cref{eqn:negLL} 

\begin{equation}
    \label{eqn:Constraints}
    G_{\rm const}(\bm{\rho}, \bm{\rho_0}, \bm{\Sigma}) = 0.5 (\bm{\rho} - \bm{\rho_{0}})^{T} {\bm \Sigma^{-1}} (\bm{\rho} - \bm{\rho_{0}})
\end{equation}

\noindent
where $\bm{\rho}$ is the vector of constrained parameters,  $\bm\rho_{0}$ is the
set of expectation values for each parameter and $\bm\Sigma$ is the covariance
matrix for the set of parameters. For each uncorrelated parameter, $\rho_i$ in
$\bm{\rho}$, \cref{eqn:Constraints} contributes the term: 

\begin{equation}
    0.5 \left(\frac{\rho_i - \rho_{i,0}}{\sigma_i} \right)^2
\end{equation}

\noindent
to \cref{eqn:negLL}, where $\rho_{i,0}$ ($\sigma_i$) is the expected value
(error) of $\rho_i$.  The constraints in the fit are as follows:

\begin{itemize}
 \item \isot{Rn}{222} decays are monitored as a function of time and found to
     occur at a rate of 3.65\(\pm\)0.37~\textmu{}Bq/kg in the xenon.  This number
     is used to constrain the following background contributions: 
     \begin{enumerate}
         \item \label{itm:RnActive} Rn in the active xenon,
         \item \label{itm:Bi214} \isot{Bi}{214} on the cathode, and
         \item \label{itm:RnInactive} Rn in the inactive xenon.  
     \end{enumerate}
     \noindent 
     \isot{Rn}{222} and its daughters decay in the bulk xenon, creating
     positive ions which drift toward the cathode. This causes a large fraction
     of the \isot{Bi}{214} decays to occur on the surface of the cathode
     (item~\ref{itm:Bi214}) rather than in the bulk (items~\ref{itm:RnActive}
     and \ref{itm:RnInactive}).  Studies of \isot{Bi}{214} to \isot{Po}{214}
     coincidences in the liquid xenon have shown this drift and demonstrated
     that 83\% of these coincidences occur on the cathode, with the remaining
     17\% in the bulk.  There is $\sim30$~kg of inactive xenon, yielding a
     total activity of $\sim110$~\textmu{}Bq in this part of the detector.  The
     total activity is divided correspondingly between the three types of
     simulated events, translated into number of events and is used as a
     correlated constraint during the fit.  This correlation is relaxed to 90\%
     to account for the estimated systematic error (10\%) on the calculated
     relative fractions of
     items~\ref{itm:RnActive}\textendash{}\ref{itm:RnInactive}. 

 \item The single-site fractions of all components ($\bm{s}$) are independently
     constrained using the 5.9\% error determined in
     \cref{sec:ss_agreement}.  This means, for each PDF the term $0.5
     \left(\left[s_i - s_{i,0}\right]/\left[0.059 s_{i,0}\right]\right)^2$ is added to
     \cref{eqn:negLL}, where $s_{i,0}$ is the single-site fraction given for
     the $i^{\rm th}$ PDF by the Monte Carlo simulation. 

\item The overall normalization, $N$, is allowed to float within constraints
    determined by the total externally estimated systematic errors in
    \cref{sec:sys_errors}.  This adds the term $ 0.5 \left([N-1]/0.0260\right)^2$ to
    \cref{eqn:negLL}.

\end{itemize}

\subsection{Summary of systematic errors}\label{sec:sys_errors}

In addition to the systematic errors arising from event selection cuts
described in \cref{sec:analysis_cuts}, several additional components
contribute systematic errors on the measured number of \twonubb{} decays.

\subsubsection{Simulation model inadequacies}
\label{sec:SimModelInadeq}

As noted in \cref{sec:shape_agreement}, there is an energy-dependent
discrepancy seen between measured and simulated energy distributions of
\thsrc{} and \cosrc{} sources.  This discrepancy was used to produce linear
skewing functions, an
example of which is shown in \cref{fig:shape_ratio}.  To estimate the general
effect of simulation model inadequacies on the \twonubb{} measurement, similar
linear skewing functions were also produced for the \twonubb{} PDF by comparing
the \twonubb{} model generated from simulation with the background-subtracted
low-background spectra produced in \cref{sec:part_rec_cut}.   

The skewing functions were then used to distort the PDFs in the fit model: 
the functions generated for \twonubb{} were used for ``$\beta$"-like PDFs
(\twonubb{}, \nonubb{}, and \isot{Xe}{135}) and the functions for sources
generated in \cref{sec:shape_agreement} were used for the remaining PDFs.  These
PDFs were used to produce $\sim1000$ toy Monte Carlo datasets by using the
expected counts for each PDF given by the low-background fit results.  These
toy Monte Carlo datasets were fit with the default, ``unskewed" PDFs, and the
best-fit \twonubb{} value compared with the true value.  With this study, it
was found that the best-fit \twonubb{} value could be biased -0.33\% with respect to
the true value. A systematic error on the \twonubb{} measurement of $\pm0.33\%$
is added to account for this bias, without applying any correction.  

\subsubsection{DAQ related}
\label{sec:Channel16}
While performing data quality control for \RunTwoA{}, it was determined
that during a portion of the run time (5.2\% of the live-time), a single U-wire
channel was not being read out.  Studies were performed to determine
the effect on the signal efficiency by comparing the default simulated-generated PDFs
with modified PDFs.  These modified PDFs were the combination of the default
PDFs and PDFs generated without the wire channel, added together in proportion
to their respective contribution to the total live-time of the \RunTwoA{}
dataset (i.e.~PDF\textsubscript{mod} = 0.948 PDF\textsubscript{def} + 0.052
PDF\textsubscript{missing chan}).  The shapes of these PDFs were compared using
a Kolmogorov-Smirnov test, finding no significant difference.  In addition, the
efficiencies for the PDFs as calculated in simulation were compared and found
to be consistent within $\sim0.1$\%.  The results of these studies imply that the
resulting systematic error on the \twonubb{} measurement due to this missing
channel is less than 0.1\%.
        
\subsubsection{Energy scale for \texorpdfstring{$\beta$}{beta}-like events}
\label{sec:BetaScale}
As noted in~\cite{Auger:2012ar}, $\beta$-like energy depositions in the
detector have a slightly different energy scale than $\gamma$-like deposits,
and are reconstructed with $\sim1$\% higher energy.  The fit to
the \twonubb{} spectrum yields a different estimate on this scale than from
studies performed using the pair-production peak of \isot{Tl}{208}.  We choose
to treat this difference as a systematic error, which contributes a 0.24\%
error to the \twonubb{} measurement. 

\subsubsection{Incomplete background model} \label{sec:SysIncompleteBackgroundModel}

To test how an incomplete background model may affect the fit, two additional
components (\isot{U}{238} and \isot{Th}{232} chains in the HFE) were added to
the model and the fit was rerun.  These two components were chosen as they
occupy the next largest amount of mass near the detector and the U and Th
chains are generally likely sources of background.  The results of this fit shifted the
best-fit value of \twonubb{} counts $<0.25$\%.  

Additional possibilities were considered, including events arising from muons
not vetoed by the muon veto.  The total expected number of counts from such
events with energy $>$700~keV over the \RunTwoA{} time period is $\lesssim20$,
indicating a $<0.1$\% effect when compared to the number of \twonubb{} counts.

The presence of the \twonubb{} decay of \isot{Xe}{134} was also considered, but
its low Q-value ($\sim820$~keV) would mean only the high-energy tail of the
spectrum would contribute above the 700~keV analysis threshold.  In addition,
cross-checks performed by increasing the analysis threshold and rerunning
the fit found no evidence for additional background (see
\cref{fig:bb2nRateVsThreshold}).  Because of this, any contribution due to
\isot{Xe}{134} was considered to be much less than 0.1\%. 

To accommodate the possibility of an incomplete background model, we include a
systematic error of 0.25\% on the \twonubb{} measurement. 

\subsubsection{Xenon related}
\label{sec:XeRelatedErrs}
The $^{136}$Xe fraction in the $\rm ^{enr}Xe$ has been measured to be
$80.672\pm 0.014$\% using dynamic dual-inlet mass
spectroscopy~\cite{Sever:2013}, where the uncertainty is systematic.  The
abundance of $^{134}$Xe is $19.098\pm 0.014$\% and other isotopes make up for
the remaining fraction of 0.230\%, dominated by 0.203\% of $^{132}$Xe.

The isotope fraction is measured using a sample from the gas introduced into the 
detector at filling.   The composition in liquid phase is expected to be the 
same as in the gas phase to $<0.01$\% based on the difference of
vapor pressures between $^{136}$Xe and $^{130}$Xe~\cite{Jancso:1974}.  The 
conservative figure of 0.01\% is used as a systematic error.   In addition some
 dilution of the $^{136}$Xe content could have occurred because of the natural Xe
used for detector commissioning desorbing over time from the plastic components
inside the TPC.  From solubility arguments this effect is estimated to be
$<0.04$\% and this figure is included as a systematic.

The temperature dependence of LXe density was measured
by~\cite{Terry:1969,Rabin:1970} and corrected to account for the small
difference between $\rm^{natl}Xe$ and $^{136}$Xe.  The stability of the
temperature of the LXe was continuously monitored by thermocouples mounted on
the cryostat during the course of data taking.  A variation of 
0.15~K RMS was observed.  The absolute temperature scale was calibrated in
a test run before the start of the experiment with a reference RTD (then
removed because
of concerns about its radioactive background) yielding $(166.6\pm0.2)$~K, which
translates into a density of $3.0305\pm0.0077~{\rm g/cm}^3$.  This absolute temperature is
consistent with that obtained as a cross-check by measuring the pressure at the
onset of condensation when the detector was filled.  The quadratic sum of the
absolute temperature uncertainty (0.2~K) and the temperature swing observed by
the thermocouples (0.15~K) results in a 0.06\% uncertainty of the LXe density.
The Xe-related systematics, listed above, contribute a $\pm0.26\%$ systematic
error to the \twonubb{} measurement.  

\Cref{tab:syserrs} summarizes
the numbers used as systematic error input to the final fit. 

\begin{table}[ht]
\centering
\begin{ruledtabular}
    \renewcommand{\arraystretch}{1.3}
\begin{tabular}{p{0.25\textwidth}r}
    Component        &   Error (\%) \\ \hline
    Failed event reconstruction (\cref{sec:ClusOverview})         & $<0.18$ \\
    Event Selection  (\cref{sec:analysis_cuts})                   &   2.53    \\
    Shape Distortion (\cref{sec:SimModelInadeq})                  & 0.33      \\ 
    Missing U-wire channel  (\cref{sec:Channel16})                & $<0.1$ \\ 
    Beta-scale  (\cref{sec:BetaScale})                            & 0.24 \\ 
    Background Model  (\cref{sec:SysIncompleteBackgroundModel})   & 0.25 \\ 
    Xe parameters  (\cref{sec:XeRelatedErrs})                     & 0.26 \\ 
    \hline 
    Total      & 2.60 \\

\end{tabular} 
\end{ruledtabular}
\caption{Summary of systematic errors determined via independent studies and
    explicitly included in the normalization term, $N$, in \cref{eqn:negLL}
    (see \cref{sec:FitConstraints}).  The total, which is dominated by the systematic
    error from event selection cuts, is produced by adding in quadrature.
 }
\label{tab:syserrs} 
\end{table}          
        
\subsection{Fit Results}

\begin{figure*}
    \centering
    \includegraphics[width=0.98\textwidth]{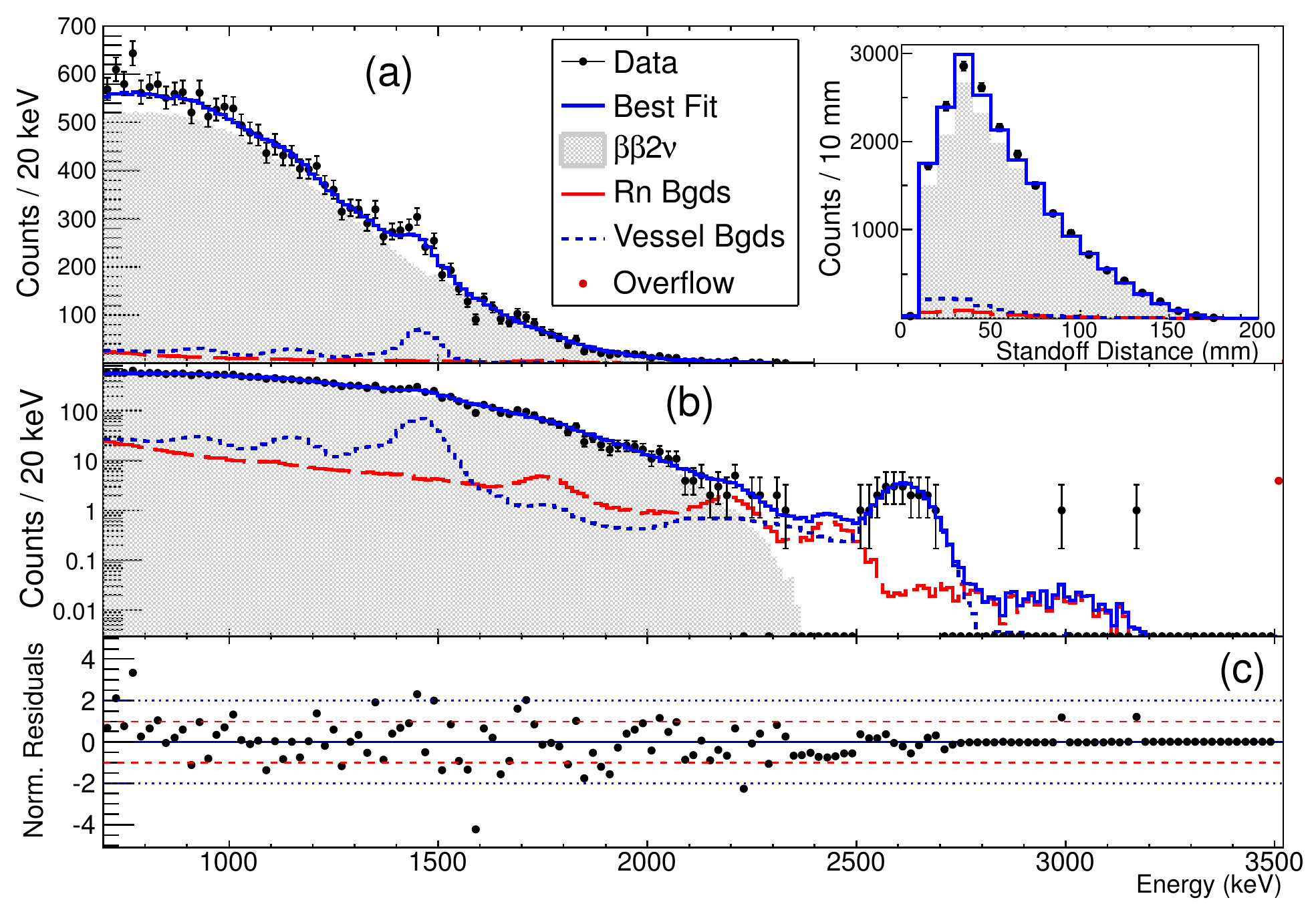}
    \caption{(Color online) Fit results.  Data and PDFs for SS energy spectra
        shown in linear (a), log (b), with residuals (c).  The
        residuals have been normalized by the bin error.  To improve
        visualization of the fit results, the energy bin widths in the plot are
        20~keV instead of the 14~keV bin size used during fitting.  SS standoff
        distribution is also shown (inset).  Backgrounds have been grouped
        together according to Rn components and
        components in or near the TPC vessel.  The best-fit counts and errors
        for each PDF are given in \cref{tab:fit_results}.  There are fewer
        events in the \nonubb{} region-of-interest than in~\cite{Auger:2012ar}
        because of the stricter fiducial volume cut.  }
\label{fig:FinalFit}
\end{figure*}

\begin{figure}
    \centering
    \includegraphics[width=0.47\textwidth]{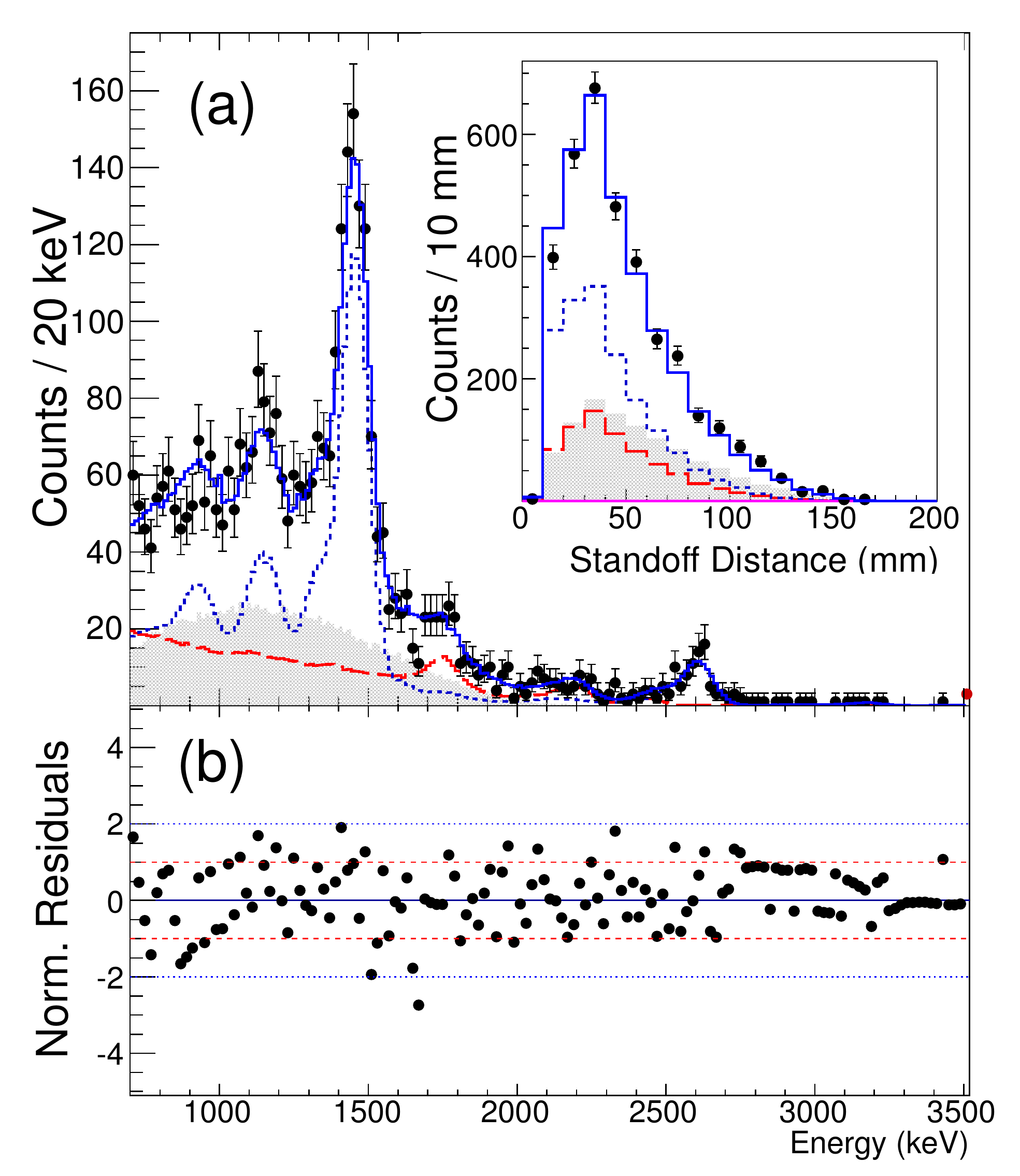}
    \caption{(Color online) Projected MS energy spectra (a) and corresponding MS standoff distance
distribution (inset) for the final fit results.  Residuals are shown in (b).  PDF components are as in
\cref{fig:FinalFit}.  }
\label{fig:FinalFitMS}
\end{figure}

To obtain the best-fit model, all parameters ($\bm{n}$, $\bm{s}$, and
$N$) were allowed to float and $-\ln L$, defined by \cref{eqn:negLL}. was minimized.  The errors on
the main parameter of interest, $n_{2\nu\beta\beta}$, were determined by
performing a profile likelihood scan.  \Cref{fig:FinalFit} shows the results of
this fit for SS events projected onto the energy axes.  The projection onto the
SD axis is provided in the inset.  The corresponding spectra for MS events are
shown in \cref{fig:FinalFitMS}.

The best-fit value for \twonubb{} corresponds to 19042 events above 700~keV.
The total error on this value is estimated by performing a profile likelihood scan,
yielding a $1\sigma$ error of 538 events, which incorporates the systematic
errors from \cref{tab:syserrs}.  The total exposure of \isot{Xe}{136} is
23.14~kg$\cdot$year and the overall detection efficiency (including the energy
spectral cut) for \twonubb{} events is 57.88\%.  This with the molar mass of
135.514~g/mol translates into a \twonubb{} half-life of 

\begin{center}
    \exomeasurement{} 
\end{center}

\begin{figure}[ht]
    \includegraphics[width=0.47\textwidth]{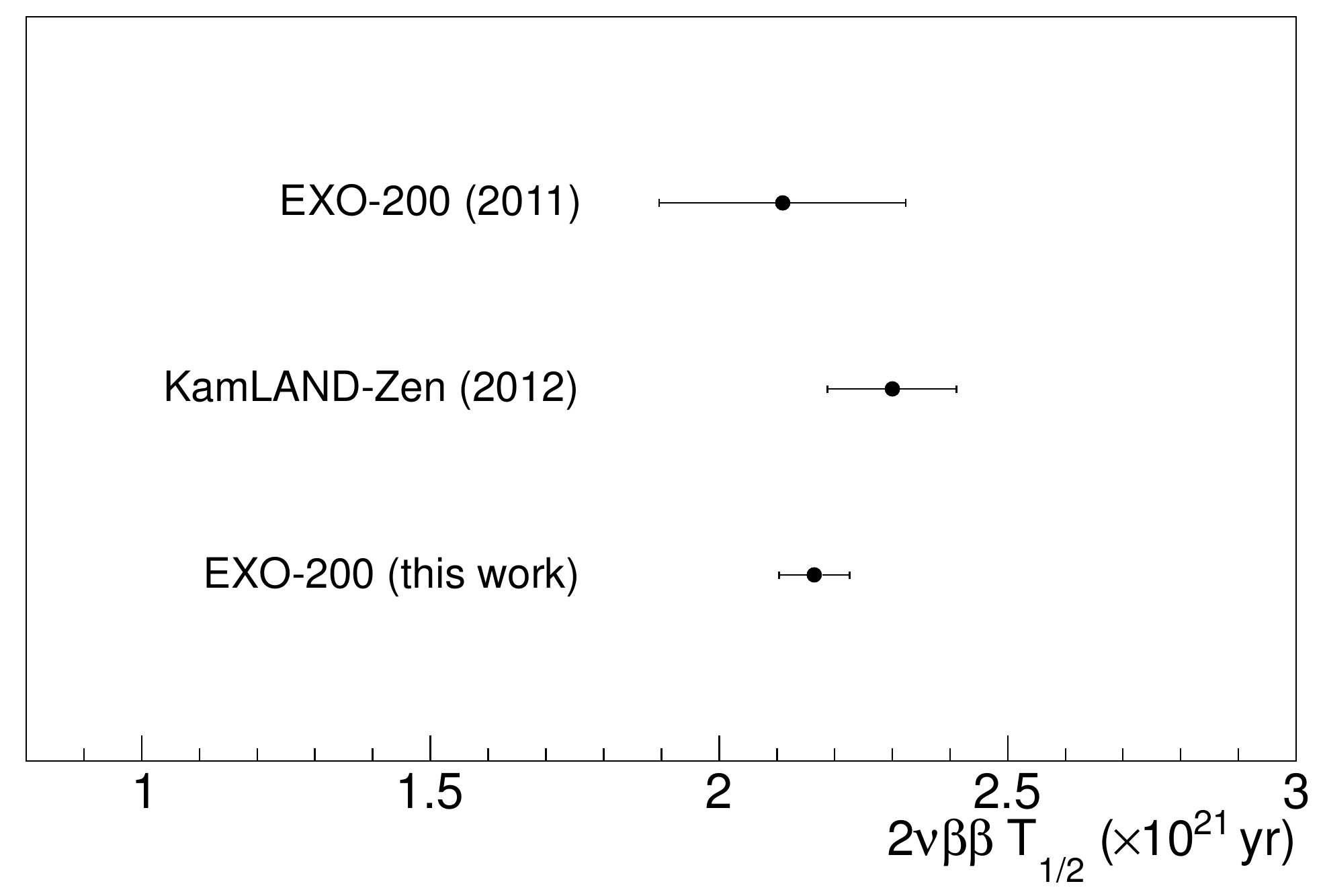} 
    \caption{A comparison of this result with 
            EXO-200 (2011)~\cite{Ackerman:2011gz}, 
            and 
            KamLAND-Zen (2012)~\cite{KamLANDZen:2012zzg}. 
            } 
\label{fig:half-life-comparison}
\end{figure}

The best fit values and associated errors for the counts of the component PDFs
are listed in \cref{tab:fit_results}~\footnote{We plan to present a full study
of the background contributions in an upcoming publication.}.  The breakdown of
the contributions from various error sources on the total error on the
\twonubb{} measurement is given
in \cref{tab:errorbreakout}. Finally, this result is compared to other
measurements of the \twonubb{} half-life of \isot{Xe}{136} in
\cref{fig:half-life-comparison}.

\begin{table}
  \centering
  \renewcommand{\arraystretch}{1.2}
  \begin{ruledtabular}
  \begin{tabular}{lrrD{,}{~\pm~}{-1}} 
  \multicolumn{3}{l}{PDF Type} & \multicolumn{1}{c}{Counts} \\ \hline
  
    \multicolumn{3}{l}{Cu vessel backgrounds} \\
    \multicolumn{3}{r}{\isot{Co}{60}} & 560,70 \\
    \multicolumn{3}{r}{\isot{K}{40}} & 1430,70 \\
    \multicolumn{3}{r}{\isot{Th}{232}} & 590,50 \\
    \multicolumn{3}{r}{\isot{U}{238}} & 90,100 \\
    \multicolumn{3}{r}{\isot{Zn}{65}} & 110,50 \\
    \multicolumn{3}{l}{Rn backgrounds} \\
    &\multicolumn{2}{l}{TPC Cathode} \\
    \multicolumn{3}{r}{\isot{Bi}{214}} & 18,1 \\
    &\multicolumn{2}{l}{Active LXe} \\
    \multicolumn{3}{r}{\isot{Rn}{222}} & 63,4 \\
    &\multicolumn{2}{l}{Air Gap} \\
    \multicolumn{3}{r}{\isot{Bi}{214}} & 1100,200 \\
    &\multicolumn{2}{l}{Inactive LXe} \\
    \multicolumn{3}{r}{\isot{Rn}{222}} & 44,3 \\

  \end{tabular}
  \end{ruledtabular}
  \caption{Summary of fit results for counts of the component PDFs in the fit
      model.  Counts are the total integrated number above 700~keV across
      \emph{both} the SS and MS spectra.  The errors quoted are estimated from
      MIGRAD and are not produced using profile-likelihood scans.
      \isot{Mn}{54} in the Cu vessel, \nonubb{} and \isot{Xe}{135} are omitted
      from this table as their best fit values are consistent with 0.  The
      division of background components is as given in 
      \cref{fig:FinalFit,fig:FinalFitMS}.}
  \label{tab:fit_results}
\end{table}

\begin{table}[ht]
\centering
\begin{ruledtabular}
    \renewcommand{\arraystretch}{1.3}
\begin{tabular}{p{0.25\textwidth}r}
    Component        &   Error (\%) \\ \hline
    Systematic errors from \cref{tab:syserrs} & 2.60 \\
    SS/(SS+MS) Fraction    &   0.77 \\ 
    Backgrounds   & 1.3 \\
    Statistical   & 0.76 \\
    \hline
    Total            & 2.83 \\
\end{tabular} 
\end{ruledtabular}
\caption{Summary of estimates of contributions to the final total error on
    \twonubb{} due to various components.  Note that this includes errors
    explicitly included in the normalization term (from \cref{tab:syserrs}) in
    addition to errors arising from other nuisance parameters (e.g.~single-site
    fraction).  The total error is taken from the profile-likelihood scan (PLL)
    and is \emph{not} a simple sum in quadrature as the components are
    correlated.  To estimate the statistical error, all nuisance parameters
    except the \twonubb{} counts are fixed to their best-fit values and the PLL
    rerun.   To estimate the error component due to other sets of nuisance
    parameters, the following procedure is followed: after the best-fit
    parameters are found, the PLL is regenerated after fixing the relevant
    nuisance parameters to their best-fit values.  The reduction of width in
    the resulting PLL from the original PLL yields the estimate on the error
    due to the particular component(s). 
    }

\label{tab:errorbreakout} 
\end{table}          
 
The cumulative \chisq{} for the projected energy and standoff distance
distributions shown in \cref{fig:FinalFit,fig:FinalFitMS} are, respectively,
100.1 and 22.8 (94.0 and 23.2) for SS (MS) events.  Since the best-fit
parameters
were determined using a ML fit, it is not possible to directly calculate the
degrees of freedom (NDF) from the number of parameters and bins (see
e.g.~\cite{Chernoff:1954}).  Instead, the NDFs for each of the 4 projection
distributions were estimated using toy Monte Carlo simulations, performed by
generating a toy data set from the best-fit parameters and rerunning the fit
5000 times.  This study estimates the NDF for energy and standoff distributions
to be, respectively, 113.7 and 17.2 (106.8 and 13.6) for SS and MS events,
yielding a reduced \chisq{} of 0.88 and 1.33 (0.88 and 1.70).

A typical quantity to compare results between experiments is the
observed signal-to-background ratio (SBR).  Because the ML fit is performed with 
2 observables (energy and standoff distance) and across 2 classes of events
(single-site and multi-site), it is difficult to define one number quantifying
this ratio.  However, the \twonubb{} signal is observed to be 95\%~SS,
consistent within the estimated systematic errors with the 98\%~SS predicted by
the EXO-200 simulation, and so we may consider the SBR quantity purely in this
class of events.  The average SBR over the SS spectra is roughly 11.  This
quantity increases to 16 (19) as one selects the inner 60\% (40\%) of the
fiducial volume, demonstrating the self-shielding of the xenon in addition to
the power of fitting over the standoff distance observable.

\subsection{Final cross-checks}

A series of cross-checks were performed on the fit result.  The rate of
\twonubb{} was binned versus time and the fit repeated for each time bin.  The results were  
found to be consistent with a constant rate.  In addition, an energy-only
(without SD) fit was performed, producing a best-fit \twonubb{} counts value
3.0\% less than the reported result.  The corresponding contribution from the
backgrounds on the total error increased slightly from 1.3\%
(\cref{tab:errorbreakout}) to 1.35\% for the energy-only fit.

In addition to the studies performed in
\cref{sec:SysIncompleteBackgroundModel}, further investigations were undertaken
to test the possibility that an unknown or unconsidered background is
affecting the results of the fit.  It is important to note that the measured
goodness-of-fit is already an indication that the chosen fit model 
describes the data well.  This suggests that, for an unknown background to affect
the \twonubb{} measurement, it would need to exhibit an energy spectrum and standoff distance
distribution similar to \twonubb{} decay.  As in~\cite{Ackerman:2011gz} we consider
two candidates satisfying these requirements, \isot{Y}{90} and \isot{Re}{188},
supported by \isot{Sr}{90} and \isot{W}{188}, which have half-lives of
28.90~yr and 69.78~d, respectively.  It is important to note that the
presence of these isotopes in the LXe is considered \emph{a priori} unlikely as
no indication of more common contaminants (e.g.~metallic components from the
U and Th chains) has been seen and the LXe is being continuously purified. 

A ML fit to the \RunTwoA{} dataset with an added time dimension was performed,
including a PDF from \isot{Re}{188} with an exponentially decaying time
component corresponding to the \isot{W}{188} half-life.  The results of this
fit found the number of counts of \isot{Re}{188} to be consistent with zero and
produced a best-fit value of \twonubb{} within 0.8\% of the quoted value.  To
investigate any effect due to \isot{Y}{90}, a ML fit was performed by adding a
\isot{Y}{90} PDF to the standard set of PDFs.  The results of this fit produced
a best-fit value of \twonubb{} within 3\% of the quoted value.

\begin{figure}[ht]
    \includegraphics[width=0.47\textwidth]{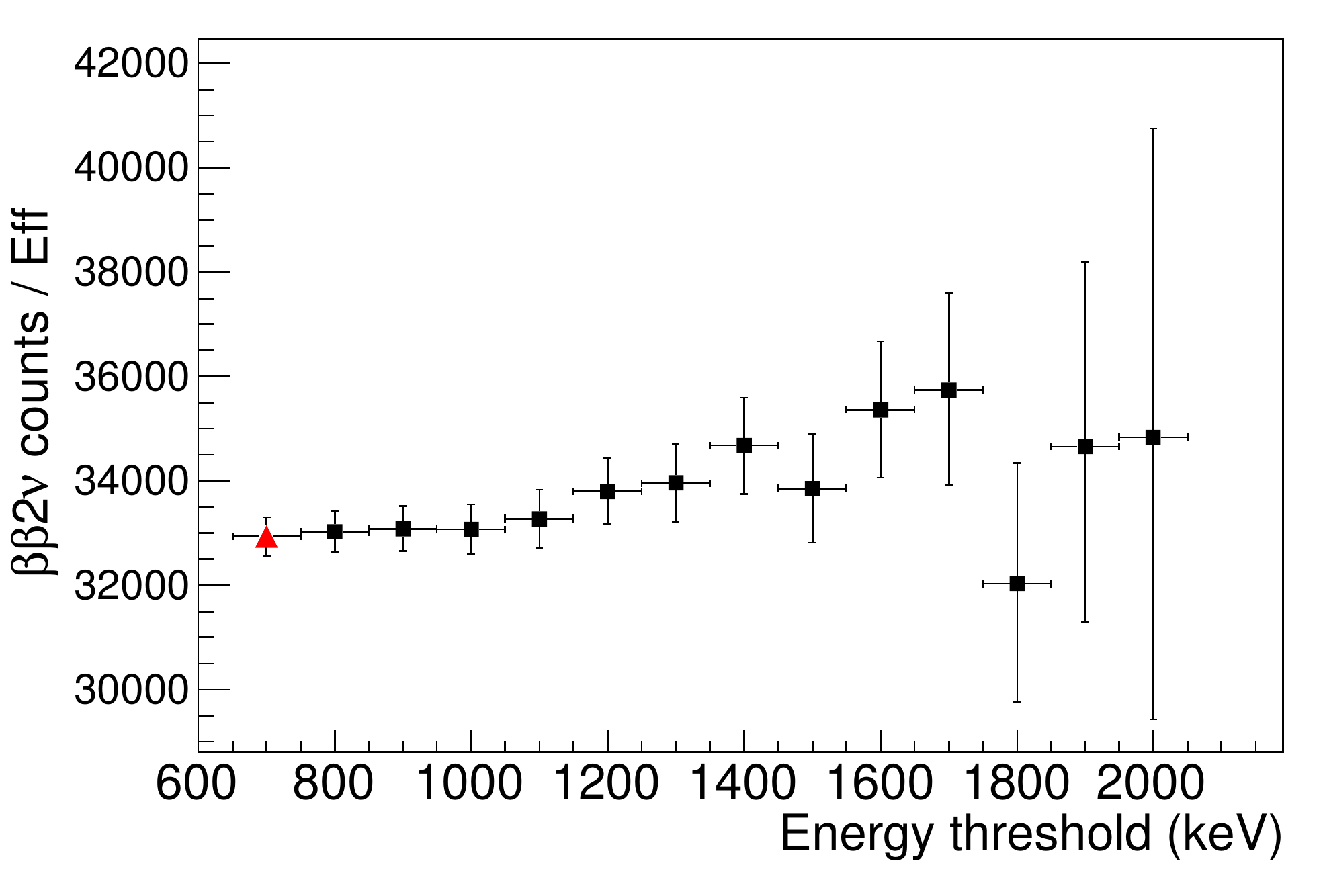} \caption{(Color
        online) The fit count rate divided by efficiency of \twonubb{} versus
        energy threshold. The main result is the (red, online) triangle at
    700 keV.}  
\label{fig:bb2nRateVsThreshold}
\end{figure}

A final cross-check consisted in performing the fit with increasing energy thresholds.
The purpose of this cross-check is to investigate the possible presence of
other unexpected backgrounds from $\beta$-decays under the \twonubb{} spectrum.
Increasing the energy threshold would change the relative contribution of any
potential background, which would manifest as a change in the fit number of
\twonubb{} events.  The results of this study are shown in
\cref{fig:bb2nRateVsThreshold}, demonstrating that the \twonubb{} measured rate
is stable under even large changes of threshold. 

\section{Conclusions}

\begin{table*}[htb]
\renewcommand{\arraystretch}{1.3}
\begin{ruledtabular}
\begin{tabular}{ l c c c c c l} 
Nuclide       & \multicolumn{1}{c}{${\rm T}_{1/2}^{2\nu\beta\beta}\pm {\rm stat} \pm {\rm sys}$ } & \multicolumn{1}{p{0.11\textwidth}}{rel.\ uncert.}  
          & \psfac{} & \multicolumn{1}{c}{\matel{}} & rel.\ uncert. & Experiment (year) \\ 
              & \multicolumn{1}{c}{[y]} & \multicolumn{1}{c}{[\%]}  
          & \multicolumn{1}{c}{$[10^{-21}$ y$^{-1}]$} & \multicolumn{1}{c}{[MeV$^{-1}$]} & \multicolumn{1}{c}{[\%]} &  \\ \hline
$^{136}$Xe    & \exomeasval[false]{}                                 & $\pm 2.83$          & 1433  & 0.0218 & $\pm 1.4$            & EXO-200 (this work)       \\
$^{76}$Ge     & $ 1.84^{+0.09+0.11}_{-0.08-0.06}\cdot 10^{21} $      & $^{+7.7}_{-5.4}$    & 48.17 & 0.129  & $^{+3.9}_{-2.8}$     & GERDA~\cite{Agostini:2012nm} (2013)    \\
$^{130}$Te    & $ 7.0\pm 0.9\pm 1.1\cdot 10^{20}              $      & $\pm 20.3$          & 1529  & 0.0371 & $\pm 10.2$           & NEMO-3~\cite{Arnold:2011gq} (2011) \\
$^{116}$Cd    & $ 2.8\pm 0.1\pm 0.3\cdot 10^{19}              $      & $\pm 11.3$          & 2764  & 0.138  & $\pm 5.7$            & NEMO-3~\cite{Barabash:2010bd} (2010) \\
$^{48}$Ca     & $ 4.4^{+0.5}_{-0.4}\pm 0.4\cdot 10^{19}       $      & $^{+14.6}_{-12.9}$  & 15550 & 0.0464 & $^{+7.3}_{-6.4}$     & NEMO-3~\cite{Barabash:2010bd} (2010) \\
$^{96}$Zr     & $ 2.35\pm 0.14\pm 0.16\cdot 10^{19}           $      & $\pm 9.1$           & 6816  & 0.0959 & $\pm 4.5$            & NEMO-3~\cite{Argyriades:2009ph}(2010) \\
$^{150}$Nd    & $ 9.11^{+0.25}_{-0.22}\pm 0.63\cdot 10^{18}   $      & $^{+7.4}_{-7.3}$    & 36430 & 0.0666 & $^{+3.7}_{-3.7}$     & NEMO-3~\cite{Argyriades:2008pr}(2009) \\
$^{100}$Mo    & $ 7.11\pm 0.02\pm 0.54\cdot 10^{18}           $      & $\pm 7.6$           & 3308  &  0.250 & $\pm 3.8$            & NEMO-3~\cite{Arnold:2005rz}(2005) \\
$^{82}$Se     & $ 9.6\pm 0.3\pm 1.0\cdot 10^{19}              $      & $\pm 10.9$          & 1596  & 0.0980 & $\pm 5.4$            & NEMO-3~\cite{Arnold:2005rz}(2005) \\
\end{tabular}
\end{ruledtabular}
\caption{Listing of the most precise measurements of
\twonubb{} half-lives for each isotope as reported in the literature. Only direct counting
experiments and ground state decays are shown here.  The results are listed
chronologically by year of publication.  Also included are phase space factors
(\psfac{}, from~\cite{Kotila:2012zza}) and nuclear matrix elements (\matel{})
as defined by \cref{eqn:halflifetwonu}.  The total relative uncertainty on the
half-life is the quadratic sum of statistic and systematic errors, as given in
the cited publications, divided by the half-life.  The uncertainty in \matel{}
is derived from the experimental uncertainty on T$_{1/2}^{2\nu}$, under the
assumption that the uncertainties in \psfac{} and $g_A$ are negligible.  In addition, these
errors are determined to be symmetric or asymmetric following the same
conventions used in assigning the half-life errors.}
\label{tab:2nbb_listing}
\end{table*}

We have reported on an improved measurement of the \twonubb{} decay of
\isot{Xe}{136} using 127.6~days of live-time collected between Sept 2011 and
April 2012.  The resulting half-life \exomeasurement{} is measured with a
total relative uncertainty of 2.83\% and is dominated by systematic
uncertainties.  This half-life corresponds to a nuclear matrix element of \matel{}
$= 0.0218 \pm 0.0003~{\rm MeV}^{-1}$, the smallest among the
isotopes measured to date.  For comparison, in \cref{tab:2nbb_listing} we have
tabulated the most precise half-life measurement of all nine nuclei for which
\twonubb{} decay has been directly observed.  The corresponding matrix elements
\matel{} are included in this table as well.  We note that \isot{Xe}{136} has
both the longest \twonubb{} half-life of any such decay, and, as reported in
this article, the most precise measurement.

We have described in some detail the data analysis methods used for this
measurement.  These methods are similar to those
employed to search for the \nonubb{} decay that, if observed, would indicate the discovery of
new physics beyond the Standard Model.  While the level of precision achieved here is not required 
for the \nonubb{} decay search, it demonstrates the
quality of the EXO-200 data and the power of a fully active, high resolution
tracking detector with very low background.

Since April 2012 EXO-200 has accumulated an
exposure several times larger than that described here. We expect to report the 
results of a new search for \nonubb{} based on this larger dataset in the near future.

\begin{acknowledgments}
    EXO-200 is supported by DOE and NSF in the United States, NSERC in Canada,
SNF in Switzerland, NRF in Korea, RFBR (12-02-12145) in Russia and DFG Cluster of Excellence
``Universe'' in Germany. EXO-200 data analysis and simulation uses resources of
the National Energy Research Scientific Computing Center (NERSC), which is
supported by the Office of Science of the U.S. Department of Energy under
Contract No.~DE-AC02-05CH11231. The collaboration gratefully acknowledges the
WIPP for the hospitality and Dr.\ J.P.\ Severinghaus of the Scripps Institution
of Oceanography for the accurate measurement of the isotopic composition of the
EXO-200 enriched xenon.

\end{acknowledgments}

\bibliography{exo_2nubb_analysis}

\appendix
\section{Reconstruction Clustering Distribution Functions}\label{sec:ClusPDFs}
Several PDFs are used when matching together U- and V-signal bundles and are
described in the following. 

\begin{figure}[ht]
  \centering
  \includegraphics[width=0.48\textwidth]{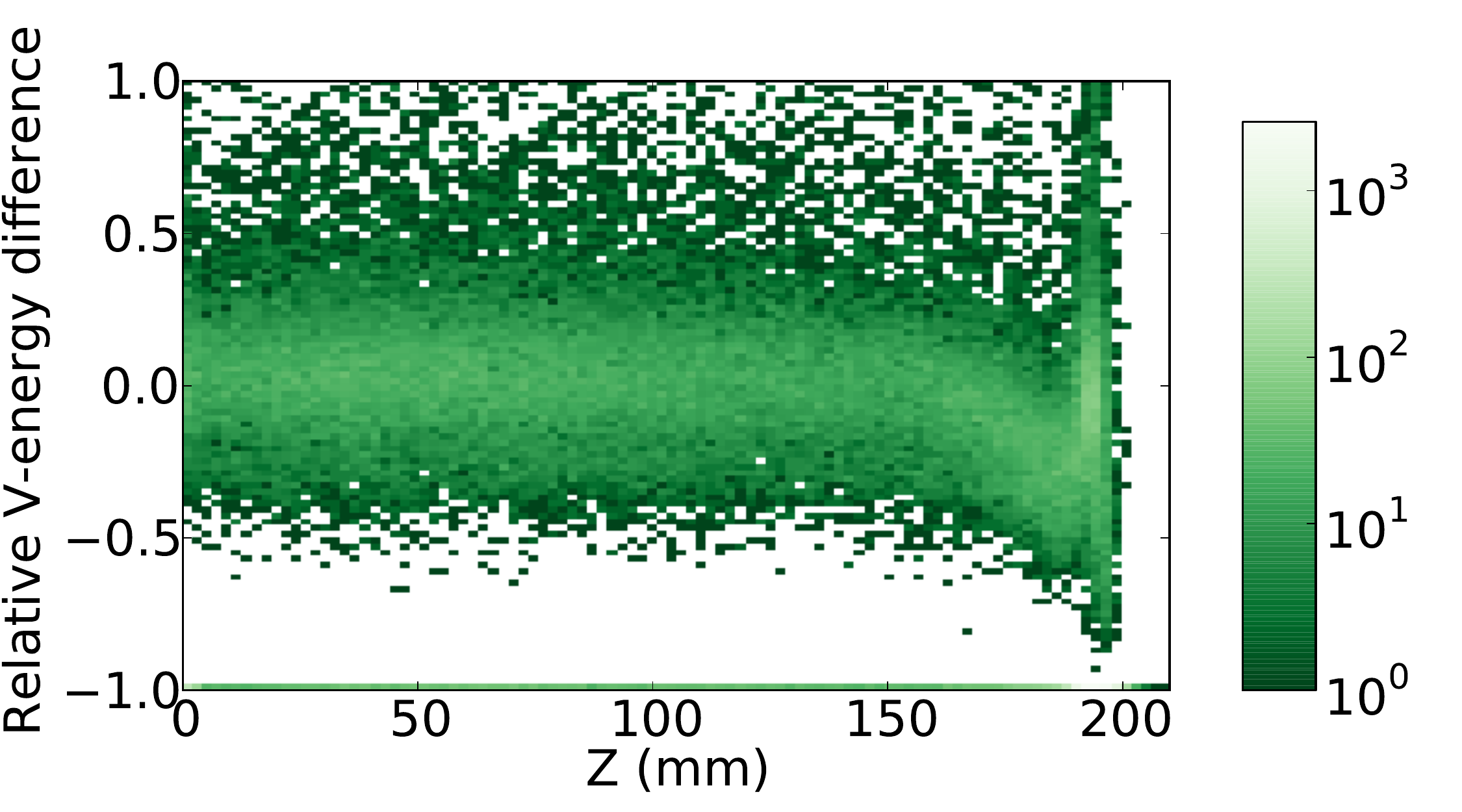}
  \caption{(Color online) Total $(E_\V-\rho_E(E_\U))/\rho_E(E_\U)$ for an event, data from Th
      source runs as well as low-background runs. There is a strong Z
      dependence for $|\Z| > 160$~mm observed, and so the energy relationship
      between U- and V-bundles is ignored for bundles in this region. 
  }
  \label{fig:UVenergyVsZ}
\end{figure}

\subsection{Energy}
The amplitude of the induced signals is directly proportional to the amplitude
of the deposition signals, and so this correlation may be used when combining
U- and V-signals.  The energy PDF, estimated from fits to source data,
quantifies this relationship:

\begin{multline}
  f_{\rm energy}(E_\U, E_\V, \Z) = \\ 
    \begin{cases}
      A       & |\Z| > 160~\text{mm}\\
      A \exp \left( -0.5 \left( \frac{\rho_E(E_\U) - E_\V}{\sigma_E(E_\U)} \right) \right)
                      & |\Z| \leq 160~\text{mm}
    \end{cases}   
  \label{eqn:ClusEnergyPDF}
\end{multline}

where $A$ is a normalization constant, , $E_{\U,\V}$ are the amplitudes of the U-
and V-bundles, respectively, in units of gain-corrected ADC counts
(i.e.~\uadc, \vadc)\footnote{The \uadc unit value is very similar to keV,
however, the data are not fully calibrated at this point of processing.}.  $\rho_E$ is the
expected value of $E_\V$ given a particular $E_\U$ and is defined as: 

\begin{equation}
  \rho_E(E_\U) = 
    \begin{cases}
        0~\text{\vadc} & E_\U < -b_E/m_E \\ 
      E_\U m_E + b_E    & E_\U \geq -b_E/m_E \\ 
    \end{cases}   
    \label{eqn:UVsV}
\end{equation}

with the constants $b_E$ and $m_E$ measured as -30.79~\vadc/\uadc\ and
0.2378~\vadc.  $\sigma_E$ from \cref{eqn:ClusEnergyPDF} is the expected
spread in $E_\V$ given a particular $E_\U$ and is defined as:

\begin{equation}
\sigma_E(E_\U) = 
  \begin{cases}
    a_E       & E_\U < 350~\text{\uadc}\\ 
    c_E E_\U + d_E \sqrt{E_\U}  & E_\U \geq 350~\text{\uadc}\\ 
  \end{cases}   
\end{equation}

$a_E$, $c_E$ and $d_E$ are constants with values of 20.22~\vadc, 0.0101~\vadc/\uadc\ and
0.892~\vadc/\uadc$^{1/2}$, respectively.  All values of the constants quoted
above are extracted from fits to calibration source data.
\Cref{fig:UVenergyVsZ} shows a comparison using calibration data of the
expected value of $E_\V$ (\cref{eqn:UVsV}) with
the true value of $E_\V$. 

At $|\Z| > 160$~mm, effects from the anode begin to distort the reconstructed
amplitudes of V-wires.  This comes from the fact that the V-wire shapes change
somewhat when charge is deposited closer to the anode and the method for
extracting the V-wire amplitudes (fitting with a signal model) uses a template
from charge deposits in the bulk. 

\subsection{Time}

\begin{figure}[ht]
  \centering
  \includegraphics[width=0.47\textwidth]{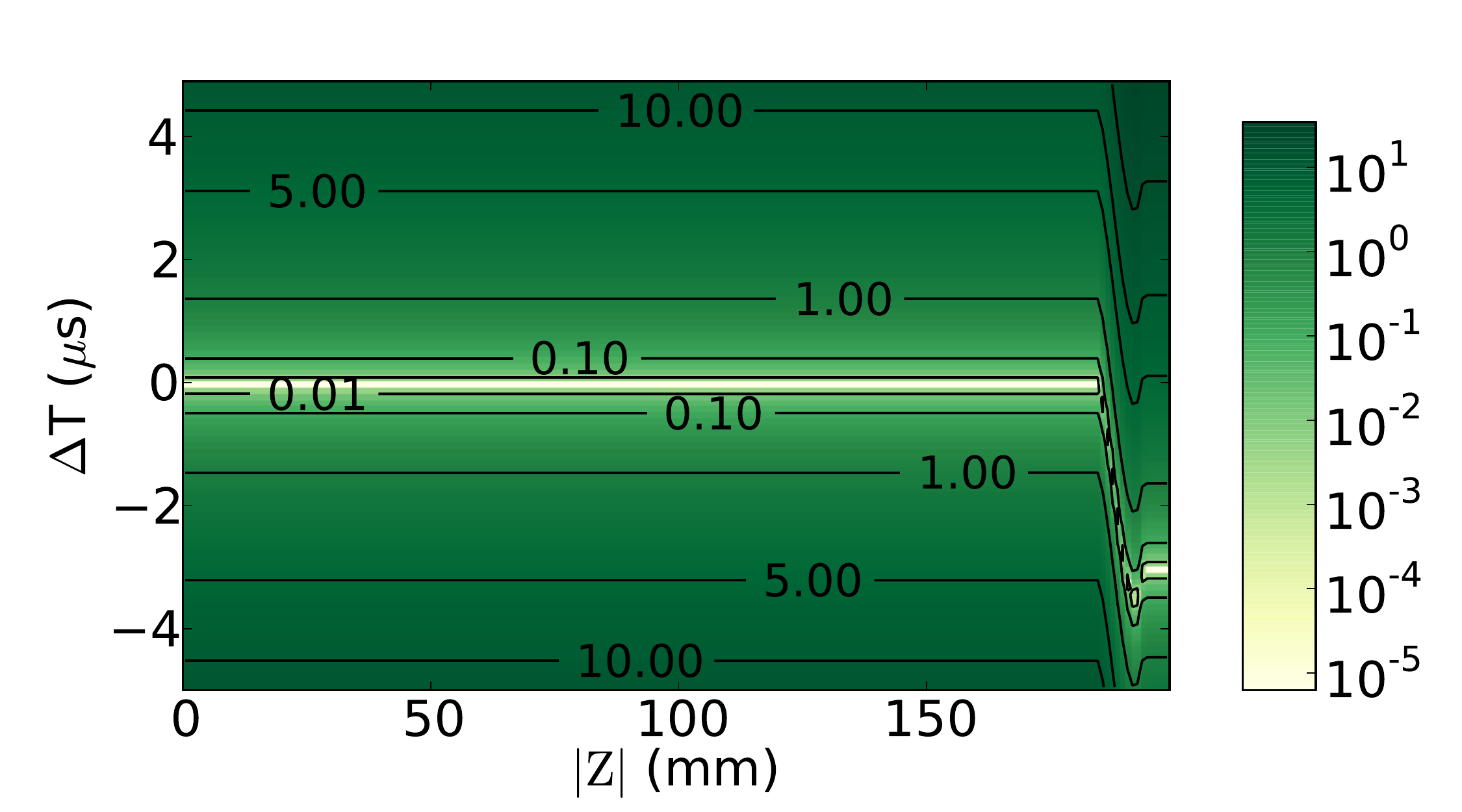}
  \caption{(Color online) Visualization of time (\cref{eqn:TimePDF}) PDF used to cluster
signal bundles together.  The Z-direction (color, contour lines) of the plot is
the negative log of the indicated PDF: a lower value denotes a higher
probability with 0 (white) being the highest probable state.  
  } 
    \label{fig:UVtimePdf}
  
\end{figure}

The relationship between the arrival times of U- and V-bundles has almost no
dependence on the Z-position of the bundles throughout most of the TPC, but
yields a Z-dependence as the bundles near the anode.  This is a relic of
reconstruction of V-wires near the anode.  The time PDF is a Gaussian with a
Z-dependent mean:

\begin{multline}
  f_{\rm time}(t_\U, t_\V, \Z) = \\
    B \exp \left( -0.5 \left( \frac{t_\U - t_\V -
    \rho_t(\Z)}{\sigma_{\rm time}} \right)^2 \right) 
    \label{eqn:TimePDF}
\end{multline}

where $B$ is a normalization constant, $t_{\U,\V}$ are the times in \mus{} of the
U and V bundles, respectively, $\sigma_{\rm time}$ is a constant (1~\mus{}), and
$\rho(\Z)$ is given by:

\begin{multline}
  \rho_t(\Z) = \\ 
  \begin{cases}
    0~\text{\mus{}}             & \frac{|\Z|}{\rm mm} \leq 185.2\\
    \rho_0 + \rho_1 \Z' + \rho_2 {\Z'}^2 - \rho_3 {\Z'}^3
                              & 185.2 < \frac{|\Z|}{\rm mm} \leq 194.1\\
    3~\text{\mus{}}             & 194.1 < \frac{|\Z|}{\rm mm} \\
  \end{cases}
\end{multline}
where $\Z' = |\Z| - 190$~mm, to accommodate the symmetry of the two TPCs and
$\rho_0$, $\rho_1$, $\rho_2$, and $\rho_3$ are 2.73~\mus{},
0.55~\mus{}~mm$^{-1}$, -0.065~\mus{}~mm$^{-2}$, and -0.013~\mus{}~mm$^{-3}$
respectively.   A visualization of the time PDF is shown in
\cref{fig:UVtimePdf}.

\begin{figure}[ht]
  \centering
  \includegraphics[width=0.47\textwidth]{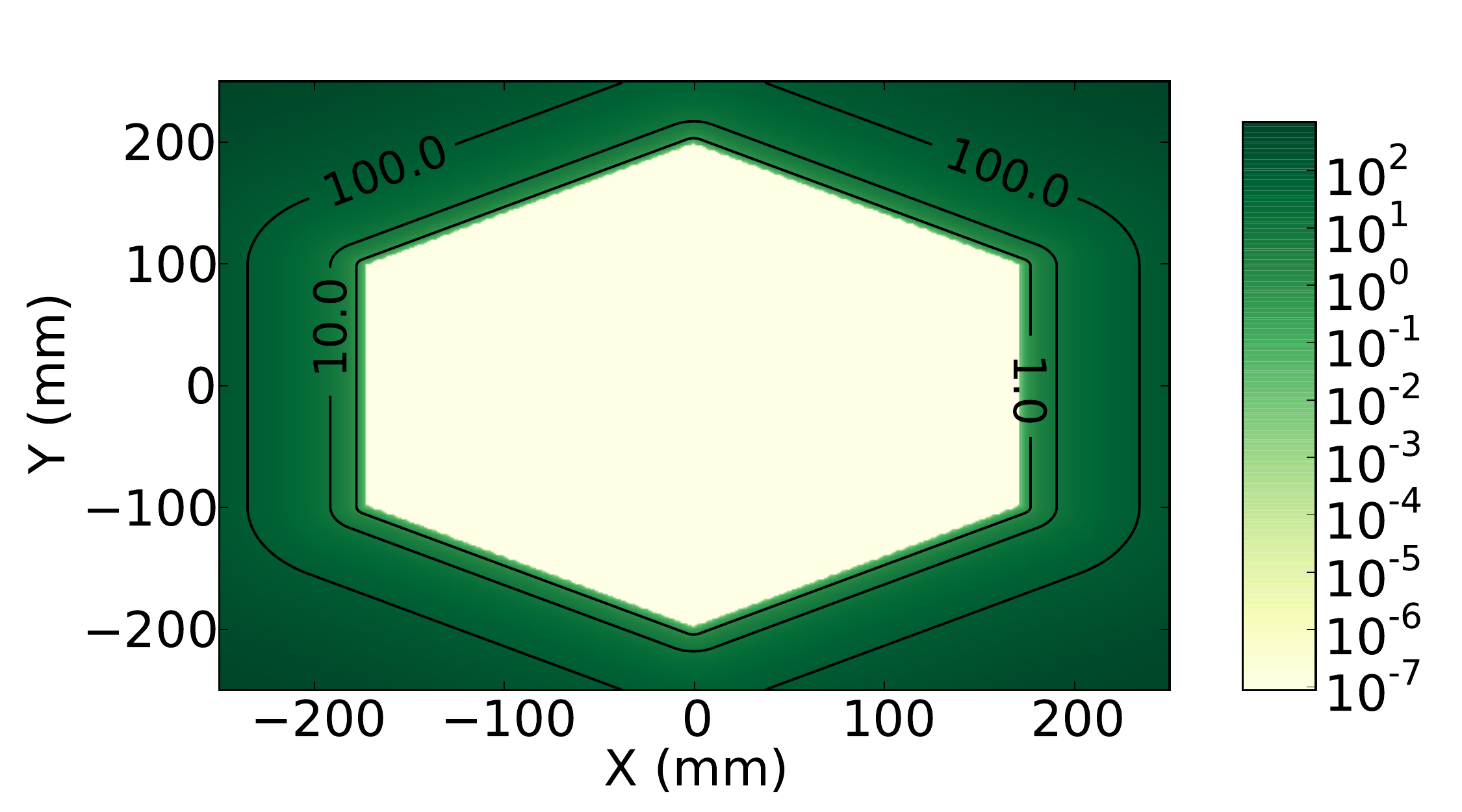}
  \caption{(Color online) As in \cref{fig:UVtimePdf} but for the position PDF
           (\cref{eqn:UVposPdf}).  } 
  \label{fig:PosPdf}
  
\end{figure}

\subsection{U,V position}
To ensure that only physically possible connections between U- and V-bundles
are created, a regular hexagon is used as position PDF. The side-to-side
diameter of the hexagon, defined by the detector geometry, is 342~mm.  The PDF
is given by:

\begin{multline}
  f_{\rm UV pos}(\U, \V) = \\ 
  \begin{cases}
    C   & \U,\V~\text{inside hexagon}\\
    C \exp \left(-0.5 \left(\frac{x_{\rm perp}}{\sigma_{\U\V}} \right)^2\right) 
        & \U,\V~\text{outside hexagon}
  \end{cases}
  \label{eqn:UVposPdf}
\end{multline}

where $C$ is a normalization constant, $x_{\rm perp}$ is the nearest distance to a
hexagon side, and $\sigma_{\U\V} = l / 2 = 4.5$~mm, where $l$ is the width of a
single wire channel (9~mm). See \cref{fig:PosPdf} for a plot of the
position PDF.

\end{document}